\RequirePackage{fix-cm}
\documentclass{svjour3}       %
\smartqed  
\usepackage{graphicx}

\usepackage{amssymb}

\usepackage{lipsum}
\usepackage{natbib}
\usepackage{lineno}
\usepackage{geometry}
\usepackage{hyperref}
\usepackage{subcaption}
\usepackage{enumitem,kantlipsum} 
\usepackage{booktabs} 
\usepackage{commath}
\usepackage{amsmath}
\usepackage{multirow}
\usepackage{soul}
\usepackage[ruled,vlined,noresetcount]{algorithm2e}

\SetCommentSty{mycommfont}
\usepackage{comment}
\usepackage{tikz}
\usepackage{tabularx}
\usepackage{colortbl}

\setstcolor{red}

\renewcommand{\vec}[1]{\mathbf{#1}}
\newcommand{\compactpara}[1]{\paragraph{\textnormal{\textbf{#1.}}}}
\newcommand{\para}[1]{\subsection*{\textnormal{\textbf{#1}}}}

\newcommand{\topM}{M(\theta_U,q_U,l_U)}

\newcommand\tabfitpagew{\resizebox{1.00\textwidth}{!}}

\newcommand{\hlt}[1]{\textnormal{#1}}
\newcommand{\hltd}[1]{\textnormal{#1}}

\newcommand{\hltIRJ}[1]{\textnormal{#1}}

\renewcommand{\cite}[1]{\citep{#1}}

\usepackage{etoolbox}
\newcommand*{\affaddr}[1]{#1} 
\newcommand*{\affmark}[1][*]{\textsuperscript{#1}}
\usepackage[misc]{ifsym}

\begin{document}

\title{Kernel Density Estimation based Factored Relevance Model for Multi-Contextual Point-of-Interest Recommendation
}

\titlerunning{KDEFRLM for Multi-Contextual POI Recommendation}        

\author{Anirban Chakraborty\affmark[1]         \and
        Debasis Ganguly\affmark[2]             \and
        Annalina Caputo\affmark[3]             \and
        Gareth J. F. Jones\affmark[3]
}

\authorrunning{A. Chakraborty et al.} 

\institute{\Letter \ \ \ \ Anirban Chakraborty \at
            \href{mailto:anirban.chakraborty@adaptcentre.ie}{anirban.chakraborty@adaptcentre.ie}
            \and
            Debasis Ganguly \at
            \href{mailto:debasis.ganguly@glasgow.ac.uk}{debasis.ganguly@glasgow.ac.uk}
            \and
            Annalina Caputo \at
            \href{mailto:annalina.caputo@adaptcentre.ie}{annalina.caputo@adaptcentre.ie}
            \and
            Gareth J. F. Jones \at
            \href{mailto:gareth.jones@dcu.ie}{gareth.jones@dcu.ie}
            \and
            \affaddr{\affmark[1] ADAPT Centre, School of Computer Science \& Statistics, Trinity College Dublin, Dublin, Ireland}\\
            \affaddr{\affmark[2] School of Computing, University of Glasgow, Glasgow, United Kingdom}\\
            \affaddr{\affmark[3] ADAPT Centre, School of Computing, Dublin City University, Dublin, Ireland}
}

\date{Received: date / Accepted: date}

\maketitle

\begin{abstract}

An automated contextual suggestion algorithm is likely to recommend contextually appropriate and personalized `points-of-interest' (POIs) to a user, if it can extract information from the user's preference history (\emph{exploitation}) and effectively blend it with the user's current contextual information (\emph{exploration}) to predict a POI's `appropriateness' in the current context. To balance this trade-off between exploitation and exploration, we propose an unsupervised, generic framework involving a factored relevance model (FRLM), constituting two distinct components, one pertaining to historical contexts, and the other corresponding to the current context. We further generalize the proposed FRLM by incorporating the semantic relationships between terms in POI descriptors using kernel density estimation (KDE) on embedded word vectors. Additionally, we show that trip-qualifiers, (e.g. `trip-type', `accompanied-by') are potentially useful information sources that could be used to improve the recommendation effectiveness. Using such information is not straightforward since users' texts/reviews of visited POIs typically do not explicitly contain such annotations. We undertake a weakly supervised approach to predict the associations between the review-texts in a user profile and the likely trip contexts. Our experiments, conducted on the TREC Contextual Suggestion 2016 dataset, demonstrate that factorization, KDE-based generalizations, and trip-qualifier enriched contexts of the relevance model improve POI recommendation.
\end{abstract}

\section{Introduction}

Finding value in the enormous volumes of online data increasingly requires the use of effective methods for contextually relevant recommendations. For example, recommending movies, articles to read, places to visit.
The 
definition of \emph{contextual recommendation} depends 
on 
the precise definition of 
\emph{context} which is being used.
Generally speaking, it can be argued that the more fine-grained the definition of the context is to a system, the greater 
is its potential for providing more personally relevant information to the user at specific points in time, specifically focused and tailored to their \emph{context} \cite{Arampatzis_ToIS2017, yu2015survey, modelingUPref_usingEmbedding_Jarana_Craig_Iadh_2016}.

To illustrate this 
point that systems addressing 
multiple 
fine-grained contexts can 
potentially be more beneficial to users, imagine two systems $A$ and $B$, where the former only keeps track of a user's geographic location, whereas the latter additionally keeps track of other qualifiers associated with the location, e.g., the specific purpose of the user to visiting that location, whether the user is alone while visiting the place or she is with her friends or family, the season, day or hour of the visit, etc.
It can be hypothesized that 
this example that this system $B$, in comparison to system $A$, could potentially provide more selective and relevant recommendations to its user about places to visit, and activities to do. 
This is because system $B$ might 
provide more reliably selected recommendations by reasoning using locational context information. For example, by reasoning that museums would not be the best place to recommend if the purpose of the user's current trip is business.
Without this locational information, 
it would be
rather difficult for system $A$ to exclude such non-relevant suggestions
because of the lack of adequately informative context.



In addition to 
context, the other source of useful information for contextual recommendation is the user's own personal \emph{history} or \emph{activity log} \cite{modelingUPref_usingEmbedding_Jarana_Craig_Iadh_2016, Liu_GepgraphicalPrefs_for_POIRecSys_KDD2013}. The rationale for using the personal historical information of the user is based on the assumption that user feedback (in the form of ratings or positive/negative comments) may help to capture her 
preferences. Consider, for example, if the user is particularly fond of live music 
(e.g., she has in the recent past favoured pubs offering live music over the ones which do not, and has also rated them positively), it is likely that suggesting a pub with live music in a new location could also be relevant to her.      
%
Specifically, a contextual recommender system could attempt to \emph{match} a user's past preferences in other contexts (e.g. locations) with the top rated points-of-interests (POIs) of the current context to suggest potentially relevant ones \cite{contentBasedRecSys_RNN_UMAP2017}.

From a general perspective, we consider that there are two broad distinct sources of information (or contexts), that a contextual recommendation system can benefit from. The first of these describes the \emph{present state} of the user at an instant of time, which is typically a combination of features with categorical values \cite{NextPOI_timeDayWeek_TOIS2019}, e.g., the location of the user (one out of a finitely many cities on Earth), purpose of a trip (e.g. leisure vs. work), current season (e.g., summer, fall, winter or spring), etc.  
The second source of information is the \emph{past state} of the user, which, acquired over a sufficient period of time, is likely to broadly capture her general preferences in particular situations. In other words, past information provides information about a user's general preferences for certain types of items over others \cite{modelingUPref_usingEmbedding_Jarana_Craig_Iadh_2016, Liu_GepgraphicalPrefs_for_POIRecSys_KDD2013}, e.g. `museums' over `beaches',
e.g. when travelling `solo' (\emph{accompanied-by} qualifier) for `leisure' (\emph{trip-type} qualifier).

To illustrate the potential usefulness of both the \emph{present state} and the \emph{past state} contexts with an example,  
consider for instance the situation when a user visits Dublin with her group of friends in early summer. Based on the user's previous preferences in other locations (e.g. the user usually loves to hangout with friends, or she is an avid draught lover, or she loves trekking or hiking), a context-aware system should seek to \emph{match} information from previous user preferences with the POI descriptors in the current context. For this example, an ideal system should recommend popular tourist destinations and activities in Dublin, that match the user preference history, such as the cliff walk in Howth, the Guinness Storehouse, Temple Bar, etc.

In addition to semantically matching the present state POI descriptors and the past preferences based on the \emph{present state} context of a given user location, an effective system should also consider the more personalized \emph{present state} context qualifiers, such as trip-type, accompanied-by etc. \cite{Aliannejadi_ToISJournal}.
Again as an example, a user's visit to Dublin for leisure with a group of friends should lead to preferring such suggestions as `lunch at cheap prices in pubs at the Temple Bar region' over the ones such as `lunch at the restaurant Avoca', because the latter is more suitable for families.

There are two fundamental differences between the location qualifier and the rest of the context qualifiers.
Firstly, the location of a POI is a universal property (irrespective of the perspective of individual users) whereas the other qualifiers, e.g. `trip type', `time of travel' etc., are intricately tied as attributes of individual users.
Secondly, the location information of a user
acts as a \emph{hard} constraint for POI recommendation because for a contextual suggestion to be meaningful and usable, the locations of the recommended POIs must be
close to the present state location of the user.
\hltIRJ{e.g., a system must make recommendations for the user's current (city) location only because a POI in a different city is obviously non-relevant.}
On the other hand, non-location qualifiers do not enforce a hard constraint, \hltIRJ{and acts as \emph{soft} constraints.} e.g., a positively rated POI in the past for a trip-type which was different from the current one (e.g. `solo' in the past vs. `with family' in the present) could still be recommended. \hltIRJ{Because a POI which is usually popular for family dinner may still be relevant or partially relevant to a solo traveller, and vice versa.}
\hlt{One may argue that the location context can also be a soft constraint}, where accurate geo-coordinates can be taken into consideration for favouring POIs that are in close proximity of the user's accurate coordinates \cite{GeoSoCa_SIGIR2015}. However, addressing this is beyond the scope of this paper and is a potential future work, possibly involving simulated users within the geographical bounding box of a city. In the scope of our work in this paper, a location context refers to a city which means that recommending POIs outside the city of the user's current (city) location is considered not to be relevant. This is also consistent with the TREC contextual suggestion (TREC-CS) task definition \cite{hashemi2016overview}, which we also follow for our evaluation framework.

\hltIRJ{Generally speaking, }
our proposed model is essentially based on semantically matching a user profile with POI description text, and is hence able to make contextual recommendations in a general scenario, i.e., with the presence of textual user profiles, POI descriptors and optional ratings.
Specifically, we assume that each user profile has a number of POIs that the user visited (either liked or disliked) in the past (say in city $X$, and $Y$), and a system needs to recommend POIs in her current city say $Z$ (also taking other non-location type constraints into consideration), that she has not visited before.
\hltIRJ{As users may be tempted to add some category tags (such as `sea food', `pub') while reviewing POIs, for our experimental setup we assume that the reviewed POIs available in a user's profile are comprised of textual description/reviews along with such tags which we will discuss in more detail in Section }\ref{sec:foundation}\hltIRJ{. We also show how these tags can be used as queries to represent user's preference history. If no tags are available, a number of keywords or important terms can be extracted from the textual description available in the profile based on their tf-idf or language model scores, which in fact is used as one of our baselines, or by employing TextRank }\cite{mihalcea2004textrank}\hltIRJ{. However, this is out of scope of this paper. In fact, the presence of such user assigned tags along with descriptive texts in the user profiles makes TREC-CS 2016 data a perfect choice for our experiments.}


\para{Key Research Challenges}

In our work, we approach the problem of contextual recommendation from an information retrieval (IR) perspective,
where POIs can be considered analogous to documents, and the information in the preference history analogous to a query. The key advantage of this approach is that it is mainly unsupervised or weakly supervised. Unsupervised approaches do not need to rely on training a model with annotated data; instead, to make recommendations they rather try to utilize the inherent semantic associations between latent features of the data itself (e.g. semantically matching the past preferences of users with the POI descriptions in the current context).
We now highlight the main research challenges in an IR-based approach to contextual recommendation.

\begin{enumerate}[label={},wide, labelwidth=!, labelindent=0pt]
\item \emph{Formulation of Query from User History} -
First, a major challenge in formulating contextual recommendation from an IR perspective is that, in contrast to the traditional IR setup, there is no notion of an explicitly entered user query \cite{FRLM_ICTIR2019}. In this case, the query needs to be automatically formulated from the information available in the user profiles, such as pieces of text describing their preferences and dislikes. This query then needs to be effectively \emph{matched} with the information of the POIs (documents) in the current context.

\item \emph{Lack of Non-location type Contextual Information in the User History} -
The second major challenge is the inevitable absence of explicit annotation of non-location type context (e.g. trip qualifiers, such as `trip purpose' etc.) in the user preference history \cite{FRLMPsi_SIGIR2020, hashemi2016overview}. To illustrate this point, consider typical user feedback in a 
Location Based Social Networks (LBSNs), such as
Foursquare\footnote{\url{https://foursquare.com}} or 
TripAdvisor\footnote{\url{https://tripadvisor.com}}.
This usually comprises a text review and an explicit rating score
(from very bad to very good).
An important point to note here is that this past information usually does not contain trip qualifier information, i.e. the context in which the POI was visited and rated thereafter.
%
Since a user's perception about a POI can be drastically different 
in changed circumstances, associating a precise context to a preference is useful to model the subtle dependence between the two, e.g. to model the situations that pubs are great for hanging out with friends only when there are no accompanying children, or hiking in the mountains is great only when it is less likely to rain. While on the one hand including this precise context as a part of the user feedback could provide additional sources of information, on the other, it is highly likely to reduce the number of users prepared to submit feedback due to the additional effort required to enter this information through a more complex interface.

\item \emph{Modeling Relevance for Non-location Contexts in the Present State (Query)} -
While user preference histories generally lack non-location or \emph{trip qualifier},
such information often forms a part of the present state of the user (i.e. the query). In contrast to the situation of a user being not prepared to enter these details every time as a part of feedback to a system, users in this case are more likely to submit such information as the type of the trip, whether they are with family or friends etc., because of their intuitive expectation that such precisely defined contextual information (in addition to the current geographic location) would enable the system to suggest more contextually relevant items (POIs).
An important research question is then how to bridge the gap between the lack of contextual information from the historical information of user feedback and the constraints imposed by them during the present context (query).

\end{enumerate}

\begin{figure}[t]
	\centering
	\includegraphics[width=0.99\textwidth]{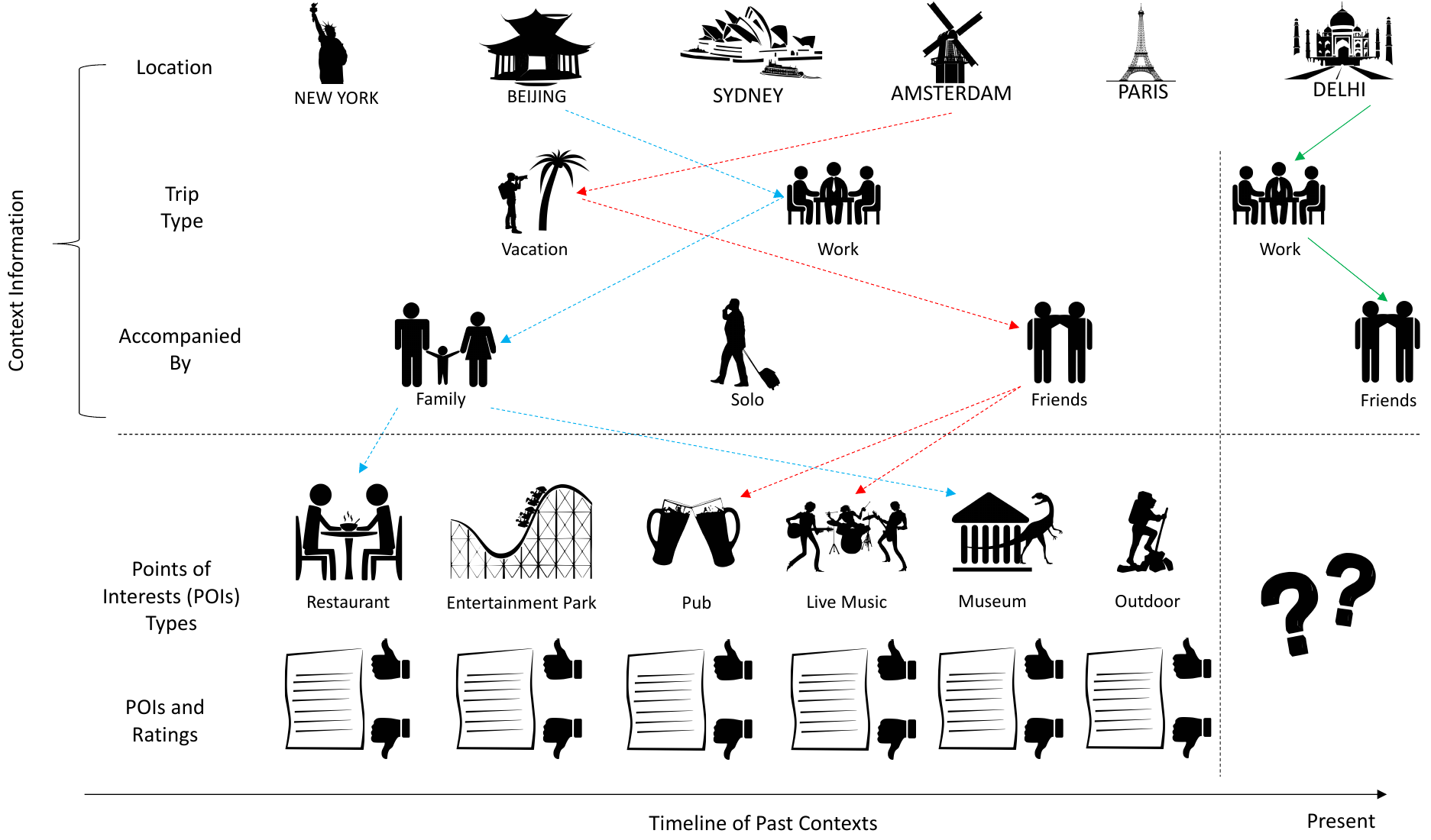}
	\caption{Schematic of Contextual Recommendation \hlt{showing the user's timeline of past and present context(s). Dotted arrows show that the non-location type contextual information (i.e. links between POIs and non-location intermediate nodes) is not present in the user's preference history while both the location and other non-location contexts are available in the present state. We estimate the likely non-location intermediate nodes by utilizing the information from the review text.}}
    \label{fig:schematic}
\end{figure}

A general approach of bridging this information gap is to employ weak supervision \cite{FRLMPsi_SIGIR2020, WSRM_CIKM2020} to associate certain topics in user feedback with a seed set of categories defining a precise context, e.g. starting with a seed set of term associations, such as `pub' being relevant to the context category `friends'. The natural language text of the reviews is also likely to be helpful in discovering more meaningful dependencies, e.g. associating `live music' with `friends', by using the semantic correlation between `pub' and `live music'.
We propose a formal framework towards this effect.


We illustrate the schematics of the overall idea of the problem and its solution in Figure \ref{fig:schematic}.
The top part of the figure shows two types of context information of a user, first, the \emph{location} of the user (specifically, a city which the user is currently visiting), and second, the more personal trip-qualifiers (non-location type) information categories which further qualify the location context, e.g. the `trip purpose' (whether vacation or work), `trip type' (i.e. whether a accompanied by family or a solo trip) etc.
The vertical line in Figure \ref{fig:schematic} separates the past context of a user from his present, e.g. the figure shows that the user's current location is Delhi, and that he has visited New York, Beijing etc. in the past.  
The bottom-left part of Figure \ref{fig:schematic}, constituting a part of a user's history, shows a list of POIs that the user rated positively (or negatively) during her different trips.
%
The main research challenge is then to estimate a likely path in the tree from a location to a number of POIs, i.e. estimate the likely non-location intermediate nodes by utilizing the information from the review text themselves.

After constructing a model of a user's preferences, the challenge in contextual recommendation is to be able to make new recommendations to the user for a new present location (that she has not visited before) with a given set of trip qualifiers, e.g., the path specified in Figure \ref{fig:schematic} with the green arrows indicates that the user's current location is `Delhi' which she is visiting for work along with her colleagues.
An effective contextual recommendation system in this scenario should seek to leverage similar situations in the past (i.e. the user's past non-solo work trips in other locations) in figuring out what type of POIs the user had previously rated positively in those situations, and then use information from these past POIs to recommend a set of similar POIs for the current location.

\para{Research Objective}
The primary objective of this paper is to investigate effective ways of addressing contextual POI recommendation
with a particular focus on obtaining high precision at top-ranks, thus accounting for the target device for delivery of content, which for the problem of contextual recommendation is most commonly a mobile with limited graphics, memory and network bandwidth resources.
%
The overall research objective can be categorized into the following three focused research questions.

\begin{itemize}[label={}, leftmargin=*]
    \item \textbf{RQ-1:} How to effectively formulate a query from a user's preference history so as to effectively retrieve the contextually relevant POI descriptors in a new location?

    \item \textbf{RQ-2:} What is a suitable way to include non-enforcing (\emph{soft}) contextual constraints such as the type of the trip etc. to further improve recommendation quality?
    
    \item \textbf{RQ-3:} To what extent, incorporating semantic association between a user's preference history and the POI descriptors of a new location, may improve the contextual POI recommendation effectiveness?
\end{itemize}

\compactpara{Our Contributions}
The 
key contributions of this paper as follows.
\begin{enumerate}[leftmargin=*]
\item
To address RQ-1, we propose
a pseudo-relevance feedback model to estimate a weighted distribution of terms from a user's preference history, which is then used as a query to topically match the content of POI descriptors in a new location.
More specifically, our proposed model is a \emph{factored} version of the standard relevance model \cite{Lavrenko_RLM2001:RBL:383952.383972}, where the first step involves enriching the user preference information
(\emph{exploitation}),
and the second step involves subsequently using the enriched information to effectively match the POI descriptors given query context
(\emph{exploration}).

\item To address RQ-2, we extend the model developed for RQ-1 with additional weights incorporated to address a set of soft constraints, related to additional contextual information of a non-enforcing nature, such as the trip type etc.
In particular, we undertake a weakly supervised approach (leveraging a small set of context-term annotations) to transform the soft constraints into term weighting functions.

\item To address RQ-3, we also incorporate term-level semantic associations within the framework of our model developed towards addressing RQ-1 and RQ-2. In particular, we use embedded vector representations of words to bridge the vocabulary gap between user preferences, POI descriptions and the trip qualifier (\emph{soft}) constraints.

\item Additionally, related to RQ-3, we also investigate the effect of different word embedding settings (in-domain vs. externally trained) on the contextual recommendation quality.

\end{enumerate}

As an extension to our previous work \cite{FRLM_ICTIR2019, FRLMPsi_SIGIR2020},
\hlt{we propose a novel word embedding} based factored relevance model (KDEFRLM) for both the location only (i.e. \emph{hard} context based) retrieval,
and the multi-contextual (i.e. \emph{hard}+\emph{soft}) recommendation.
The inclusion of word embedding within the framework (i.e., KDEFRLM) is able to achieve significant improvements over a number of IR-based, and recommender system (RecSys) based baselines.
We also investigate the choice of \emph{different word embedding techniques (in-domain vs. externally trained)} in the effectiveness obtained with our proposed model.
This paper presents a coherent synthesis of this complete line of research.
In addition, this paper contains more experiments and analysis, as outlined below.

\begin{itemize}[label=\textbullet]
    \item It introduces detailed analysis of the results including the differences between our initial model(s) and the newer ones.
    
    \item It provides detailed sensitivity analysis of our models with different contextual constraints.
    
    \item It presents comparative analysis with additional state-of-the-art baselines.
    
    \item It includes experiments with a per-query based variation in the \emph{exploitation-exploration} parameter.
\end{itemize}

The rest of the paper is organized as follows. In Section \ref{Sec:RelatedWork}, we survey existing work on contextual recommendation.
In Section \ref{sec:foundation}, we formally describe a generic IR setup as a common foundation for
the specific approaches that we propose in the subsequent sections.
Section \ref{sec:frlm} describes our proposed approach towards contextual recommendation using a factored relevance model (FRLM) that addresses the location constraint only, and
Section \ref{sec:kdefrlm} generalizes the model to include word semantic information. Following this, Section \ref{sec:frlm_psi} further generalizes FRLM to the multi-contextual case by incorporating term preference weights corresponding to trip qualifier (\emph{soft}) constraints. 
We describe the setup of our experiments in Section \ref{sec:setup}, which is followed by the results and discussions in Section \ref{sec:results}. Finally, Section \ref{sec:conclusions} concludes the paper with directions for future work.

\section{Related Work} \label{Sec:RelatedWork}


The problem of contextual recommendation has been investigated by a number of studies from the point of view of matching the contents of a user profile (query) representation and the POI (document) representation.
Among these, the studies in \cite{yang2012exploration,jiang2013pitt} combined similarities between POI categories and user profile content. Generally speaking,
for the POI categories, these approaches made use of external tag information from location-based social networks (LBSNs), such as Yelp or Foursquare, to match past user preferences and POIs in the current context. Note that contextual suggestion systems based on this thread of work primarily rely on \emph{exploiting} the available preferential knowledge of users from their profiles.
A different thread of work \cite{cheng2012fused,Griesner:2015:PRT:2792838.2799679} makes use of rating-based collaborative filtering, i.e. information from other users, to estimate a POI's popularity in a current context with the hypothesis that POIs with frequent positive ratings from other users could also be appropriate to the current user. In contrast to \emph{exploitation}, this thread of work based on collaborative filtering, mainly relies on \emph{exploring} the POIs in the current context.

The contextual suggestion track\footnote{\url{https://sites.google.com/site/treccontext/}} (TREC-CS) provides a common evaluation platform for researchers working on the contextual recommendation problem.
\hltIRJ{Generally speaking, given a set of example POIs which reflect the user's past preferences, and some contextual information such as temporal, geographical and personal contexts, the task was to return a ranked list of most relevant POIs given the user profile and current context. The task in the TREC-CS track was to test if a system can recommend POIs effectively in a new city, say New York, when the system has the previous knowledge of user's preferences in other cities, such as Seattle or Detroit.}
\hltIRJ{Over the years, TREC organizers experimented with a number of different experimental setups including the use of open web or a variant of ClueWeb as the corpus} \cite{dean2012overview, dean2013overview}\hltIRJ{. However, it is shown that the approaches exploiting ClueWeb tend to receive worse results} \cite{Arjen_IRJ2016_reproducibilityVsRepresentativeness}\hltIRJ{. Finally, to overcome the dynamic nature of the open web, TREC released a static web crawl}\cite{hashemi2016overview}.
A very popular approach among the task participants was to retrieve POIs from different LBSNs such as Google Place, Foursquare or Yelp based on geographical context, and then to apply some heuristics such as ``night club will not be preferred in morning'' or ``museum will be closed at night'' to filter out POIs that do not match the given temporal context \cite{dean2012overview, hashemi2016overview}.
Arampatzis and Kalamatianos \cite{Arampatzis_ToIS2017} experimented with different content-based, collaborative filtering based and hybrid approaches on TREC-CS, and found that the content-based approaches performed better than other approaches.

Most TREC-CS participants formulated the task as a content-based recommendation problem \cite{yang2012exploration,jiang2013pitt,roy2013simple,li2014user}. A common approach was to estimate a user profile based on the POIs that the user preferred in the past, and then rank the candidate POIs by their similarities to the estimated profile, on the assumption that users would prefer POIs that are similar to those they liked before. Some of these studies used the descriptive information of the POIs and/or the web pages of the preferred POIs to build user profiles \cite{Aliannejadi_ToISJournal, Chakraborty:SIGIR2018DC}, and then used several similarity measures to rank the POIs \cite{yang2012exploration,jiang2013pitt}. The authors of \cite{li2014bjut,li2014user} explored the use of LBSNs' category information for user modeling and POI ranking.
\hltIRJ{A distinguishing characteristic of TREC-CS 2016 framework is that the user profiles are assigned with user tags or endorsements, which were not available in earlier TREC-CS tracks.
A number of studies explored standard word/category embedding techniques to measure the similarity between the tags available in a user profile and the tags available in the content of a candidate POI }\cite{FUMIRlab_TRECCS2016, ExPoSe_dehghaniTRECCS2016, UAmsterdam_hashemiTRECCS2016, Aliannejadi_ToISJournal}\hltIRJ{. One drawback of these approaches is their reliance on the existence of matching tags in both the user profile and the candidate POIs. In addition, the set of category tags available in one LBSN (say Foursquare) may not be same in another LBSN (say Yelp). In contrast, our proposed approach does not rely on tag-matching and essentially performs a query expansion by selecting a number of contextually appropriate terms that better represent a user profile. Unlike Hashemi et al. }\cite{UAmsterdam_hashemiTRECCS2016} \hltIRJ{who explore a supervised approach of learning users' preferences based on tag embedding by taking all the user assigned tags, our proposed approach is unsupervised and estimates a term weight distribution based on a subset of user assigned tags. In particular, to accurately represent user's preference history, we make use only of positive tags i.e. the tags that are associated with the POIs liked by a user, while discarding those associated with the POIs that are not liked by the user.}
\hltIRJ{A recent work by Aliannejadi and Crestani}
\cite{Aliannejadi_ToISJournal}, applied linear interpolation and learning-to-rank to combine multiple scores such as review-based score and tag matching score for context-aware venue suggestion.
\hltIRJ{The motivation behind using a review-based score was to better understand the user's motivation behind rating a POI (liked or disliked). They trained a binary SVM classifier by considering review texts from positively rated POIs as positive samples, and review texts from negatively rated POIs as negative samples. On the other hand, tag matching score contributed to a similarity measure between POIs by making use of Foursquare, and Yelp category tags. It is becoming increasingly popular among researchers to make use of online user reviews in different ways for contextual recommendation, such as by learning the importance of user ratings, by learning the latent topic, or contexts present in the review text} \cite{chen2015recommender}\hltIRJ{. Musat et al.} \cite{Musat_RecSys_textOpinion} \hltIRJ{made use of weighted ratings. Specifically they consider the topics mentioned in both the candidate POI's review text, and the review text present in the user profile. The similarity between these topics were then used for ranking.}
%
%
The study in \cite{chen2015recommender} leverages users' opinions about a POI based on reviews that are available online \cite{DBLP:conf/fdia/Chakraborty17}. Use of a single LBSN may not be sufficient to capture the information about all POIs and/or all the available types of information about the POIs. Recently, Aliannejadi et al. \cite{aliannejadi2017personalized} reported that the amalgamated use of a user's current context and the ratings and reviews of previously rated POIs from multiple LBSNs improve recommendation quality. This thread of work for contextual recommendation is mainly based on \emph{exploiting} the user's existing preference history information and essentially performs content matching between the POIs in the user's preference history and the candidate POIs.

Recommendation system (RS) based algorithms mainly involve applying rating-based collaborative filtering approaches that are based on finding features that are common among multiple users' interests, and then recommending POIs to users who share similar preferences.
Matrix factorization, a standard technique that represents both users and items in a latent space,
forms the core of most of these recommendation based approaches. It is common to make use of the check-in information collected from LBSNs for recommending POIs \cite{cheng2012fused,Griesner:2015:PRT:2792838.2799679}.
Generally speaking, collaborative filtering based techniques often suffer from the data sparsity problem \cite{Arampatzis_ToIS2017, Bayomi:SAC2019_CoRE}. This problem is even worse for POI recommendation where a single user can only visit (and rate) a small
number of the POIs available in a city.
As a result, the user-item matrix \cite{Gemulla:2011} becomes very sparse \cite{yu2015survey} which leads to poor recommender system performance.
Due to this data sparsity problem, it can be difficult for purely recommendation based approaches to yield effective outcomes for POI recommendation.
%
In the context of our problem, an RS approach is likely not to be effective, firstly because of the \emph{lack of sufficient data for training} standard RS approaches \cite{Arampatzis_ToIS2017} in learning the user-item associations, e.g., by factorizing
a user-item matrix \cite{Gemulla:2011}, and secondly because there may be \emph{no ratings available for the POIs in query locations (contexts)}, which is specifically true for our experimental setup.


Some existing work \cite{Ye_SIGIR2011, Yuan_SIGIR2013} has addressed this data sparsity problem of collaborative filtering by incorporating supplemental information into the model. Specifically, Ye et al. \cite{Ye_SIGIR2011} argued that the spatial influence of locations affects users' check-in behaviour. They incorporated spatial and social influence to build a unified location recommender system. On the other hand, the system developed by Yuan et al. \cite{Yuan_SIGIR2013} which is a time-aware collaborative filtering model, recommends locations to users at a certain time of the day by leveraging other users' historical check-in information. To address the cold-start situation for hotel recommendation, Levi et al. \cite{Levi:2012:FNH:2365952.2365977} designed a context-aware recommender system. They constructed context groups based on user reviews and regarded the user's preferences in trip intent i.e. the purpose of the trip, and the similarity of the current user with other users such as their nationality. They also consider user preferences for different hotel features in their model. Fang et al. \cite{Fang_2016_TIST} developed a model that consider use of both spatial and temporal context information to handle the data sparsity problem.

Existing research that use time as a context includes the ones reported in \cite{Gao_RecSys2013,Deveaud:2015:EVM:2806416.2806484}. Deveaud et al. \cite{Deveaud:2015:EVM:2806416.2806484} designed a time-aware venue suggestion system which modeled popularity or appropriateness of venues (POIs) in the immediate future with the help of time series. In contrast to \emph{exploitation}, this thread of work for contextual recommendation primarily relies on \emph{exploring} the candidate POIs using the current contextual information. 

\section{IR Setup Foundation} \label{sec:foundation}
Unlike the traditional IR setup, there is no explicit user query in the contextual recommendation (CR) task that we address in this paper. The primary objective in CR is rather to match the user's preference history with the POI descriptors (analogous to documents) in the user's current context(s).
This contrasts with an IR-based approach where an explicit query can be formed from bits of information from the user profile.

\begin{figure}[t]
	\centering
	\includegraphics[width=0.99\textwidth]{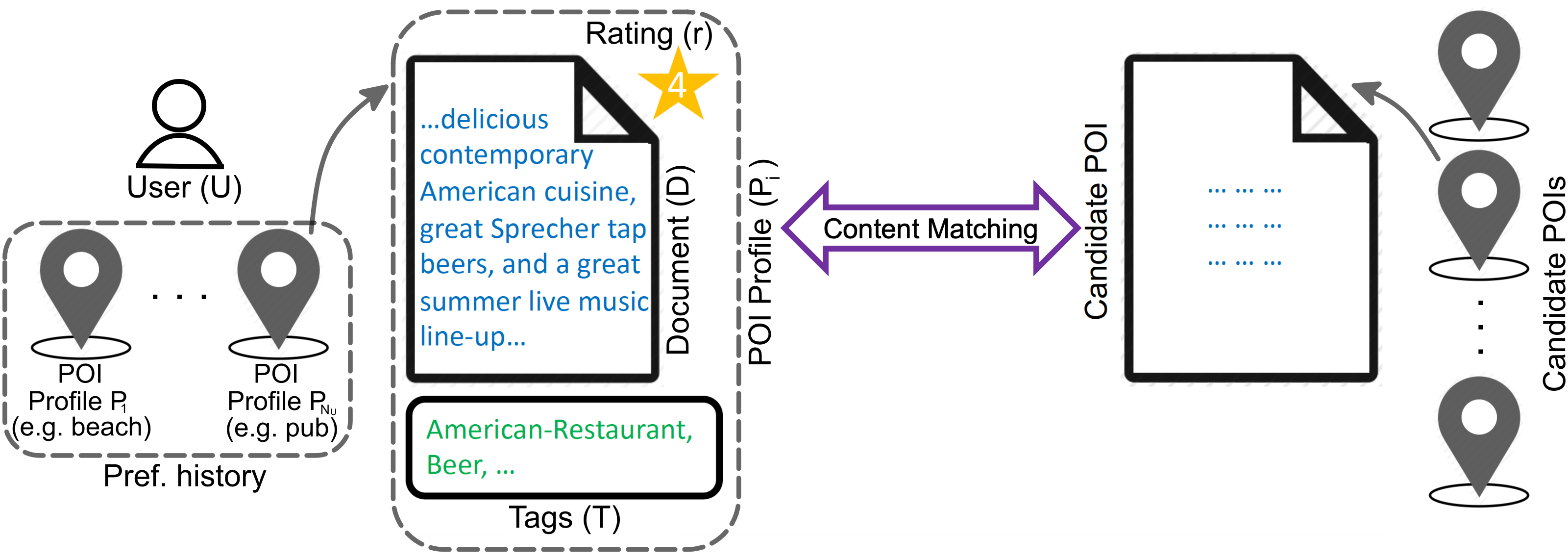} 
	\caption{Pictorial depiction of the IR setup for contextual suggestion\hlt{, which essentially involves matching the content between a
	candidate document to be retrieved (i.e. a POI description) and a textual representation of a user profile of the form
	$P_i = (D, T, r) \in U$ in the user's preference history.}}
    \label{fig:IRSetup_pictorial}
\end{figure}

\subsection{Notations for User Profile}  \label{ss:irsetupforcs}

A user profile is comprised of a descriptive text, a set of tag terms added to it and a score (see the bottom part of Fig. \ref{fig:schematic}). It should be noted that a document representation in a user profile does not have information about trip qualifiers, as indicated by the dotted arrows from the upper part of Figure \ref{fig:schematic} into each review.
The current context of a user forms a part of the query comprised of a pair of trip qualifiers of the form $(L, Q)$, where $L$ is the location (\emph{hard}) context, and $Q=Q_1\times\ldots Q_c$ is a combination of $c$ non-location (\emph{soft}) contexts.
\hlt{The general definition allows $c$} to be any finite integer. As per our experiments with the TREC-CS 2016 dataset \cite{hashemi2016overview} \hltd{the available number of such non-location qualifiers is $c=3$, i.e. the value of $c$ specifically for our experiments is 3. In particular, $Q_1$=\texttt{trip-type}, e.g. vacation, $Q_2$=\texttt{trip-duration}, e.g. day-trip, and $Q_3$=\texttt{accompanied-by}, e.g. solo or with friends.}
Each non-location type context $q_U$ is hence a $3$-dimensional categorical vector.

\hltIRJ{Although other contextual information such as the age, gender, season are available as a part of the TREC-CS dataset, the reason we take into account only the three non-location type constraints, namely the trip-type, trip-duration and accompanied-by, is because our proposed method being weakly supervised depends on a resource released by} \cite{Aliannejadi_crowdsourceData}.
\hltIRJ{Unlike other previous approaches} \cite{TRECCS2015overview} \hltIRJ{that did not follow a unified framework to incorporate these \emph{soft} contextual constraints, we, in fact, make use of this information as a part of the working principle of our model (to be discussed in Section} \ref{sec:frlm_psi}).
\hltIRJ{We also assume that all soft contexts are equally important, and show that the incorporation of these three contexts, in combination, improve retrieval effectiveness.}

\hltIRJ{
Generally speaking, in our model it should be possible to include} any number of
contextual constraints such as \hltIRJ{age, gender,} geographical influence \cite{Ye_SIGIR2011}\hltd{, time of the day} \cite{Yuan_SIGIR2013}, road traffic or availability of transportation, current weather etc. 
as a part of the non-location type constraints (i.e. use a value of $c$ higher than that of $3$). However, we restrict the scope of our current investigation to the aforementioned three specific non-location type attributes only, and leave the other attributes for a possible future extension of this work.


From a general IR point-of-view, we assume that a user profile $U$ is composed of a set of $N_U$ profile $P_i$'s
and an instance of the user's current context specified by the location and trip qualifiers $(l_U, q_U) \in (L,Q)$. 
Each profile $P_i$ is a 3-tuple consisting of a document ($D$ which belongs to a static collection $\mathcal{D}$), a set of user assigned tags ($T$ which is a subset of a controlled tag vocabulary $\mathcal{T}$), and a user provided rating ($r$ normalized within $[0, 1]$, higher the better). This is stated formally in Equation \ref{eq:profile}.
\begin{equation}
U = \cup_{i=1}^{N_U} \{P_i: P_i = (D, T, r) \in \mathcal{D}\times \mathcal{T} \times [0,1]\} \label{eq:profile}
\end{equation}

The objective of a tag $t \in \mathcal{T}$ is to express a POI as a set of single words or short phrases that best represents the POI, real instances of which are `beer', `American Restaurant', etc. assigned to the POI e.g. a restaurant. The document representation of the POI is composed of the text description accumulated from the POI's home page, customers' reviews on social networks etc.
The definition of every each document in the collection is assumed static.
For the sake of convenience in referring back to the notations,
we define them
in Table \ref{Table:Notations}. 
\begin{table}[t]
\centering
\small
\caption{List of the notations used in this paper.}
\label{Table:Notations}
\begin{tabularx}{\columnwidth}{lX}
    \toprule
    Notation & Implication\\
    \midrule
    $\mathcal{D}$ & Overall collection of documents (POI descriptors).\\
    $U$ & User profile\\
    $N_U$ & No. of POIs available, as preference history, in user profile $U$\\
    $D$ & Document (bag-of-words) representation of a POI, $D \in \mathcal{D}$\\    
    $P$ & $3$-tuple representation of a POI, $(D, T, r)$\\
    $T$ & A set of user created tags, a subset of $\mathcal{T}$\\
    $r$ & User assigned rating for $D$, $r \in [0, 1]$\\
    $\mathcal{T}$ & Overall (controlled) vocabulary of tags used across the user profiles\\
    $l_U$ & \emph{Hard} location constraint of $U$, $l_U \in L$\\
    $q_U$ & \emph{Soft} contextual constraint(s) or trip qualifier(s) of $U$, $q_U \in Q$\\
    $Q=Q_1\times\ldots Q_c$ & Overall set of non-location (\emph{soft}) trip-qualifiers comprised of $c$ trip qualifier types across the  collection\\
    $Q_i$ & A particular non-location type constraint \\
    $L(d)$ & Location of a POI $d$\\
    $\topM$ & Top set of $M$ documents (location constrained to $l_U$) retrieved with the query with term distribution $\theta_U,q_U$\\        
    $S(l_U)$ & Set of POIs constrained to (\emph{hard}) location, $l_U$\\
    $\phi(P, d)$ & Text-based content matching between a candidate POI $d$, and a POI $P = (D, T, r) \in U$\\
    $\mathcal{S}(d, U)$ & Text-based content matching between a candidate POI $d$, and the user profile $U$\\
    $\psi_s(w, q_U)$ & Contextual appropriateness measure
    of the term $w \mapsto [0, 1]$ for a \emph{single context}, $q_U$\\
    $\psi_j(w, q_U)$ & Contextual appropriateness measure
    of the term $w \mapsto [0, 1]$ for a \emph{joint context}, $q_U$\\
    \bottomrule
\end{tabularx}
\end{table}

\subsection{Retrieval with the Location Constraint}
The objective then is to \emph{rank} a set of POIs (hard constrained by $L=l_U$) in decreasing order of their estimated relevance scores within the current context.
A simple way to estimate the relevance scores is to first restrict the set of candidate POIs to only the ones in the specific location (by employing the hard constraint), i.e. $S(l_U)=\cup\{d: L(d)=l_U\}$ ($L$ denoting the location attribute of a POI). The next step then makes use of the text in the user profile, $U$, and this candidate set of POI descriptors $S(l_U)$ to estimate the relevance scores,
\begin{equation}
    \phi: U \times S(l_U) \mapsto \mathbb{R},\,\,
    S(l_U)=\cup\{d: L(d)=l_U\},
\label{eq:phi}
\end{equation}
where the output of the function, $\phi$ (e.g. with BM25 or a pseudo-relevance feedback method), does not depend on the non-location type qualifiers $q_U \in Q$.

A simple content matching technique is then to employ a standard ranking function, e.g. BM25, or language model
computing the similarity between a candidate POI $d: L(d)=l_U$ and all POIs in the user profile,
\begin{equation}
\mathcal{S}(d,U) = \sum_{P=(D,T,r) \in U} \phi(P,d),\,\, d \in S(l_U), \label{eq:simqd}
\end{equation}
where $\mathcal{S}(d,U)$ is the text-based content matching score between a candidate POI $d$, and the user profile $U$.
Each POI in the current location context can then be sorted in descending order of their similarity scores and presented to the user.

Figure \ref{fig:IRSetup_pictorial} shows a pictorial representation of our proposed IR setup for contextual suggestion where each POI is represented as a document (bag-of-words). A sample profile $P_i = (D, T, r)$ for a user's preference history is shown as a collection of three components (tuples): the document representation ($D$) of the POI, a set of tags ($T$) and the rating ($r$), provided by the user, for the POI. From the ranking perspective, we then need to perform content matching between a candidate document (representation of a candidate POI) $d: L(d) = l_U$ and every document (representation of profile $P_i = (D, T, r) \in U$) in the user's preference history.

\section{A Factored Relevance Model} \label{sec:frlm}
The key idea of our proposed methodology for contextual recommendation (CR) is to make use of a pseudo-relevance feedback based framework to effectively balance the trade-off between exploitation and exploration. In this section, we first introduce the general concept of the relevance model. We then provide a general description of the IR setup for contextual recommendation and discuss how pseudo-relevance feedback in the form of a generalized relevance model
can be applied in our problem context.

\subsection{Relevance Model for IR} \label{RelevanceModelRevisited}
The relevance model (RLM) \cite{Lavrenko_RLM2001:RBL:383952.383972} is a relevance feedback method which estimates the importance of terms for relevance feedback by using the co-occurrence information between a set of given query terms and those occurring in the top-ranked documents. RLM assumes that the terms frequently co-occurring with a query term are semantically associated to the information need and, hence, could be used to enrich the query with supplemental information.

Formally speaking, given a query $Q=\{q_1,\ldots,q_n\}$, RLM involves estimating a term weight distribution from a latent relevance model $R$, $P(w|R) \approx P(w|Q)$, from a set of $M$ top-retrieved documents $\mathcal{M}=\{D_1,\ldots,D_M\}$, as shown in Equation \ref{eq:traditional_rlm}.
\begin{equation} \label{eq:traditional_rlm}
	P(w|Q) = \sum_{D \in \mathcal{M}} P(w|D) \prod_{q \in Q} P(q|D)
\end{equation}
From Equation \ref{eq:traditional_rlm}, it can be noted that higher $P(w|Q)$ values, i.e. RLM term weights are achieved for a term $w$, when it occurs frequently in a top-retrieved document, i.e. $P(w|D)$ is large, in conjunction with the frequent occurrence of a query term $q \in Q$ such that $P(q|D)$ is also a large value.
\hltIRJ{In other words, RLM assumes that there is a latent relevance model, which needs to be estimated based on the evidence that the terms in the query and those in the relevant documents are generated by this distribution. The RLM approach essentially estimates the probability of sampling a term $w$ along with the query terms by the joint probability of observing $w$ along with the query terms. This joint probability estimation of $P(w|Q)$ is based on \emph{independent and identically distributed (i.i.d.)} sampling.}

In literature, this version of the relevance model is commonly known as `RM1'. `RM1' does not consider the original query terms while estimating the density function, which often results in a query drift \cite{ComparativeRLM}. In \cite{ComparativeRLM}, it is shown that a mixture model of the other term weights' estimated density in conjunction with the original query terms yields better feedback results.
This mixture model, widely known as `RM3' \cite{NasreenJaleel_RM3}, is shown in Equation \ref{eq:RM3}.
\begin{equation} \label{eq:RM3}
	P'(w|R) = \lambda P(w|R) + (1-\lambda) P(w|Q)
\end{equation}
Each mention of `relevance model' or `RLM' in this work is to be considered as its more effective mixture model variant, i.e. `RM3'.

\subsection{User Profile based RLM}  \label{ss:profile_rlm}

The primary challenge in matching a user profile with a POI descriptor in the current context (Equation \ref{eq:simqd}) is to extract a set of contextually relevant terms from the documents and tags of the user profile.
A naive way to compute the similarity scores in Equation \ref{eq:simqd} is to consider each document along with the user tags as a simple bag-of-words representation. This could potentially lead to noisy similarity estimation. To be more precise, there are two likely reasons that this naive similarity estimation may be ineffective. First, the information present in a user profile may be quite diverse in nature with only a specific aspect of it being likely to be useful in the current context, e.g. a user is likely to visit many different locations under different contexts in her past, however only a small number of them would be relevant within a present context).
Second, it is often the case that the POI descriptors are long documents likely to introduce noise in the estimated similarities. Instead, focusing on relevant parts of these documents that are contextually related with the query rather than the whole document may lead to better similarity estimation.

With this motivation, we propose to employ a RLM to estimate a weighted distribution of terms extracted from the user profile, and use this term distribution $\theta_{U, q_U}$ to rank the POIs (documents) in the current context, $(l_U, q_U) \in (L,Q)$, where $l_U$ is user's current location qualifier and $q_U$ is the non-location type trip qualifier.

\begin{figure}
	\centering
	\includegraphics[width=0.6\textwidth]{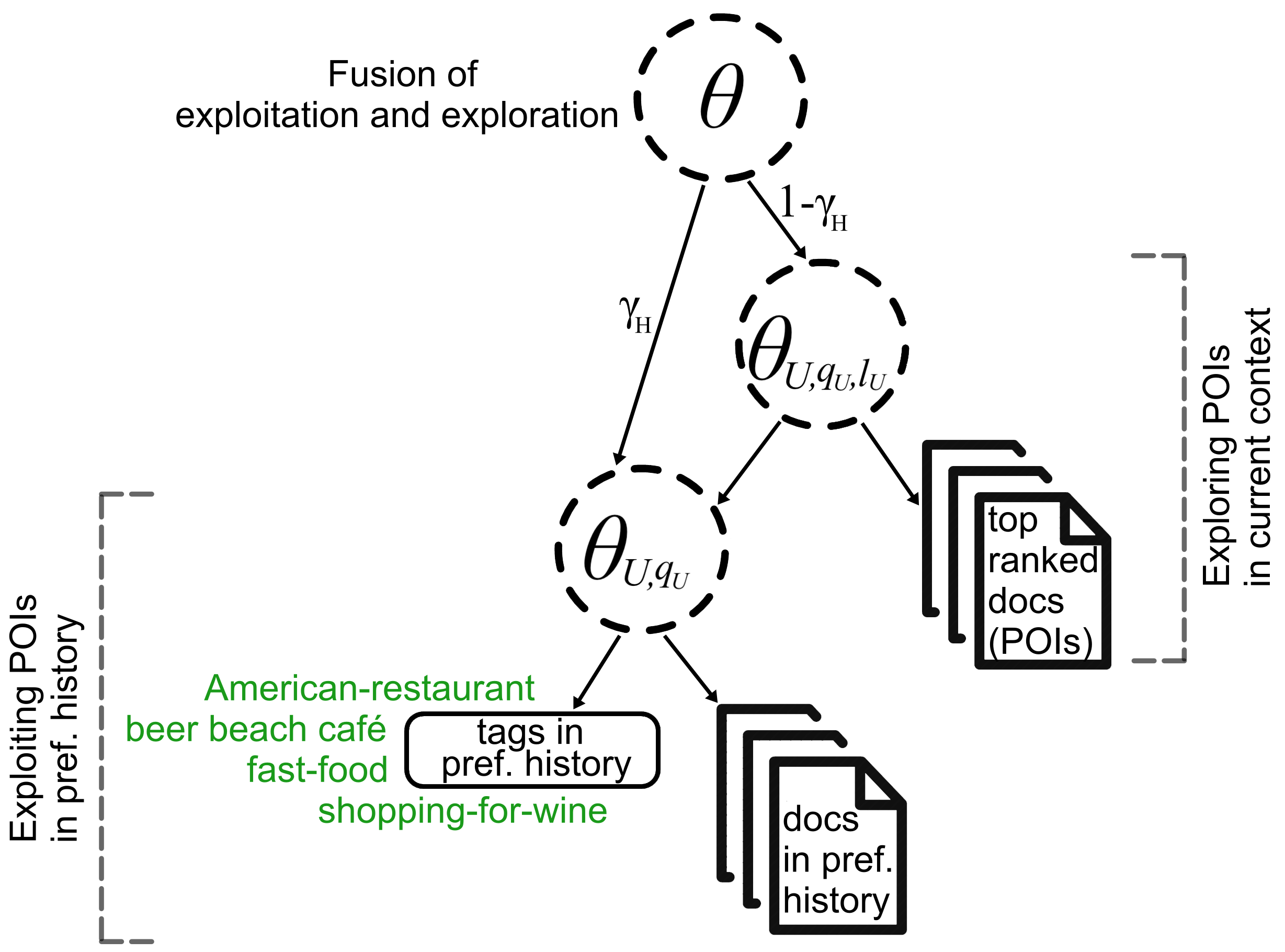} 
	\caption{Schematic diagram of a Factored Relevance Model (FRLM). \hlt{The FRLM first estimates a relevance model based on the user's preference history (\emph{exploitation}). Then it estimates another relevance model based on both the initial model and the top retrieved POIs in the current context (\emph{exploration}). Finally these two models are linearly combined (\emph{fusion}).}}
    \label{fig:frlm}
\end{figure}

To estimate a relevance model based on a user profile $U$, we consider the set of tags in a POI descriptor $P=(D,T,r) \in U$ (Equation \ref{eq:simqd}) as the observed or known terms (which are analogous to query terms in the IR framework of RLM). Let $T'$ be the set of user assigned tags, i.e. union of all $T$s from the set of tuples $(D, T, r) \in U$. A sample set $T' = \{$\texttt{American-restaurant, beer, beach, caf\'{e}, fast-food, shopping-for-wine}$\}$ is shown in Fig. \ref{fig:frlm}. The set of top ranked documents on this occasion is the provided set of documents in the user preference history, i.e. union of all $D$s from the set of tuples $(D, T, r) \in U$. Formally,
\begin{equation}
    P(w|\theta_{U, q_U}) = \sum_{(D,T,r) \in U} r P(w|D) \prod_{t \in T'}P(t|D) \label{eq:rlmprofile},
\end{equation}
where the estimated RLM captures the semantic relationship between a user specified tag and a term presented in the documents, by co-occurrence corroboration from the user profile.


\hlt{The rating values are used as confidence scores} for the co-occurrences allowing the relevance model to assign higher weights to terms that co-occur more frequently with the user assigned tags within a POI with a high rating. Although it may seem at a cursory glance that the use of user assigned ratings in an RLM framework makes it supervised, we would like to emphasize that these rating scores are not used as \emph{labels} in a supervised setting to optimize an objective function. In the degenerate case, i.e. when no ratings are available, our RLM-based feedback model would use a constant confidence value of $1$, i.e. it would assign uniform weights to all POIs in the user profile.



\subsection{Factored RLM for Contextual Relevance} \label{ss:factoredrlm}

To impose the \emph{hard} constraint of the location qualifier $l_U$, we estimate another relevance model $\theta_{U,q_U,l_U}$, by making use of both the user profile based relevance model estimated only with the soft constraints (Equation \ref{eq:rlmprofile}) and the selected subset of location-specific POIs (documents). This time the terms estimated in the user profile based RLM
$\theta_{U, q_U}$
are considered to be the observed terms. \hltIRJ{In fact, the set of terms considered to be the `observed' ones in Equation} \ref{eq:flm} \hltIRJ{are restricted to be those with the highest weights (probability values) computed as per Equation} \ref{eq:rlmprofile}\hltIRJ{. The number of terms selected is, in fact, controlled by a parameter $\tau$ in our experiments.}

Further, the set of top ranked documents,
denoted by $\topM$, refer to the top $M$ documents retrieved in response to the query. In this case, the query is constrained to be satisfying the \emph{hard} location constraint $l_U$. This is stated formally in Equation \ref{eq:flm}.
\begin{equation}
    P(w|\theta_{U, q_U, l_U}) = \sum_{d \in \topM} P(w|d) \prod_{t \in \theta_{U, q_U}} P(t|d) \label{eq:flm}.
\end{equation}
Equation \ref{eq:flm} is a factored relevance model in which estimating $\theta_{U,q_U,l_U}$ needs $\theta_{U,q_U}$ to be estimated first, which acts as the factor model. This factored relevance model \emph{explores} the potentially relevant POIs in the user's current location context $l_U$, to achieve a better ranking of the POIs.

As a generalization, we use a linear combination of the two relevance models of Equation \ref{eq:rlmprofile} (\emph{exploitation} part) and Equation \ref{eq:flm} (\emph{exploration} part), into a combined model,
\begin{equation}
    P(w|\theta) = \gamma_H P(w|\theta_{U, q_U}) + (1-\gamma_H)P(w|\theta_{U, q_U, l_U}), \label{eq:frlm_linearcomb}     
\end{equation}
where $\gamma_H$ is the trade-off parameter to control the relative importance of the two relevance models. We call this version of our proposed model the Factored ReLevance Model (FRLM).

\section{Factored Relevance Model with Word Semantics} \label{sec:kdefrlm}
The user profile based RLMs as presented 
in Section \ref{sec:frlm}
($\theta_{U, q_U}$ of Equation \ref{eq:rlmprofile} or its factored
version, $\theta_{U, q_U, l_U}$, of Equation \ref{eq:flm}) 
can take into account only the document level co-occurrence of terms (ignoring any semantic associations between them). In this section, we generalize this proposed factored relevance model of Section \ref{sec:frlm} by employing the concept of kernel density estimation. In the context of our specific problem, this favours those terms which in addition to exhibiting local (top-retrieved) co-occurrence, are also semantically related to the query terms. Before describing our generalized model, we outline the existing work on kernel density based relevance models \cite{DoiKDERLM_CIKM16}.


\subsection{Kernel Density Estimation based RLM}
Kernel Density Estimation (KDE) is a non-parametric method to estimate the probability density function of a random variable. Formally, let $\{x_1, \dots, x_n\}$ be independent and identically distributed (i.i.d.) samples drawn from a distribution. The shape of the density function, $f$, from which these points are sampled can be estimated as
\begin{equation} \label{eq:weightedKDE}
	\hat f_\alpha(x) = \frac{1}{nh} \sum_{i=1}^n \alpha_i K \Big(\frac{x-x_i}{h}\Big),
\end{equation}
where $x_i$ is a given data point (commonly known as the pivot point), $\hat f_\alpha(x)$ is the estimated value of the true density function $f(x)$,
$\alpha_i$ is the relative importance of the $i^{th}$ data point with the constraint that $\sum_i \alpha_i = 1$,
and
$K(.)$ is a kernel function scaled by a bandwidth parameter $h$.
By definition, a kernel function is a monotonically increasing function of the distance between two points (vectors). A common choice of a kernel function is a Gaussian function.

Roy et al. \cite{DoiKDERLM_CIKM16} observed that since the relevance model estimates a distribution of (real-valued) weights over terms, the concept of KDE can be applied to define this distribution in a generalized way (the model being called Kernel Density Estimation based RLM or KDERLM for short). The basic idea to define the relevance model distribution this way is to treat the query terms as a set of pivot terms (analogous to the $x_i$'s of Equation \ref{eq:weightedKDE}). Rather than treating terms as independent, the distance between the vector representation
(obtained by applying a word embedding method such as \texttt{word2vec} \cite{Word2vec_NIPS2013}) of a pivot (query) term with that of a term occurring in the top-ranked documents is then used to define 
the kernel function. This results in the influence of a query term to \emph{propagate} to other terms that have similar (close) vector representations in the embedded space. 
Formally, assuming that the query terms $Q = \{q_1, \dots, q_n\}$ are embedded as vectors, the probability density function estimated with KDE is
\begin{equation} \label{eq:KDERLM_1D}
    f(w) = \frac{1}{nh} \sum_{i=1}^n P(w|\mathcal{M}) P(q_i|\mathcal{M}) K \Big(\frac{w-q_i}{h}\Big),
\end{equation}
where the kernel function $K$ is a function of the distance between the word vectors of a term $w$ (within a top-ranked document) and a query term $q_i$.
The set of top-ranked documents is considered as a single document model $\mathcal{M}$.
Moreover, $P(w|\mathcal{M}) P(q_i|\mathcal{M})$ acts as the weight associated with this kernel function (thus incorporating the local RLM effect in addition to the global term semantics from the embedded space). In other words,
the closer the word $w$ is to a query term $q_i$ in conjunction with a high RLM term weight, the higher becomes the value of the KDERLM weight $f(w)$.

\subsection{KDE based RLM on User Profiles}

In the context of the POI recommendation problem,
the KDERLM model potentially assigns higher importance to
a word $w$ from a POI descriptor if it is semantically associated to a tag (query) term $t$ (as per the embedding space). 
We can imagine that the discrete probabilities $P(w|\theta_{U, q_U})$ of the user profile based RLM (Equation \ref{eq:rlmprofile}) are smoothed out to form a continuous probability density function $f(w)$.
The shape of this density function is controlled by a set of pivot points comprising the tag terms in a user's profile.
%
%
%
Concretely, for a user profile $U =\cup_{i=1}^{N_U} \{P_i: P_i = (D, T, r)\}$ with the set of unique tag terms,
$T'$,
the probability density function estimated by KDE (with a Gaussian kernel) is given by
\begin{equation} \label{eq:kderlmprofile_1}
f_\alpha(w) = \frac{1}{nh} \sum_{t \in T'} \alpha_t K \Big(\frac{w-t}{h}\Big)
= \sum_{t \in T'} \alpha_t \frac{1}{\sigma \sqrt{2 \pi}} \exp (-\frac{(\mathbf{w}-\mathbf{t})^T (\mathbf{w}-\mathbf{t})}{2\sigma^2 h^2}),
\end{equation}
where $\vec{w}$ and $\vec{t}$ denote the vectors for the word $w$ and the tag $t$, and $\alpha_t$ is the weight assigned to the tag term which we describe how to compute next.

Considering the set of all documents (reviews or POI descriptions) of a user profile, i.e. D belonging to some tuple in
$U = \cup_{i=1}^{N_U} \{P_i: P_i = (D, T, r)\}$,
as a single document model $\mathcal{M}$,
the estimation of our previously proposed user profile based RLM (Equation \ref{eq:rlmprofile})
can be reduced as shown in Equation \ref{eq:rlmprofile_1D}.
\begin{equation}
    P(w|\theta_{U, q_U}) = P(w|\mathcal{M}) \prod_{t \in T'} P(t|\mathcal{M}) \label{eq:rlmprofile_1D}
\end{equation}
Then maximum likelihood estimates (MLE) of $P(w|\mathcal{M})$ and $P(t|\mathcal{M})$ ensure that to maximize $P(w|\theta_{U, q_U})$, both $P(w|\mathcal{M})$ (i.e. the normalized term frequency of a word $w$ in the set of documents in the user's preference history, or in other words, the set of terms a user generally prefers, e.g., `friends', `pubs' etc.), and $P(t|\mathcal{M})$ (i.e., the normalized term frequency of the tags in the set of documents in the user's preference history) are both maximized, i.e., Equation \ref{eq:rlmprofile_1D} captures the local
co-occurrences between a tag and a term within a user profile.
%
We then assign
$\alpha_t = P(w|\mathcal{M}) P(t|\mathcal{M})$
and substituting it in Equation \ref{eq:kderlmprofile_1}, yields Equation \ref{eq:KDERLM_1D_fullForm}.
\begin{equation} \label{eq:KDERLM_1D_fullForm}
    f(w) = \sum_{t \in T'} P(w|\mathcal{M}) P(t|\mathcal{M}) \frac{1}{\sigma \sqrt{2 \pi}} \exp (-\frac{(\mathbf{w}-\mathbf{t})^T (\mathbf{w}-\mathbf{t})}{2\sigma^2 h^2})
\end{equation}

In Equation \ref{eq:KDERLM_1D_fullForm}, we consider all documents in the user's preference history as a single document model and ignored document level user rating. To incorporate the document level importance of a term $w$ in the estimation of the probability density function, we introduce the document level user rating while computing $P(w|\mathcal{M})$. We compute document-level user rating based relevance weights, $P(w|\mathcal{M})$ as shown in Equation \ref{eq:KDERLM_2D}.
\begin{equation} \label{eq:KDERLM_2D}
    P(w|\mathcal{M}) = \sum_{(D,T,r) \in U} r P(w|D)
\end{equation}
Plugging this into Equation \ref{eq:KDERLM_1D_fullForm} yields Equation \ref{eq:KDERLM_2D_fullForm}.
\begin{equation} \label{eq:KDERLM_2D_fullForm}
    P(w|\theta_{U, q_U};h,\sigma) = \sum_{t \in T'} \Big(\!\!\!\!\!\!\sum_{\ \ (D,T,r) \in U} \!\!\!\! r P(w|D)\Big) P(t|\mathcal{M}) \frac{1}{\sigma \sqrt{2 \pi}} \exp (-\frac{(\mathbf{w}-\mathbf{t})^T (\mathbf{w}-\mathbf{t})}{2\sigma^2 h^2})
\end{equation}
Similar to our previous version of the user profile base RLM (Equation \ref{eq:rlmprofile}), the rating values in Equation \ref{eq:KDERLM_2D_fullForm} are used as confidence scores for the co-occurrences, which allows the relevance model to
preferentially weigh the term co-occurrences of across POIs that are high rated in a user profile.

\subsection{A Factored version of KDERLM}
We argued in Section \ref{ss:factoredrlm} (Figure \ref{fig:frlm}) that a factored version of the RLM is particularly suitable for the task of contextual POI recommendation because it is useful to enrich the initial query (comprised of tag terms) with additional relevant terms from the user profile (review text/POI descriptors).
Since term weights estimated from Equation \ref{eq:KDERLM_2D_fullForm} yield a set of such potentially relevant terms, we make use of the term weight distribution estimated from Equation \ref{eq:KDERLM_2D_fullForm} to estimate another relevance model for the retrieval step with the hard location constraint, i.e., this time the term weights are useful to effectively match the information need (weighted query estimated from a user profile) with the documents that are to be retrieved (specified by the set of documents from the collection satisfying the location constraint). More formally,
\begin{equation}
P(w|\theta_{U, q_U, l_U};h,\sigma) = \sum_{d \in \topM}
\frac{1}{\sigma \sqrt{2 \pi}}
P(w|d) \prod_{t \in \theta_{U, q_U}} P(t|d)
\exp (-\frac{(\mathbf{w}-\mathbf{t})^T (\mathbf{w}-\mathbf{t})}{2\sigma^2 h^2}),
\label{eq:flmkde}
\end{equation}
where we make use of the set of POIs of the current location (constrained by $L(d)=l_U$) to estimate the KDERLM corresponding to the \emph{exploration} mode (similar to Equation \ref{eq:KDERLM_1D_fullForm}).

Similar to FRLM, where we combine both the models corresponding to exploitation and exploration, we can create a combined version of this KDE based model as shown in Equation \ref{eq:kdefrlm_linearcomb}.
\begin{equation}
    P(w|\theta;h,\sigma) = \gamma_H P(w|\theta_{U, q_U};h,\sigma) + (1-\gamma_H)P(w|\theta_{U, q_U, l_U};h,\sigma) \label{eq:kdefrlm_linearcomb}     
\end{equation}
The trade-off parameter $\gamma_H$ controls the relative importance of the two relevance models. We call this version of our proposed model Kernel Density Estimation based Factored ReLevance Model (KDEFRLM), scaled with kernel bandwidth $h$, and standard deviation $\sigma$.

\section{Multi-Contextual Generalization of Factored Relevance Model} \label{sec:frlm_psi}
Until this point our proposed models, the factored relevance model (FRLM) and its KDE based variant, have been able only to address the location (\emph{hard}) constraint in POI recommendation. In this section,
%
we propose a multi-contextual extension to our proposed models so as to additionally take into account a set of \emph{soft} (trip-qualifier) constraints.
%

\subsection{Weakly Supervised approach for addressing Trip Qualifier (\emph{Soft}) Constraints} \label{ss:weakSupervision_forSoftContext}
To incorporate non-location type qualifiers, one needs to learn an association between a word from the review text or the tag vocabulary of a user profile, and the likely (historical) context (trip-type, duration, etc.) that leads to creating the review text in the first place. As an example, it should be possible for humans (with their \emph{existing knowledge}) to infer that a review about a pub frequently mentioning phrases, such as `friends', `good times', `tequila shots' etc. is most likely associated with accompaniment by friends on vacation (i.e. \texttt{trip-type}=\texttt{vacation} and \texttt{accompanied-by}=\texttt{friends}).

\begin{table}[tp]
    \centering
    \small
    \caption{Soft constraint categories
with their values. \label{Table:softConstraints}
}
    \begin{tabular}{ll}
    \toprule
    Categories &  Values \\
    \midrule
    $Q_1$: \texttt{trip-type} & \{\texttt{business}, \texttt{holiday}, \texttt{other}\}\\
    $Q_2$: \texttt{trip-duration} & \{\texttt{day-trip}, \texttt{longer}, \texttt{night-out}, \texttt{weekend-trip}\}\\
    $Q_3$: \texttt{accompanied-by} & \{\texttt{alone}, \texttt{family}, \texttt{friends}, \texttt{other}\}\\
    \bottomrule
    \end{tabular}
\end{table}

\begin{table}[ht]
\centering
\small
\caption{Crowd sourced contextual appropriateness data for \emph{single context} \cite{Aliannejadi_crowdsourceData}. Table \ref{Table:softConstraints} lists the categorical values corresponding to the three trip qualifiers.}
\label{Table:CrowdsourcedContextualRelevanceSingle}
\begin{tabular}{llll}
    \toprule
    \#Assessors & Appropriateness & Term/Phrase & Single Context ($Q_i$)\\
    \midrule
    12 & 1.00 & American Restaurant & \texttt{trip-duration=weekend-trip}\\
    7 & 0.71 & American Restaurant & \texttt{trip-duration=longer}\\
    12 & -0.48 & Nightlife Spot & \texttt{trip-type=business}\\
    7 & -1.0 & Nightlife Spot & \texttt{accompanied-by=family}\\
    \bottomrule
\end{tabular}
\end{table}

\begin{table}[ht]
\centering
\small
\caption{Crowd-sourced contextual appropriateness data for \emph{joint context} \cite{Aliannejadi_crowdsourceData}.}
\label{Table:CrowdsourcedContextualRelevanceJoint}
\begin{tabular}{@{}cclc@{}}
    \toprule
    \multirow{2}{*}{\#Assessors} & \multirow{2}{*}{Appropriateness} & \multirow{2}{*}{Term/Phrase} &  Joint Context $Q=Q_1\times Q_2 \times Q_3$\\
    & & & (\texttt{trip-type}, \texttt{trip-duration}, \texttt{accompanied-by})\\
    \midrule
    3 & 1.0 & Movie Theater & `\texttt{holiday, day-trip, friends}'\\
    3 & 1.0 & Irish Pub & `\texttt{holiday, night-out, friends}'\\
    3 & 1.0 & Steakhouse & `\texttt{business, longer, family}'\\
    3 & -1.0 & Bar & `\texttt{holiday, weekend-trip, family}'\\
    3 & 1.0 & Bar & `\texttt{holiday, weekend-trip, alone}'\\
    3 & -1.0 & Grocery Store & `\texttt{business, day-trip, alone}'\\
    
    \bottomrule
\end{tabular}
\end{table}

A computational approach to automatically constructing this association requires
the use of
a knowledge base (e.g. a seed set of term-category associations). One such knowledge resource was compiled in \cite{Aliannejadi_crowdsourceData}, which is composed of the following two different types of manually assessed information.

\begin{enumerate}[leftmargin=*]
\item List of pairs constituting a term and a \emph{single} non-location trip-qualifier with manually judged relevance scores of the form $(t,q,a)$, where $t$ is a term (e.g. food), $q$ is a single category (e.g. holiday) and $a\in [0,1])$ is a manually judged \emph{appropriateness score}. An example of a non-relevant pair is (\texttt{nightlife, business, 0.1}) with a lower score. Table \ref{Table:CrowdsourcedContextualRelevanceSingle} shows more examples of this sort.

\item List of pairs of a term with a \emph{joint} context (a $3$-dimensional vector of categories) along with a manually assessed binary label (1/0) indicating whether the term is relevant in the given \emph{joint context} or not. As an example, the word `pub' is assessed to be non-relevant in the joint context of `(\texttt{holiday, family, weekend})', whereas it is relevant in the context `(\texttt{holiday, friends, weekend})'.
Table \ref{Table:CrowdsourcedContextualRelevanceJoint} shows more examples of this sort.

\end{enumerate}

We formally denote these two types of knowledge resources (Tables \ref{Table:CrowdsourcedContextualRelevanceSingle} and \ref{Table:CrowdsourcedContextualRelevanceJoint}) as
\begin{equation}
\begin{split}
\kappa_s: (w, q) & \mapsto [0,1], w \in V, q \in Q_i, i \in \{1,\ldots,c\} \\
\kappa_j: (w, q) & \mapsto \{0,1\}, w \in V, q \in Q=Q_1\times\ldots Q_c,
\end{split}
\label{eq:kappa}
\end{equation}
where $Q$ denotes the set of \emph{joint} non-location type contexts (\emph{soft} constraints), $Q_i$ denotes a \emph{single} context category, and $V$ denotes the vocabulary set of the review text and tags.

A seed set of such labeled examples of term-context (single or joint) association pairs can then be used to define a modified similarity score function $\psi$. In contrast to the text-based function of Equation \ref{eq:phi}, this also takes into account the information from the \emph{soft} constraints of the query context.
In particular for a given \emph{soft} constraint vector $q_U$ in the user query, we use embedded word vector representations to aggregate the similarities of each word in the review text/tag of a user profile with the seed words assessed as relevant for a single or a joint context $q_U$.
Formally, $\forall w \in U$ we define two functions of the form $\psi:(w, q_U) \mapsto \mathbb{R}$, one each for the addressing the single and the joint contexts, as shown in Equation \ref{eq:psi}.
\begin{equation}
\begin{split}
\psi_s(w, q_U) & = \max (\vec{w}\cdot \vec{s}),\,s \in \cup \{t:\kappa_s(t, q_U)>0\}\\
\psi_j(w, q_U) & = \max (\vec{w}\cdot \vec{s}),\,s \in \cup \{t:\kappa_j(t, q_U)=1\}
\label{eq:psi}     
\end{split}
\end{equation}
Equation \ref{eq:psi} shows that for each word $w$ (embedded vector of which is represented as $\vec{w}$) contained in the text from the historical profile of a user, we compute its maximum similarity:
\begin{itemize}[label=\textbullet]
\item In the case of single context ($\psi_s$), over all seed words, and
\item In the case of the joint context ($\psi_j$), over a subset of seed words relevant only for the given context, i.e., the words for which $\kappa(q_U,s)=1$.
\end{itemize}
We use \texttt{word2vec} \cite{Word2vec_NIPS2013}, to embed the vector representation of a word (similar to the KDEFRLM approach described in Section \ref{sec:kdefrlm}).

The reason for using the maximum as the aggregate function in Equation \ref{eq:psi} is that a word is usually semantically similar to a small number of seed
words relevant to a given context.
To illustrate this with an example, 
for the query context `holiday, day-trip, friends', the relevant seed set constitutes words such as `base-ball stadium', `beer-garden', `salon', `sporting-goods-shop', etc. However, a word such as `pub' is similar to only one member of this seed set, namely `beer-garden', which means that other aggregation functions, such as averaging, can lead to a low aggregated value, which is not desirable in this case.

\subsection{Factored Relevance Model with \emph{Soft} Constraints}

To incorporate the multi-contextual appropriateness measure into our proposed factored relevance model (FRLM), we combine both the text-based similarity $\phi$ (Equation \ref{eq:phi}), and the trip context driven similarity function $\psi$ ($\psi_s$ or $\psi_j$ of Equation \ref{eq:psi}) into our proposed relevance models. Specifically, the user profile based RLM of Equation \ref{eq:rlmprofile} is generalized as shown in Equation \ref{eq:rlmprofile_withPsi}.
\begin{equation}
    P(w|\theta_{U, q_U}) = \sum_{(D,T,r) \in U} r P(w|D) \psi(w, q_U) \prod_{t \in T'}P(t|D) \label{eq:rlmprofile_withPsi}
\end{equation}
In addition to addressing the semantic relationship between a user assigned tag and a term present in the POI description, this relevance model of Equation \ref{eq:rlmprofile_withPsi} also takes into account the trip-qualifier based contextual appropriateness of a term $w$ by the use of the $\psi(w, q_U)$ factor. A higher value of this factor indicates that either $w$ is itself one of the seed words in an existing knowledge base or its embedded vector is close to one of the seed words, thus indicating its likely contextual appropriateness.
It is worth noting that substituting an identity function for $\psi(w, q_U)$, i.e., $\psi_l: (w, q) \mapsto 1$, degenerates the general case to the particular case of \emph{location-only} user-profile based RLM of Equation \ref{eq:rlmprofile}.

In a similar manner, the \emph{exploration} part of the model (Equation \ref{eq:flm}) is generalized as shown in Equation \ref{eq:flm_withPsi}.
\begin{equation}
    P(w|\theta_{U, q_U, l_U}) = \sum_{d \in \topM} P(w|d) \psi(w, q_U) \prod_{t \in \theta_{U, q_U}} P(t|d) \label{eq:flm_withPsi}
\end{equation}
\hltIRJ{More specifically, the \emph{soft} constraint similarity function, for which we use the generic notation $\psi$, is in fact, substituted with three different functions, namely $\psi_l$, $\psi_s$, and $\psi_j$, respectively modeling the location constraint only, a single-context, and a joint-context.}


The word-semantics enriched relevance models (KDEFRLM) can also
be generalized by incorporating the $\psi$ function within them to further generalize them to address multiple contexts.
Similar to the non-semantic version of the factored relevance model, the multi-contextual appropriateness measure, $\psi(w, q_U)$, is incorporated into the KDE based FRLM model as a part of the kernel function weights $\alpha_t = P(w|\mathcal{M}) \psi(w, q_U) P(t|\mathcal{M})$ in Equation \ref{eq:KDERLM_1D_fullForm}, as shown in Equation \ref{eq:KDERLM_2D_fullForm_withPsi}.
\begin{equation} \label{eq:KDERLM_2D_fullForm_withPsi}
    P(w|\theta_{U, q_U};h,\sigma) = \sum_{t \in T'} \Big(\!\!\!\!\!\!\sum_{\ \ (D,T,r) \in U} \!\!\!\! r P(w|D)\Big) \psi(w, q_U) P(t|\mathcal{M}) \frac{1}{\sigma \sqrt{2 \pi}} exp (-\frac{(\mathbf{w}-\mathbf{t})^T (\mathbf{w}-\mathbf{t})}{2\sigma^2 h^2})
\end{equation}

%
Finally, the \emph{exploration} side of the model is generalized as shown in Equation \ref{eq:kdeflm_withPsi}.
\begin{equation}
P(w|\theta_{U, q_U, l_U};h,\sigma) = \sum_{d \in \topM}
\frac{1}{\sigma \sqrt{2 \pi}}
P(w|d) \prod_{t \in \theta_{U, q_U}}
P(t|d)
\psi(w, q_U)
\exp (-\frac{(\mathbf{w}-\mathbf{t})^T (\mathbf{w}-\mathbf{t})}{2\sigma^2 h^2})
\label{eq:kdeflm_withPsi}
\end{equation}
Equation \ref{eq:kdeflm_withPsi} is the most general among our proposed family of models, the contributing factors being
\begin{enumerate}[leftmargin=*]
    \item $\theta_{U, q_U}$, which takes into account an enriched user profile while matching against POIs of the current location,
    \item $\exp (-\frac{(\mathbf{w}-\mathbf{t})^T (\mathbf{w}-\mathbf{t})}{2\sigma^2 h^2})$, which addresses the semantic association between tags and document terms (both user profile and POI descriptors of the current location), and
    \item $\psi(w, q_U)$, which factors in the trip-qualifier based contextual appropriateness.
\end{enumerate}

\section{Experimental Setup} \label{sec:setup}

Our experiments are conducted with the TREC Contextual Suggestion (TREC-CS) 2016 Phase-1 task \cite{hashemi2016overview}. The task requires a system to return a ranked list of 50 POIs (from a pre-defined collection) that best fit the user preference history and the user's current context.
The (query) context is comprised of a \emph{hard} location constraint, and $c=3$ different non-location type \emph{soft} qualifiers outlined in Table \ref{Table:softConstraints}.

We now first define the POI and user profile representation, followed by a detailed description of the data sets used for our experiments. We then describe the methods investigated in our experiments, following which, we present the results and their analysis.

\subsection{Representation of POIs and user profiles} \label{ss:poi-profile}
In our experiment setup, each document $D \in \mathcal{D}$ is represented as a bag-of-words which is comprised of descriptive information about the POI (available as a part of the crawled TREC web corpus) and other available information such as review texts collected from a location based social network (LBSN), viz. Foursquare. The combined use of the web crawl and content collected from LBSN as a static corpus complies with the standard experimental setup of most systems which participated in the TREC contextual suggestion (TREC-CS) tracks over a number of years \cite{hashemi2016overview}.

\hltIRJ{We note at this point that the crawled web content is likely to have been substantially different} across different systems participating over a number of years in the TREC-CS tracks (primarily due to the dynamic nature of the content present in different LBSNs, and also because of changes in the APIs used to obtain the data). Consequently, the results reported by different TREC-CS participating systems are somewhat difficult to compare against one another.
Instead of directly comparing against the reported results from the TREC-CS track overview papers, to ensure reproducibility and fairness in comparison of results, we apply a number of approaches within the same experimental framework.

Moreover, a majority of the TREC-CS participating systems made use of external data, such as ratings from other users, category information, external review texts etc. for their experimental setup. These systems, therefore, depend heavily on a number of different LBSN data sources, such as Trip Advisor, Yelp, Foursquare etc., which again makes the results difficult to compare due to the dynamic nature of the data and the APIs.
To overcome reproduciblity and fairness concerns,
our experiment setup makes use of a static data collection of POI contents. Moreover, while it may be argued that applying a combination of post-processing techniques such as rule based heuristics developed from external knowledge resources \cite{MostafaTRECCS16}, may further enhance the effectiveness of the methods investigated (including our proposed approaches), we do not employ any post processing techniques in our experiments. This is primarily because the purpose of our experiments is to investigate the effectiveness of different POI retrieval approaches under a data-driven controlled setup, and relying on a set of pre-existing rules defeats the purpose, because these rules are prone to changes with changes in the data, thus making such rule-based approaches not scalable.  


For all our experiments, we only use a part of the user profile information, specifically, the POIs with
a user-assigned rating higher than a threshold value. In the TREC-CS 2016 data, ratings are integers within $[-1, 4]$. As per the general user profile representation (Equation \ref{eq:profile}), each rating value is normalized within $[0, 1]$ (by min-max normalization). We then apply a threshold of $0.8$ to define the \emph{relevant} set of POIs for a user, i.e., these are the ones that are eventually used to construct the user profile for FRLM and KDEFRLM estimation. Formally speaking, in our experiments, the user profile $U$ (Section \ref{ss:irsetupforcs}) is comprised of only those triples, of the form $(D, T, r)$, where $r\geq 0.8$.
\hltIRJ{According to the TREC-CS task description, POIs with ratings $3$ or higher (where TREC-CS ratings are within $[-1, 4]$) in a user profile are considered to be positive or relevant (i.e., liked by the user). As we have normalized the ratings within $[0, 1]$ by min-max normalization, threshold value for selecting positively rated POIs i.e., $3$ maps to $0.8$.}



\subsection{Dataset} \label{Data}

\begin{table}[t]
\centering
\small
\caption{TREC-CS 2016 collection statistics.}
\label{Table:Dataset}
\begin{tabular}{ll}
    \toprule
    Information & Value\\
    \midrule
    Total number of POIs in corpus & 1,235,844\\
    Number of cities per user profile & 1 or 2\\
    Number of rated POIs per user profile & 30 or 60\\
    Total number of candidate cities & 164\\
    Number of candidate cities used by TREC & 48\\
    Maximum number of POIs per city & 23,939\\
    Minimum number of POIs per city & 1,070\\
    Average number of POIs per city & 4,543.54\\
    Total number of user profiles & 438\\
    Number of user profiles used by TREC & 61\\

    \bottomrule
\end{tabular}
\end{table}

\compactpara{TREC-CS 2016 Data}

\hltIRJ{One of the reasons why we follow the TREC-CS 2016 framework} \cite{hashemi2016overview} \hltd{is that this framework, unlike others, facilitates a Cranfield-style evaluation with pool based relevance judgements, which is a key component in \emph{Cranfield tradition} research paradigm} \cite{Cranfield_Cleverdon}\hltd{, makes it a better choice for our experiments over other frameworks/datasets such as Yelp dataset\footnote{\url{https://www.kaggle.com/yelp-dataset/yelp-dataset}}.}
A static web crawl of the TREC-CS 2016 collection has been released by TREC. There are around 1.2 million POIs in the TREC-CS 2016 collection that are based on 164 seed cities, out of which 48 of these seed cities were officially considered by TREC for experiments.
Although the collection has a total of 438 user profiles, TREC officially used 61 profiles for the Phase-1 task, and released corresponding relevance assessments for these 61 user profiles. Table \ref{Table:Dataset} shows a brief statistics of the TREC-CS 2016 collection. In each user profile, preference history is available for 1 or 2 seed cities with 30 or 60 POIs (i.e. 30 POIs per city), that have been rated by the user. Technically, a system needs to make contextual suggestion from a total of 48 seed cities for those 61 user profiles.
\begin{figure*}[t]
\centering
\begin{subfigure}[t]{0.49\textwidth}
\centering
    \includegraphics[width=\textwidth]{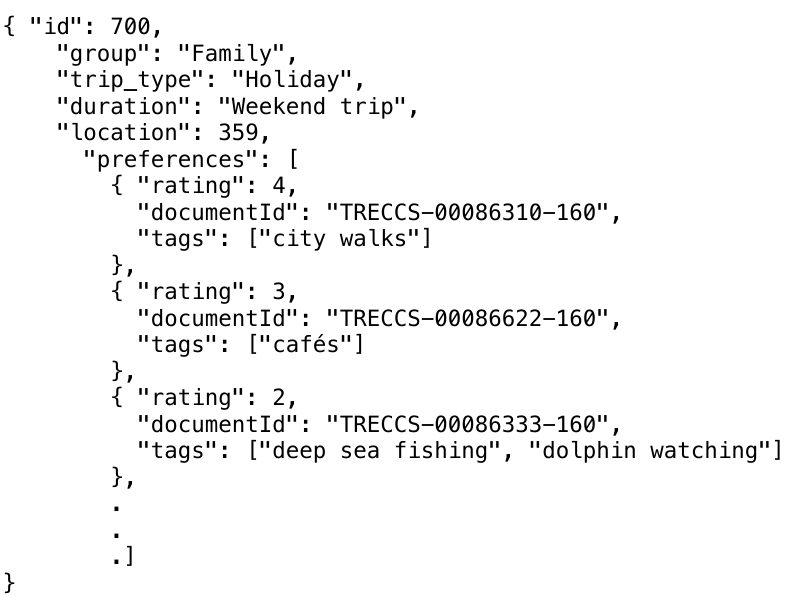} 
     \caption{User profile (ID: 700)}
     \label{fig:userProfileExample}
\end{subfigure}
\begin{subfigure}[t]{0.49\textwidth}
\centering
     \includegraphics[width=\textwidth]{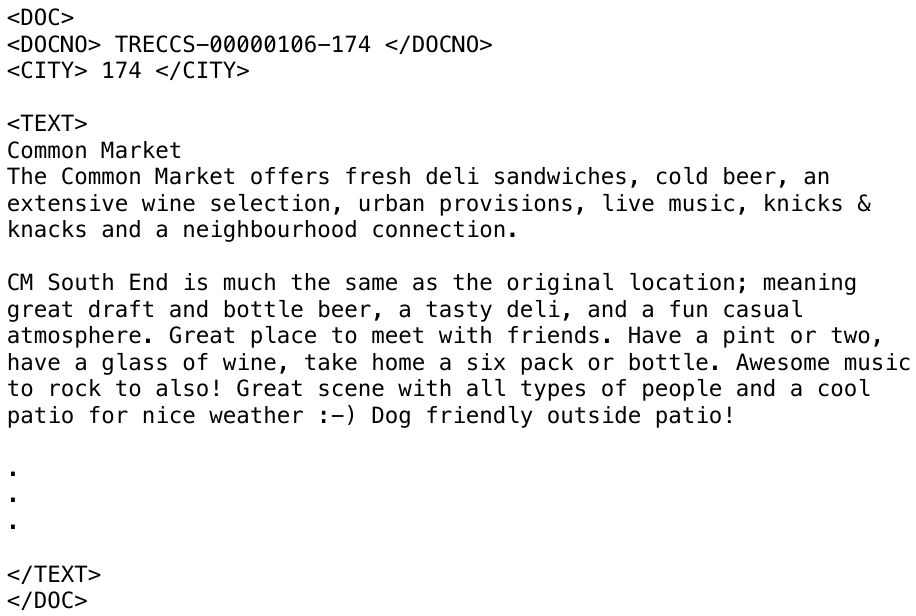} 
     \caption{Document representation of POI (ID: TRECCS-00000106-174)}
     \label{fig:poiAsDoc}
\end{subfigure}
\begin{subfigure}[t]{0.49\textwidth}
\centering
    \includegraphics[width=\textwidth]{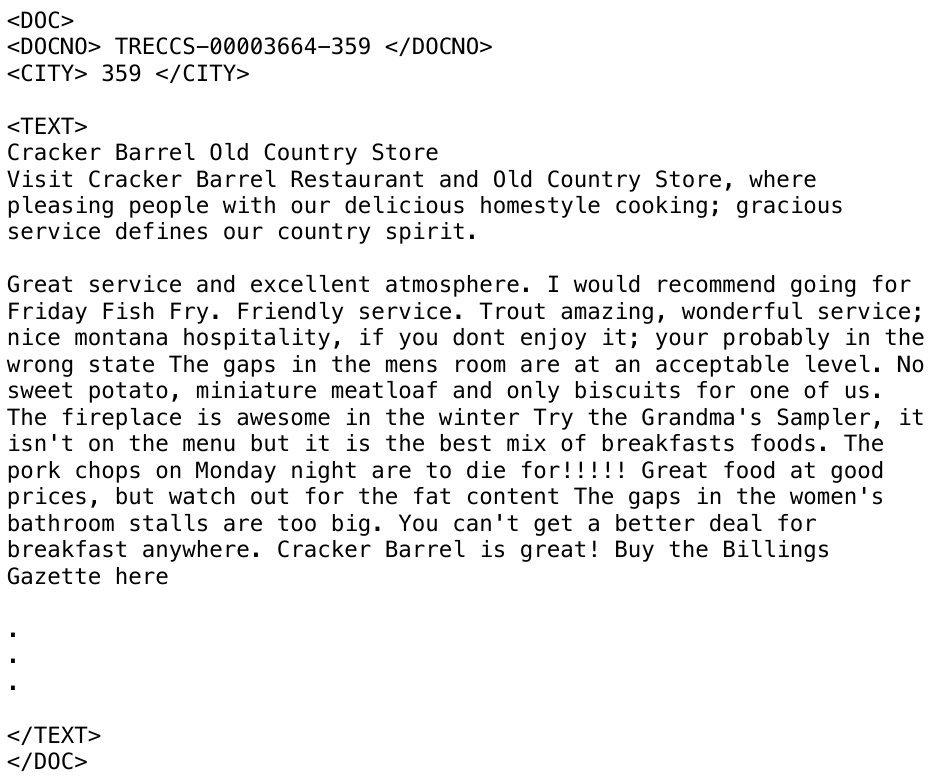} 
     \caption{POI (ID: TRECCS-00003664-359) relevant to user (ID: 700)}
     \label{fig:poiAsDoc_Rel}
\end{subfigure}
\begin{subfigure}[t]{0.49\textwidth}
\centering
     \includegraphics[width=\textwidth]{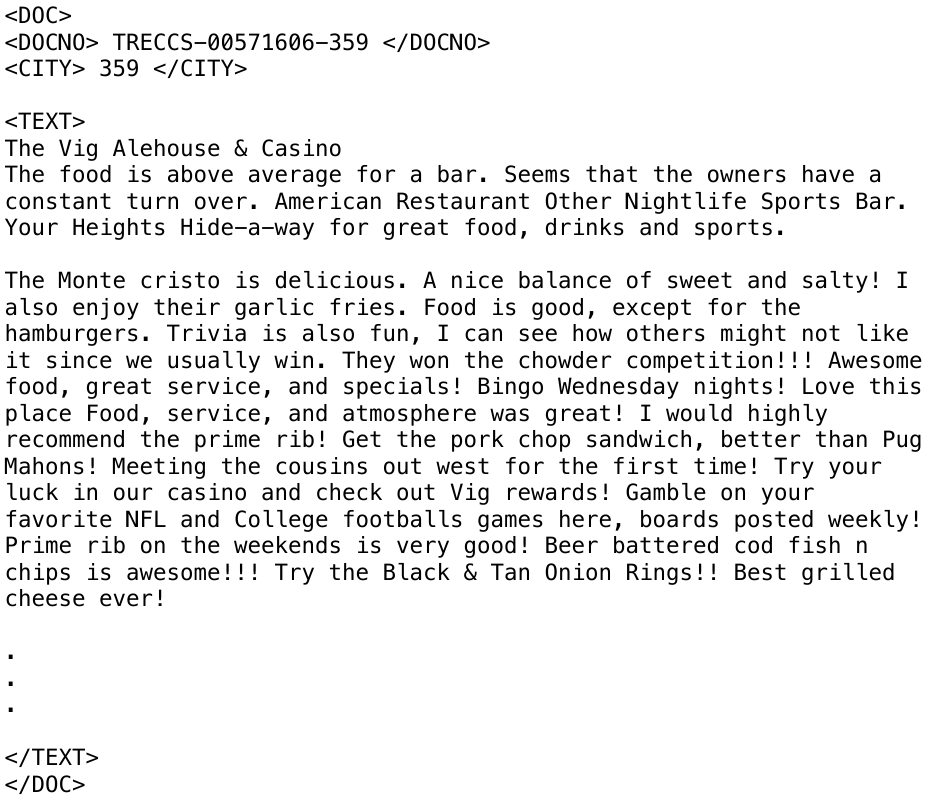} 
     \caption{POI (ID: TRECCS-00571606-359) not relevant to user (ID: 700)}
     \label{fig:poiAsDoc_NonRel}
\end{subfigure}
\caption{\hlt{Sample user profile (user ID: 700), and document representation of POIs from TREC-CS 2016 collection.}}
\label{fig:DataExample}
\end{figure*}

\hlt{Fig. }\ref{fig:userProfileExample} \hlt{shows a sample user profile} with user ID.: $700$, which has a set of POIs that the user visited in the past. The user rated the POI `TRECCS-00086310-160' with rating $4$, and assigned a tag ``city walks'', for instance. This user profile also contains the user's current contextual information such as the city identifier $359$, which maps to city Billings, MT, USA, as the \emph{hard} location context, and other non-location type \emph{soft} contexts such as \texttt{trip-type=holiday}, \texttt{trip-duration=weekend-trip}, and \texttt{accompanied-by=family}. A sample document representation of a POI is shown in Fig. \ref{fig:poiAsDoc}\hltd{, which follows the traditional TREC document format} \cite{harman1996TREC4overview}\hltd{. Each document is constituted of a `DOCNO' field representative of its unique document (POI) identifier, and a `CITY' field containing the city identifier of the POI.
The city identifier $174$ (of this example) as per the dataset} \cite{hashemi2016overview} \hltd{maps to Charlotte, NC, USA, which is the actual location of the POI.
A `TEXT' field representing the description of the POI
contains the main content i.e. the descriptive texts about the POI, and/or the available review texts. Similarly, each rated POI available in the user profile such as the POI with ID `TRECCS-00086310-160' in Fig.} \ref{fig:userProfileExample} \hltd{has its unique document representation.}

\hlt{Fig.} \ref{fig:poiAsDoc_Rel}, and \ref{fig:poiAsDoc_NonRel} \hlt{show a sample relevant (contextually appropriate)}, and a sample non-relevant document, respectively for the user profile (ID.: $700$). The non-relevant POI (ID: TRECCS-00571606-359), which is essentially a bar and/or casino, is possibly more appropriate when the user is with her friends. Note that both the relevant POI (ID: TRECCS-00003664-359), and the non-relevant POI (ID: TRECCS-00571606-359) are in the same city (ID: 359). However, the POI (ID: TRECCS-00000106-174) is in a different city (ID: 174), hence obviously non-relevant for the user of this example.

\compactpara{Details of the resource for modeling \emph{soft} contextual constraints}
We noted earlier in Section
\ref{ss:weakSupervision_forSoftContext}
that Aliannejadi et al. \cite{Aliannejadi_crowdsourceData} released a manually assessed dataset\footnote{Available at \url{https://www.inf.usi.ch/phd/aliannejadi/data.html}} comprising two different types of knowledge bases for
information corresponding to a seed set of term-context associations. Single context based appropriateness scores of some instances of association between a term or a short phrase and a single context are shown in Table \ref{Table:CrowdsourcedContextualRelevanceSingle}. The appropriateness scores lie within $[-1, +1]$, $-1$ being completely inappropriate and $+1$ being completely appropriate. The first row of Table \ref{Table:CrowdsourcedContextualRelevanceSingle} shows that 12 assessors agreed that `American Restaurant' is appropriate for the `trip-duration=Weekend-trip' context. The average appropriateness score (from 7 assessors) for `American Restaurant' is $0.7142$ when the context is `trip-duration=Longer'. `Nightlife Spot', as expected, is judged to be inappropriate for `accompanied-by=Family'.

For the joint context based appropriateness measure (Table \ref{Table:CrowdsourcedContextualRelevanceJoint}), the scores are either $-1$ or $+1$, $+1$ being contextually appropriate and $-1$ being contextually inappropriate. It can be seen that an `Irish pub' or a `Movie Theater' is very appropriate (appropriateness score of 1.0), when a user is accompanied by her friends on a holiday trip. Similarly a `Steakhouse' is appropriate when the joint context is business trip (trip-type), family (accompanied-by) and longer trip (trip-duration). Although a `bar' is appropriate for the joint context ``Holiday, Alone, Weekend trip'', it is judged to be inappropriate in the context of a weekend trip with family.

The contextual appropriateness data contains a total of 11 different contextual categories - $3$ instances of `trip-type' context (business trip, holiday or other trip), $4$ instances of `trip-duration' context (day trip, longer, night out or weekend trip), and $4$ instances of `accompanied-by' context (alone, family, friends or other). Assessments are available for $179$ most frequent Foursquare category tags and $27$ unique combinations of three contextual constraints.
For our experimental setup, as shown in Equation \ref{eq:kappa}, we normalized the contextual appropriateness scores for both single and joint context, within $[0, 1]$.

\subsection{Methods Investigated}   \label{ss:methods}

We employ a number of standard IR based and recommender system (RecSys) based methodologies as baselines for comparison against our proposed models.
In addition to investigating the overall effectiveness of alternative approaches, with respect to our proposed models, we explore the following.
%
\begin{itemize}[label=\textbullet]
    \item Finding an optimal trade-off between a user's preference history (\emph{exploitation}) and the information about the POIs constrained to a \emph{hard} contextual constraint such as `location' (\emph{exploration}) for contextual POI recommendation.
    \item Finding the most effective way to include \emph{soft} contextual constraints such as `trip-type', `accompanied-by' of a given user profile into the POI recommendation framework with a particular focus to improve the precision at top ranks.
\end{itemize}

With respect to the second objective above, the choice of the \emph{soft} constraint similarity function $\psi=\{\psi_l, \psi_s, \psi_j\}$ yields three different versions for each method investigated, corresponding to: i) not using the soft constraints (i.e. location-only based retrieval), ii) using the single-context, iii) using the joint-context based similarities, respectively. In our results reported in Table \ref{Table:ResultsCombined2_psi_l} and \ref{Table:ResultsCombined2_psi_j}, we denote this choice of our model instantiation by an additional parameter for the function $\psi$. The function corresponding to only location (\emph{hard}) constraints corresponds to the constant function $\psi_l: (w, q) \mapsto \{1\}$.



\subsubsection{IR Baselines} \label{ss:IRbaselines}
To acquire the comparative effectiveness of our proposed approaches
we choose a number of baselines based on ablations of components/factors from our proposed models. 
The IR baselines are enlisted below.
\begin{enumerate}[leftmargin=*]
    \item \textbf{BL1 - BM25}:
    We employ the standard BM25 retrieval model as the similarity measure function of Equation \ref{eq:simqd}. We select user assigned tags ($T'$, as we used in Equation \ref{eq:rlmprofile}) from the set of tuples $(D, T, r)$, where $r \geq 0.8$ to form a weighted query, where for each query term $t \in T'$, we include the value of $\psi(t, q_U)$ as the weight of that term in the query. BM25 parameters $k, b$  are optimized by grid search with respect to nDCG@5.
    
    \item \textbf{BL2 - Term Selection}:
    Since our proposed models estimate a weighted term distribution, we apply a method of extracting a set of terms
    from the set of documents from the set of tuples $(D, T, r)$, where $r \geq 0.8$
    (based on BM25 weights) as one of the baselines. Note that the parameter settings of $k$ and $b$ for BM25 remain the same as that of BL1. We optimize the number of selected terms to $25$ by grid search. Additionally, similar to BL1, we include the value of $\psi(t, q_U)$ as the weight of each selected term $t$, in the query. This model is able to take into account exploitation by selecting terms from user profile.
    
    \item \textbf{BL3 - BM25 with Term Selection:}
    Since we combine both the user preference history and information about the POIs within a current context for FRLM and KDEFRLM estimation, we apply a CombSUM \cite{Fox94combSUM} technique to merge the two ranked lists obtained with BL1 (BM25) and BL2 (Term Selection). This offers a naive method of combining two sources of information, i.e. user preference history and the POI content in current contexts.

    \item \textbf{BL4 - RLM}:
    Since, at its core, our proposed approach relies on estimating a factored relevance model, we select the traditional relevance model (RLM) of Equation \ref{eq:traditional_rlm} as a baseline. Similar to BL1 (BM25), we consider the user assigned tags from the user profile with ratings $r \geq 0.8$ as observed terms (analogous to a query). We then estimate a relevance model (RLM) to rank the POIs within the current context. To incorporate the \emph{soft} contextual constraints into the traditional RLM framework,
    we include the weights obtained from the $\psi$ function (external knowledge resource) as weights into the standard RLM equation (Equation \ref{eq:rlmprofile}).
    In contrast to the factored relevance model, this baseline model only makes use of the \emph{exploration} part while formulating the query, i.e., with respect to the standard RLM \cite{Lavrenko_RLM2001:RBL:383952.383972}, the set of tags in a user history acts as the query and the RLM term weights are computed using the local co-occurrences from the top-retrieved POI descriptors constrained to a given user-specified location.
    
    
    \item \textbf{BL5 - KDERLM}: We choose word vector compositionality based relevance feedback using kernel density estimation \cite{DoiKDERLM_CIKM16} (Equation \ref{eq:KDERLM_1D}) as another baseline. This baseline corresponds to a KDE based generalized version of traditional RLM (the factored part corresponding to an enriched matching between the user profile and POIs in a current location being ablated).
    Similar to BL4 (RLM), in this baseline we also use the tags from a user profile with ratings higher than or equal to $0.8$ as observed terms (analogous to a query), and then estimate a KDE-based RLM to score POIs within a current context.
    Again, the \emph{soft} contextual constraints are incorporated within KDERLM (Equation \ref{eq:KDERLM_1D}) as weighting factors
    computed with the $\psi$ function.

\end{enumerate}

Parameters for each method were independently tuned with the help of a grid search. \hltIRJ{Since our proposed models are unsupervised} (without involving any parameter learning with the help of gradient descent updates), we do not employ a setup involving separate training and test splits of the data.
\hltIRJ{It is important to note that for the fair comparisons, each method is tuned to its optimal setting given the common experimental setup. Thus, our experiments 
compare the best possible results that can
be obtained by each method, which as is, in fact, a
common practice for in standard IR tasks }\cite{crossLingualRLM_Lavrenko_SIGIR2002, PassageRetrieval_LM_LiuCroft_CIKM2002, IRfromNoisy_IPM2016}.
\hltIRJ{This setting is particularly applicable in cases when no standard partitioning of a dataset exists, e.g., the partition of the TREC-Robust query set into TREC-6, TREC-7 etc.} \cite{harman1996TREC4overview, Voorhees1999TREC8}.
\hltIRJ{Moreover, the fact that the relevance model based approaches are not too sensitive to the values of hyper-parameters such as the number of top-documents or the number of feedback terms} \cite{Lavrenko_RLM2001:RBL:383952.383972}, \hltIRJ{suggests that the trends that we observed in our results are likely to be similar with experiment setups involving hyper-parameter tuning on development sets, e.g. those involving cross-validation or with held-out development and test sets of queries}.  

The two common parameters to all the relevance feedback models are the number of feedback documents, $M$, and the number of feedback terms, $\tau$.
It was found after a grid search that RLM and FRLM yielded optimal results with the values $5$ (\#documents) and $25$ (\#terms). Similarly for KDERLM, $M$ and $\tau$ were optimized to the values $3$ and $80$, whereas for KDEFRLM, the optimal values of $M$ and $\tau$ were found to be $2$ and $100$, respectively.

\subsubsection{Recommender System Baselines} \label{ss:RecSysbaselines}

%

In the absence of other users' ratings, it is not possible to apply standard recommender system (RecSys) approaches, such as collaborative filtering, directly to predict the relevance of a POI (considered as an item in RecSys research). However, a disparate analogy allows us to employ standard RecSys methodologies as a pre-processing step in our experimental setup. Specifically, one may imagine that the contents in user profiles are analogous to users in RecSys terminology, whereas the set of user-assigned tags used to describe POIs are analogous to items. This user-item analogy allows us to learn semantic associations between a user profile and the tag vocabulary. Given a user profile, it is thus possible to enrich the set of tags (analogous to suggesting more items for a user in the traditional framework of RecSys research).
Following this general setup for our RecSys based experiments, we now explain the details of each RecSys based baseline approach.

\begin{enumerate}[leftmargin=*]
\setcounter{enumi}{5}
    \item \textbf{BL6 - Most Popular K}:
    A simple (but effective) RecSys methodology is the recommendation of the \emph{most popular} items based on overall ratings across all users, with the expectation that these items will be appropriate to the new user as well \cite{mostpop}. With respect to our experimental setup, we extract the $K$ most popular tags across each user's preference history. We then use these selected tags to form the query for each user. For instance, if the tag `beer' is one of the most popular tags in the tag vocabulary across all users, suggesting pubs as candidate POIs for a new user is likely to be a good recommendation.
    
    After formulating an enriched query based on the most popular tags, we apply the standard BM25 retrieval model as the similarity matching function (Equation \ref{eq:simqd}) with the same settings of $k, b$, as that in BL1. $K$ (the number of popular tags to extract for enriching the query) is optimized based on the average rating of tags across the set of all users. The threshold for this average rating was set to $0.8$. A \emph{soft} constraint based variant of this baseline includes the value of $\psi(t, q_U)$ as the weight of each selected tag $t$ in a query.

    \item \textbf{BL7 - Profile Popular K}:
    In contrast to the previous approach of finding the globally most popular tags across all users, this approach restricts the selection of the most popular tags to each user profile only. It can be argued that this approach extracts tags in an entirely personalized manner. For instance, this method selects the tag `seafood' as a query term if it is one of the most popular tags in the preference history of only the current user. Similar to BL6 (Most Popular K), BM25 is used as the similarity function (Equation \ref{eq:simqd}) with the same settings of $k, b$, as that in BL6. $K$ is optimized based on the user profile specific average rating of tags and the cut-off for average rating is set to $0.8$. Again, \emph{soft} constraints are included as $\psi(t, q_U)$ weights associated with each tag $t$ in the query.

    
    
    \item \textbf{BL8 - NeuMF}:
    We used a state-of-the-art neural network based matrix factorization method \cite{NCF_WWW2017}, which makes use of a fusion of generalized matrix factorization (GMF) and multi-layer perceptron (MLP) to better model the complex user versus item interactions (in our case, an item corresponding to a tag). Similar to the Popular-K baselines (both collaborative and personalized), the $K$ most likely tags, as predicted by the NeuMF model, are then used to construct a weighted query, using the $\psi$ function as the weight values similar to the previously described approaches.

    \item \textbf{BL9 - Bayesian content-based recommendation}:
    A standard text classification based content matching technique, widely used in recommender systems, is employing a Bayesian classifier \cite{bayesian_classifier}. As per the requirement of a supervised binary classification approach, we consider the set of all positively rated documents in a user profile, i.e. all $D$s from the set of tuples $(D, T, r)$, where $r \geq 0.8$, as the `positive' class, whereas the set of all negatively rated documents in a user profile, i.e. all $D$s from the set of tuples $(D, T, r)$, where $r < 0.8$, are considered to define the `negative' class. We then train a binary Naive-Bayes classifier. During recommendation, for each POI that is classified as `positive', we consider the posterior likelihood value of the classifier as the score of the POI. We then present the ranked list by sorting the POIs in decreasing order of these likelihood scores.

    Since BL9 (Bayesian) is primarily a text classification based approach, and there is no direct notion of weighted query with varying term importance, we limit use of this baseline to our \emph{hard} constraint only experiments. 
    
\end{enumerate}

\subsubsection{Hybrid baselines}

\begin{enumerate}[leftmargin=*]
\setcounter{enumi}{9}

    \item \textbf{BL10 - Content + Tag Matching}:
    As mentioned earlier in Section \ref{ss:poi-profile}, due to the use of external data resources by the 
    TREC-CS participating systems, the
    results reported therein are not directly comparable with our results (in terms of the absolute values of the measured metrics). We therefore conduct experiments with the recorded best performing method of TREC-CS 2016 within our setup. This method involves a
    hybrid of content and tag matching \cite{Aliannejadi_ToISJournal}.
    More precisely speaking, the similarity matching function of this method is a combination of
    %
    query words/tags and document (POI) words/tags similarity (Content + Tag score) with a predicted likelihood score of the relevance between a query word and a given non-location \emph{soft} constraint category.
    
    As per
    \cite{Aliannejadi_ToISJournal},
    we trained an SVM-based binary classifier on the joint-context knowledge resource \cite{Aliannejadi_crowdsourceData} (with relevance labels 0/1) using as inputs the scores for the single contexts. While testing (i.e., at query time), the distance of a $3$-dimensional joint context input from the classifier boundary is added to the text (tag-word) matched score (higher the distance, the higher is the likelihood of a tag to be appropriate to the given joint context). We employ this approach as a baseline and denote it by `Content + Tag + SVM'.
    Additionally, we also investigate the method of adding the scores obtained from the $\psi_s$ and $\psi_j$ functions in conjunction with the `Content + Tag' approach.
    
    
    \item \textbf{BL11 - Hybrid}: We employ a CombSUM \cite{Fox94combSUM} of the two ranked lists obtained with the best performing IR-based baseline BL5 (KDERLM), and another strong baseline BL10 (Content + Tag Matching), which allows provision for an ensemble of content and tag matching.
    
\end{enumerate}

\subsection{Word Embedding Settings} 
\label{embedding}

\hlt{In this section, we discuss about word embedding setup} which is required for our embedding based model KDEFRLM, and in modeling multiple \emph{soft} contextual constraints for both FRLM and KDEFRLM. Since different choices in an embedding method, such as the embedding objective function or the collection on which the embedding model is trained on etc., may influence the retrieval effectiveness \cite{embeddingsForIR_Doi_CIKM2018}\hltd{, we explore four different ways for generating the embedded word vectors. In fact, we report the performance variation of our proposed model KDEFRLM as obtained with a number of different embedding methodologies in Table} \ref{Table:ResultsOutdomain_psi_l} and \ref{Table:ResultsOutdomain_psi_j}.

Distances between word vectors are used in
the kernel density based approaches and in modeling the soft constraints. Specifically, for our experiments (Table \ref{Table:ResultsCombined2_psi_l} and \ref{Table:ResultsCombined2_psi_j}) the embedded space of word vectors is obtained by executing skipgram \cite{Word2vec_NIPS2013} with default values for the parameters of window-size (5) and the number of negative samples (5), as set in the \texttt{word2vec} tool\footnote{\url{https://github.com/tmikolov/word2vec}}. Skipgram was trained on the collection of the POI descriptors in the TREC-CS collection. \hltd{We mention this version of word embeddings as \texttt{word2vec-In} i.e. \emph{In-domain}} (Table \ref{Table:ResultsOutdomain_psi_l} and \ref{Table:ResultsOutdomain_psi_j}) \hltd{ as it is trained on the target corpus.}


\hlt{As an alternative to training word vectors on the target collection}, we also explore pre-trained word vectors trained on large external corpora, which is a common practice for supervised NLP downstream tasks \cite{embeddingsForIR_Doi_CIKM2018, BERT_devlin} \hltd{. Specifically, we employ two word embedding methodologies \texttt{word2vec} (Out-domain), and \texttt{GloVe} (Out-domain)} \cite{GloVe}\hltd{. We also employ \texttt{BERT} (Out-domain)} \cite{BERT_devlin} \hltd{which is a contextual embedding method that uses masked language models.}

\hlt{While the $300$ dimensional \texttt{word2vec} pre-trained vectors} that we used were trained on Google news dataset\footnote{Available at \url{https://code.google.com/archive/p/word2vec/}}\hltd{, the $300$ dimensional \texttt{GloVe} vectors that we used were trained on the Common Crawl}\footnote{Available at \url{https://nlp.stanford.edu/projects/glove/}}\hltd{. The transformer based contextual vectors that we used for our experiments uses the pre-trained \texttt{RoBERTa}} \cite{RoBERTa} \hltd{model, which is an optimized version of the original BERT model. Given a word, the RoBERTa model outputs a $768$ dimensional vector. Since the objective of this set of experiments is to investigate the effect of different embedding approaches on the effectiveness of our proposed model, the remaining parameters for KDEFRLM method such as the number of feedback documents, $M$, and the number of expansion terms, $\tau$, were set to their optimal values as tuned on the \texttt{word2vec-In} experiments.}

\section{Results and Discussion}   \label{sec:results}

We first report the results of our set of experiments and summarize the overall observations. Then we investigate the sensitivity analysis of our models with different contextual constraint settings. \hlt{Finally, we discuss about the effect of different embedding techniques on the effectiveness of our proposed model.}

\subsection{Overall Observations}

Table \ref{Table:ResultsCombined2_psi_l} and \ref{Table:ResultsCombined2_psi_j}
show the results obtained by each contextual recommendation approach that we investigated, as outlined in Section \ref{ss:methods}.
Each method was separately optimized with grid search on the nDCG@5 metric, the official metric to rank systems in the TREC-CS task.
\hltIRJ{Since the effectiveness of a particular approach (e.g. FRLM) in comparison to a baseline (e.g. RLM) is comparable across the same setting (i.e., location-only or location + soft constraints), we present the optimal results for location only setting (i.e., $\psi_l$) in Table }\ref{Table:ResultsCombined2_psi_l}\hltIRJ{, and the optimal results for location +\emph{soft} constraints setting (i.e., $\psi_s$ and $\psi_j$) in Table }\ref{Table:ResultsCombined2_psi_j}.


In summary, from Table \ref{Table:ResultsCombined2_psi_l} and \ref{Table:ResultsCombined2_psi_j} we can see that the word semantics based extension of our proposed factored relevance model, i.e., KDEFRLM, outperforms all other methods for both the location-only (hard) and `location + trip-qualifier' (hard and soft) constrained contextual POI recommendation tasks.
A paired $t$-test showed that the improvements in nDCG@5, nDCG@10, nDCG, P@5, P@10, and MAP with KDEFLRM were statistically significant (95\% confidence level) in comparison to the three strongest baselines: BL5 (KDERLM), BL11 (Hybrid) and BL1 (BM25).
We now highlight and comment on the key observations from our set of experiments.

\compactpara{Factored models (exploration and exploitation) outperform the other approaches}
The superior performance of the factored models (FRLM and KDEFRLM) in comparison with BL2 (Term Selection) indicates that the probability distribution of weighted terms, as estimated by the factored models, is a more effective way to select candidate terms for query formulation. Although BL3 (BM25 + Term selection) takes both the preference history of the user (term selection based exploitation) and the top ranked POIs (BM25 based exploration) into account, the superior performance of both FRLM and KDEFRLM indicates that such information turns out to be more
effective when intricately integrated within the framework of a relevance based model, leveraging information from
both preference history and the top retrieved POIs
(rather than the ad-hoc way of first retrieval and then term selection for query expansion).

\begin{figure*}[tp]
\centering
\begin{subfigure}[t]{0.49\textwidth}
\centering
    \includegraphics[width=\textwidth]{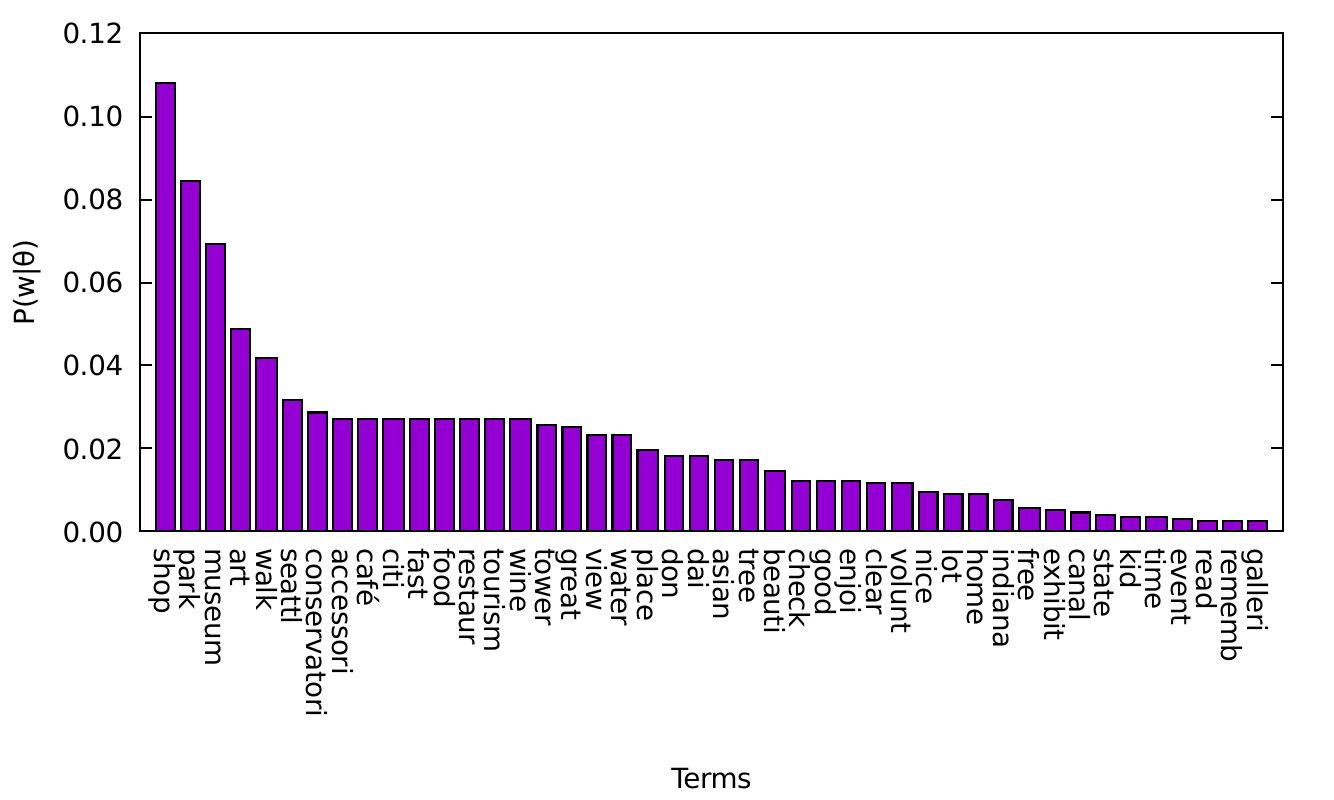} 
     \caption{FRLM ($\psi_l$)}
     \label{fig:FRLM_term_distribution_psi_l}
\end{subfigure}
\begin{subfigure}[t]{0.49\textwidth}
\centering
     \includegraphics[width=\textwidth]{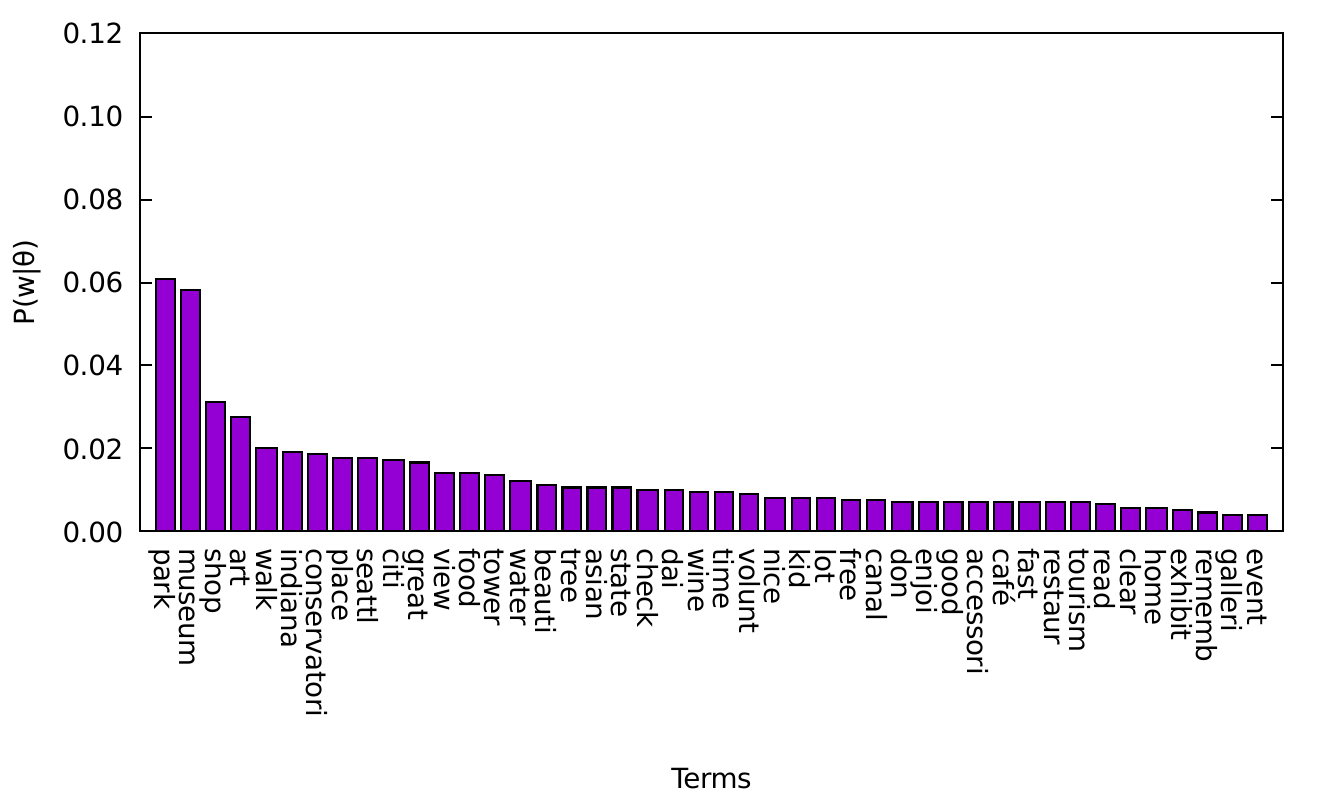} 
     \caption{KDEFRLM ($\psi_l$)}
     \label{fig:KDEFRLM_term_distribution_psi_l}
\end{subfigure}
\begin{subfigure}[t]{0.49\textwidth}
\centering
    \includegraphics[width=\textwidth]{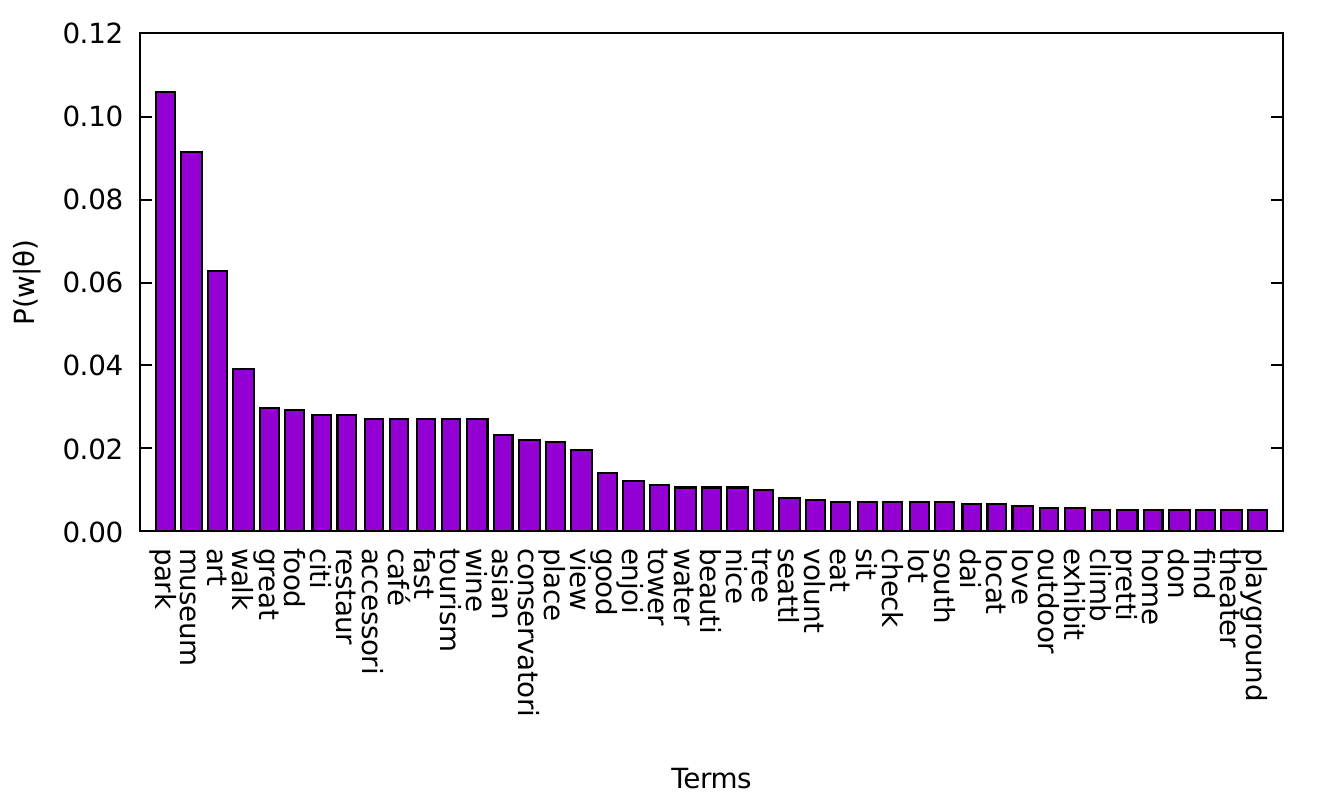} 
     \caption{FRLM ($\psi_j$)}
     \label{fig:FRLM_term_distribution_psi_j}
\end{subfigure}
\begin{subfigure}[t]{0.49\textwidth}
\centering
     \includegraphics[width=\textwidth]{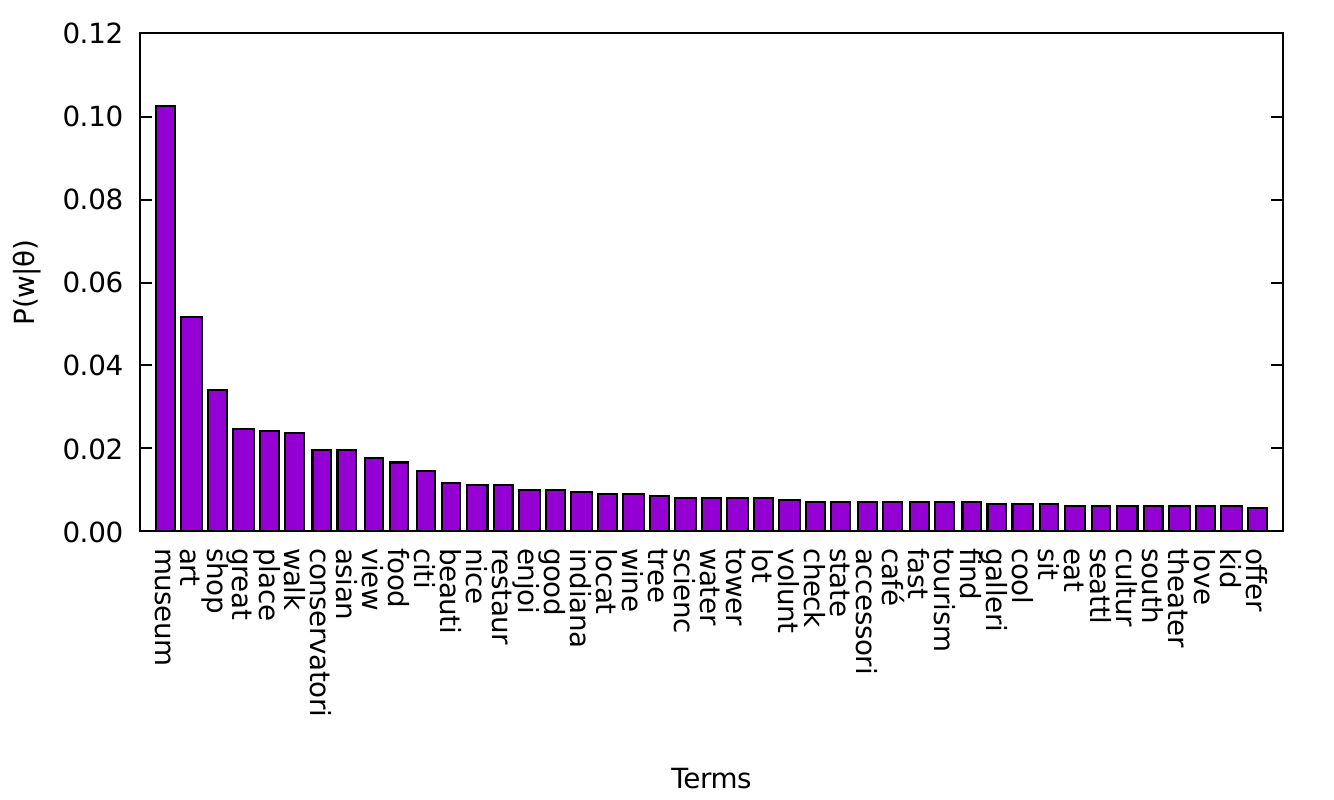} 
     \caption{KDEFRLM ($\psi_j$)}
     \label{fig:KDEFRLM_term_distribution_psi_j}
\end{subfigure}
\caption{Comparisons of term distribution weights (sorted from highest to lowest) between FRLM and KDEFRLM on
single (location) and multiple contexts (joint modeling with 
$\psi_j$).
For location-only modeling, FRLM ($\psi_l$) uses $M=5$ (number of top-retrieved documents for feedback as per the $\topM$ notation of Table \ref{Table:Notations}) and $\tau=25$ (number of top-scoring terms in the estimated RLM distributions). KDEFRLM ($\psi_l$) uses $M=2$ and $\tau=$ 80.
FRLM with joint context ($\psi_j$) uses parameters $M=5$ and $\tau=35$, whereas the results for KDEFRLM with the joint context ($\psi_j$) were obtained with $(M,\tau)=(2,100)$.
}
\label{fig:FRLM_Vs_KDE_term_distribution}
\end{figure*}

Figure \ref{fig:FRLM_Vs_KDE_term_distribution} shows the comparison of relative term distributions (common terms) between FRLM and KDEFRLM for a user request (user ID 763) where $T' = \{$\texttt{art, city-walks, caf\'{e}s, fast-food, museums, parks, restaurants, tourism, shopping-for-wine, shopping-for-accessories}$\}$. Both FRLM and KDEFRLM assign higher weights to terms such as `park', `museum', which are clearly relevant for this particular example. Indeed, both these models are also successful at capturing other relevant terms such as `view', `tree', `canal' etc.

\compactpara{Incorporating term semantics improves POI effectiveness}
We observe from Table \ref{Table:ResultsCombined2_psi_l} and \ref{Table:ResultsCombined2_psi_j} that the KDE extended versions of the factored models (for both single and multi-contexts)
mostly outperform their non-semantic (non-KDE) counterparts.
This shows that leveraging underlying term semantics of a collection in the form of an embedded space of vectors helps to retrieve more relevant POIs at better ranks.
Figure \ref{fig:FRLM_Vs_KDE_term_distribution} shows that KDEFRLM is able to capture the semantic relationship between terms better than FRLM. For example, the semantic relationship between the term `histori' (stemmed form of `history') and `museum' was successfully captured by the KDE-based variant of FRLM. This demonstrates that KDEFRLM is able to successfully leverage the semantic association between terms, in addition to those of the term-based statistical co-occurrences only. 
Table \ref{Table:W2VnearestTerms} shows a few terms whose word vectors are in close proximity of the user assigned tags in the embedded space.

\begin{table}[tp]
\centering
\caption{Comparisons between POI retrieval approaches\hltIRJ{ in location only setting ($\psi_l$).}
The notations, `$^*$', `$^\dagger$' and `$^\ddagger$' denote significant (paired t-test with 95\% confidence) improvements over the three strongest baselines - BL5 (KDERLM), BL11 (Hybrid) and BL1 (BM25), respectively.}
\label{Table:ResultsCombined2_psi_l}
\tabfitpagew
{
\begin{tabular}
    {@{}l@{\ \ }l@{\ \ \ }l@{\ }l@{\ }l@{\ }ll@{\ }l@{\ }l@{\ \ }r@{}}
    \toprule
    & & Context & \multicolumn{3}{c}{Graded Evaluation Metrics} & \multicolumn{4}{c}{Binary Evaluation Metrics}\\
    \cmidrule(r){4-6}
    \cmidrule(l){7-10}
    \multicolumn{2}{c}{Method} & $(\psi)$ & nDCG@5 & nDCG@10 & nDCG & P@5 & P@10 & MAP & MRR \\
    \midrule
    \multicolumn{2}{c}{\textbf{IR-based approaches}} & & & & \\
    \midrule
    BL1 & BM25 & $\psi_l$ & 0.2747 & 0.2484 & 0.2889 & 0.3934 & 0.3066 & 0.1326 & 0.6539\\
    BL2 & Term Sel. & $\psi_l$ & 0.2484 & 0.2383 & 0.3034 & 0.3639 & 0.3066 & 0.1466 & 0.6148\\
    BL3 & BM25 + Term Sel. & $\psi_l$ & 0.2411 & 0.2332 & 0.3143 & 0.3672 & 0.3115 & 0.1530 & 0.5607\\
    BL4 & RLM
    & $\psi_l$ & 0.2615 & 0.2453 & 0.3091 & 0.3574 & 0.3033 & 0.1437 & 0.6441\\
    BL5 & KDERLM
    & $\psi_l$ & 0.2829 & 0.2682 & 0.3191 & 0.3967 & 0.3361 & 0.1495 & 0.6539\\
    \midrule
    \multicolumn{2}{c}{\textbf{RecSys based approaches}} & & & & \\
    \midrule
    BL6 & Most Popular K & $\psi_l$ & 0.1861 & 0.1926 & 0.2580 & 0.2787 & 0.2705 & 0.1016 & 0.4154\\
    BL7 & Profile Popular K & $\psi_l$ & 0.2488 & 0.2409 & 0.2811 & 0.3410 & 0.3016 & 0.1280 & 0.6486\\
    BL8 & NeuMF
    & $\psi_l$ & 0.1626 & 0.1655 & 0.2480 & 0.2361 & 0.2344 & 0.0937 & 0.4314\\
    BL9 & Bayesian & $\psi_l$ & 0.2170 & 0.1774 & 0.1816 & 0.3082 & 0.2082 & 0.0672 & 0.5831\\
    \midrule
    \multicolumn{2}{c}{\textbf{Hybrid approaches}} & & & & \\
    \midrule
    BL10 & Content + Tag.
    & $\psi_l$ & 0.2499 & 0.2411 & 0.2800 & 0.3967 & 0.3377 & 0.1330 & 0.5390\\
    BL11 & Hybrid (BL5 + BL10) & $\psi_l$ & 0.2805 & 0.2667 & 0.3329 & 0.3902 & 0.3311 & 0.1583 & 0.6514\\
    \midrule
    \multicolumn{2}{c}{\textbf{Proposed approaches}} & & & & \\
    \midrule
    & FRLM ($\gamma_H=0.8$) & $\psi_l$ & 0.2919 & $0.2810^{\ddagger}$ & $0.3418^{* \dagger \ddagger}$ & 0.3934 & $0.3443^{\ddagger}$ & $0.1616^{* \ddagger}$ & \textbf{0.6786}\\
    & KDEFRLM ($\gamma_H=0.6$) & $\psi_l$ & $\textbf{0.2996}^{\dagger \ddagger}$ & $\textbf{0.2868}^{\dagger \ddagger}$ & $\textbf{0.3490}^{* \dagger \ddagger}$ & $\textbf{0.4295}^{\dagger \ddagger}$ & $\textbf{0.3656}^{\dagger \ddagger}$ & $\textbf{0.1725}^{* \dagger \ddagger}$ & 0.6553\\
    \bottomrule
\end{tabular}
}
\end{table}

\begin{table}[tp]
\centering
\caption{Comparisons between POI retrieval approaches\hltIRJ{ in location + \emph{soft} constraints setting ($\psi_j$).}
The notations, `$^*$', `$^\dagger$' and `$^\ddagger$' denote significant (paired t-test with 95\% confidence) improvements over the three strongest baselines - BL5 (KDERLM), BL11 (Hybrid) and BL1 (BM25), respectively.}
\label{Table:ResultsCombined2_psi_j}
\tabfitpagew
{
\begin{tabular}
    {@{}l@{\ \ }l@{\ \ \ }l@{\ }l@{\ }l@{\ }ll@{\ }l@{\ }l@{\ \ }r@{}}
    \toprule
    & & Context & \multicolumn{3}{c}{Graded Evaluation Metrics} & \multicolumn{4}{c}{Binary Evaluation Metrics}\\
    \cmidrule(r){4-6}
    \cmidrule(l){7-10}
    \multicolumn{2}{c}{Method} & $(\psi)$ & nDCG@5 & nDCG@10 & nDCG & P@5 & P@10 & MAP & MRR \\
    \midrule
    \multicolumn{2}{c}{\textbf{IR-based approaches}} & & & & \\
    \midrule
    \multirow{2}{*}{BL1} & \multirow{2}{*}{BM25} & $\psi_s$ & 0.2609 & 0.2441 & 0.2889 & 0.3869 & 0.3164 & 0.1335 & 0.5967\\
    & & $\psi_j$ & 0.2641 & 0.2464 & 0.2916 & 0.3639 & 0.3033 & 0.1355 & 0.6565\\
    \multirow{2}{*}{BL2} & \multirow{2}{*}{Term Sel.} & $\psi_s$ & 0.2424 & 0.2411 & 0.3039 & 0.3607 & 0.3148 & 0.1458 & 0.6186\\
    & & $\psi_j$ & 0.2539 & 0.2447 & 0.3099 & 0.3705 & 0.3197 & 0.1514 & 0.6419\\
    \multirow{2}{*}{BL3} & \multirow{2}{*}{BM25 + Term Sel.} & $\psi_s$ & 0.2462 & 0.2471 & 0.3207 & 0.3672 & 0.3344 & 0.1578 & 0.6095\\
    & & $\psi_j$ & 0.2530 & 0.2429 & 0.3195 & 0.3869 & 0.3328 & 0.1557 & 0.6191\\
    \multirow{2}{*}{BL4} & \multirow{2}{*}{RLM}
    & $\psi_s$ & 0.2583 & 0.2466 & 0.3107 & 0.3475 & 0.3016 & 0.1443 & 0.6441\\
    & & $\psi_j$ & 0.2692 & 0.2514 & 0.3189 & 0.3639 & 0.3131 & 0.1496 & 0.6544\\
    \multirow{2}{*}{BL5} & \multirow{2}{*}{KDERLM}
    & $\psi_s$ & 0.2839 & 0.2668 & 0.3236 & 0.3902 & 0.3902 & 0.1530 & 0.6639\\
    & & $\psi_j$ & 0.2772 & 0.2666 & 0.3287 & 0.3869 & 0.3311 & 0.1594 & 0.6623\\
    \midrule
    \multicolumn{2}{c}{\textbf{RecSys based approaches}} & & & & \\
    \midrule
    \multirow{2}{*}{BL6} & \multirow{2}{*}{Most Popular K} & $\psi_s$ & 0.1765 & 0.1894 & 0.2579 & 0.2590 & 0.2689 & 0.1015 & 0.4055\\
    & & $\psi_j$ & 0.1877 & 0.1844 & 0.2590 & 0.2656 & 0.2475 & 0.1010 & 0.4247\\
    \multirow{2}{*}{BL7} & \multirow{2}{*}{Profile Popular K} & $\psi_s$ & 0.2529 & 0.2381 & 0.2861 & 0.3639 & 0.3016 & 0.1321 & 0.6296\\
    & & $\psi_j$ & 0.2568 & 0.2487 & 0.2908 & 0.3574 & 0.3098 & 0.1362 & 0.6500\\
    \multirow{2}{*}{BL8} & \multirow{2}{*}{NeuMF}
    & $\psi_s$ & 0.1491 & 0.1601 & 0.2466 & 0.2131 & 0.2344 & 0.0935 & 0.3969\\
    & & $\psi_j$ & 0.1698 & 0.1834 & 0.2457 & 0.2393 & 0.2525 & 0.0923 & 0.4300\\
    \midrule
    \multicolumn{2}{c}{\textbf{Hybrid approaches}} & & & & \\
    \midrule
    \multirow{3}{*}{BL10} & \multirow{3}{*}{Content + Tag.}
    & $\psi_s$ & 0.2623 & 0.2496 & 0.2841 & 0.4066 & 0.3492 & 0.1383 & 0.5982\\
    & & $\psi_j$ & 0.2688 & 0.2651 & 0.2979 & 0.4000 & 0.3656 & 0.1484 & 0.6260\\
    & & SVM & 0.2656 & 0.2476 & 0.2833 & 0.3770 & 0.3262 & 0.1330 & 0.5850\\
    \multirow{2}{*}{BL11} & \multirow{2}{*}{Hybrid (BL5 + BL10)} & $\psi_s$ & 0.2777 & 0.2612 & 0.3420 & 0.3869 & 0.3230 & 0.1648 & 0.6540\\
    & & $\psi_j$ & 0.2771 & 0.2615 & 0.3471 & 0.3902 & 0.3246 & 0.1716 & 0.6586\\
    \midrule
    \multicolumn{2}{c}{\textbf{Proposed approaches}} & & & & \\
    \midrule
    \multirow{2}{*}{} & \multirow{2}{*}{FRLM ($\gamma_H=0.8$)} & $\psi_s$ & 0.2956 & 0.2806 & 0.3435 & 0.4033 & 0.3443 & 0.1637 & 0.6922\\
    & & $\psi_j$ & $0.3075^{* \dagger \ddagger}$ & $0.2935^{* \dagger \ddagger}$ & $0.3498^{* \dagger \ddagger}$ & 0.4098 & 0.3541 & $0.1687^{* \ddagger}$ & 0.7098\\
    \multirow{2}{*}{} & \multirow{2}{*}{KDEFRLM ($\gamma_H=0.7$)} & $\psi_s$ & 0.3079 & 0.2852 & 0.3502 & 0.4361 & 0.3557 & 0.1729 & 0.6648\\
    & & $\psi_j$ & $\textbf{0.3199}^{*\dagger \ddagger}$ & $\textbf{0.2980}^{*\dagger \ddagger}$ & $\textbf{0.3645}^{* \dagger \ddagger}$ & $\textbf{0.4426}^{* \dagger \ddagger}$ & $\textbf{0.3623}^{*\dagger \ddagger}$ & $\textbf{0.1824}^{* \dagger \ddagger}$ & \textbf{0.7143}\\
    \bottomrule
\end{tabular}
}
\end{table}

\begin{table}[tp]
    \centering
    \caption{Words (stemmed version shown) whose vectors are closest, with respect to the \texttt{word2vec} (in-domain) embedded space, to a number of sample user assigned tags. \label{Table:W2VnearestTerms}}
    \begin{tabular}{ll}
    \toprule
    Tags &  Semantically close terms \\
    \midrule
    \texttt{beer} & \texttt{tap}, \texttt{draft}, \texttt{craft}, \texttt{microbrew}, \texttt{draught}, \texttt{\texttt{ipa}}, \texttt{pint}, \texttt{breweri}, \texttt{hefeweizen}, \texttt{delirium}, \texttt{lager}\\
    \texttt{beach} & \texttt{oceanfont}, \texttt{ocean}, \texttt{lifeguard}, \texttt{pier}, \texttt{beachfront}, \texttt{\texttt{sand}}, \texttt{surfer}, \texttt{pismo}, \texttt{murrel}, \texttt{seasid}, \texttt{vacat}\\
    \texttt{seafood} & \texttt{shellfish}, \texttt{oyster}, \texttt{crab}, \texttt{fish}, \texttt{shrimp}, \texttt{\texttt{triggerfish}}, \texttt{restaur}, \texttt{fisherman}, \texttt{swordfish}, \texttt{lobster}\\
    \texttt{pub} & \texttt{irish}, \texttt{gastropub}, \texttt{bar}, \texttt{fado}, \texttt{behan}, \texttt{\texttt{sport}}, \texttt{british}, \texttt{linkster}, \texttt{mccool}, \texttt{mcgregor}, \texttt{alehous}\\
    \texttt{family} & \texttt{oper}, \texttt{kid}, \texttt{pantuso}, \texttt{parent}, \texttt{niec}, \texttt{\texttt{children}}, \texttt{orient}, \texttt{sicilli}, \texttt{fun}, \texttt{home}, \texttt{yohan}\\
    \bottomrule
    \end{tabular}
\end{table}

%

\compactpara{IR approaches outperform collaborative/personal RecSys ones}
A common and sometimes very useful recommendation approach is BL6 (Most Popular K).
The poor performance of this method demonstrates that globally popular items (across a number of different users)
do not work well for the POI retrieval task. The likely reason for this being that personal choices in this case are more important.
The fact that BL7 (Profile Popular K) performs better than BL6 is consistent with this hypothesis of emphasizing personal preferences more than the global ones.

However, it can be seen that the effectiveness of this RecSys based approach (BL7) is inferior to that of BL1 (BM25), which is a standard IR based approach making use of the information in the set of tags from POIs with ratings higher than $0.8$. This shows that user ratings are more important than the popularity (relevance likelihood) of tags created by a user. A frequently used tag may have been used to create negative reviews by a user, in which case assigning importance to these tags may introduce noise into POI recommendation.

\hltIRJ{It is important to note that in the absence of rating from other users, it is not possible to directly apply standard collaborative filtering based RecSys approaches for this task of personalized POI recommendation.
Although for the sake of comparison, we employ a number of RecSys based approaches, it is to be noted that these approaches are particularly suitable for situations where there is a large volume of data from a number of users with similar interests.}

\compactpara{Unsupervised approaches outperform supervised ones}
Supervised approaches, namely BL8 (NeuMF) and BL9 (Bayesian), do not perform well. This is most likely due to the lack of sufficient training data. One of the problems of a supervised approach is that it involves learning a hard decision during the training phase to classify POIs as either relevant (with rating values higher than a threshold) or non-relevant (otherwise). The advantage of our proposed models is that they do not involve hard decision steps during any stage of their working procedure. Moreover, the primary advantage of an unsupervised approach is that it can work in situations where user preference data (for training) is sparse or even non-existent.

Another observation from the comparisons between KDEFRLM and the matrix factorization based technique BL8 (NeuMF) is that representation learning over words (which is trained on unannotated document collections available in large quantities) is more beneficial than the matrix factorization based joint representation learning of users and POIs in a latent space (which requires large quantities of training data in the form of user-item associations).
Moreover, the POI recommendation problem is more of a personalized retrieval problem, where information from other users (which is what happens in a user-item matrix factorization based collaborative setup such as NeuMF) may in fact turn out to be ineffective. This is also reinforced by our previously reported observation that `Profile Popular K' (personalized retrieval) outperformed the `Most Popular K' (collaborative retrieval).

%

\compactpara{A combination of content and tags is more effective than tag-matching alone}
The BL10 (Content + Tag matching) baseline involves a hard classification step, and then an aggregation over tag matching scores. In contrast, our proposed models (FRLM and KDEFRLM) do not involve hard selections for either documents or tags/terms, which means that they are able to selectively leverage the information from each source. 
 
\compactpara{Joint context modeling is better for modeling soft constraints}

From Table \ref{Table:ResultsCombined2_psi_l} and \ref{Table:ResultsCombined2_psi_j}, we observe that including trip-qualifier based information in the form of joint context ($\psi_j$) generally improves POI retrieval effectiveness, e.g. improvements are observed for RLM, NeuMF, etc. (compare the results between $\psi_j$ from Table \ref{Table:ResultsCombined2_psi_j} and $\psi_l$ from Table \ref{Table:ResultsCombined2_psi_l} for each method).
%
%
%
Standard approaches do not benefit much from the inclusion of the trip-qualifiers in the form of single-context driven scores, a plausible reason for which can be attributed to the fact that relevant single-context matches may not lead to the conjunctive relevance for the joint context. However, including even the single context based similarity scores as part of the query term weights in standard IR and RS (recommender system) approaches tends to improve the recall. E.g. effectiveness measures such as MAP and nDCG mostly improve at the cost of a decrease in nDCG@5 or P@5.
%

It can be seen that using soft constraint scores as a part of a model is usually more effective than a simple post-hoc combination of these scores with content matching scores (e.g. the relative improvements in FRLM as compared to that of Popular K or Content + Tag).

Additionally, in contrast to a parametric approach, such as SVM, the proposed similarity function $\psi_j$ (Equation \ref{eq:psi}) works better. This is because supervised approaches typically require
large quantities of training data to work well. Moreover, the SVM based approach of \cite{Aliannejadi_ToISJournal} did not take into account the semantic similarities between words to estimate the trip-qualifier based appropriateness. It is observed that computing similarities with the embedded word vectors turns out to be more effective.
%

Finally, it can be observed that the best results are obtained when the joint-context based similarity function is incorporated into the factored models. Incorporating term semantics in combination with the soft constraints (KDEFRLM with joint context modeling, $\psi_j$) further improves the results. 

\compactpara{Better precision-oriented and recall-oriented retrieval}
In addition to the aforementioned observations, we also note that KDEFRLM results in the best nDCG@5 value (a precision-oriented metric). This indicates that the model is able to retrieve documents assessed to be most relevant towards the top ranks in comparison to the other baselines. This is particularly beneficial from a user satisfaction point-of-view because a user does not need to scroll-down a list of retrieved suggestions to find her likely best matches.
It is particularly worth noting the considerable improvements
in the nDCG values (which is both a precision and a recall oriented measure) obtained with KDEFRLM. This indicates that KDEFRLM achieves high recall, in addition to achieving high precision. The high recall implies that, in real-life situations, it is also beneficial for patient users who are prepared to explore a list of recommendations to find a set of likely matching venues.

\begin{figure*}[htp]
\centering
\begin{subfigure}[b]{0.49\textwidth}
\centering
    \includegraphics[width=\textwidth]{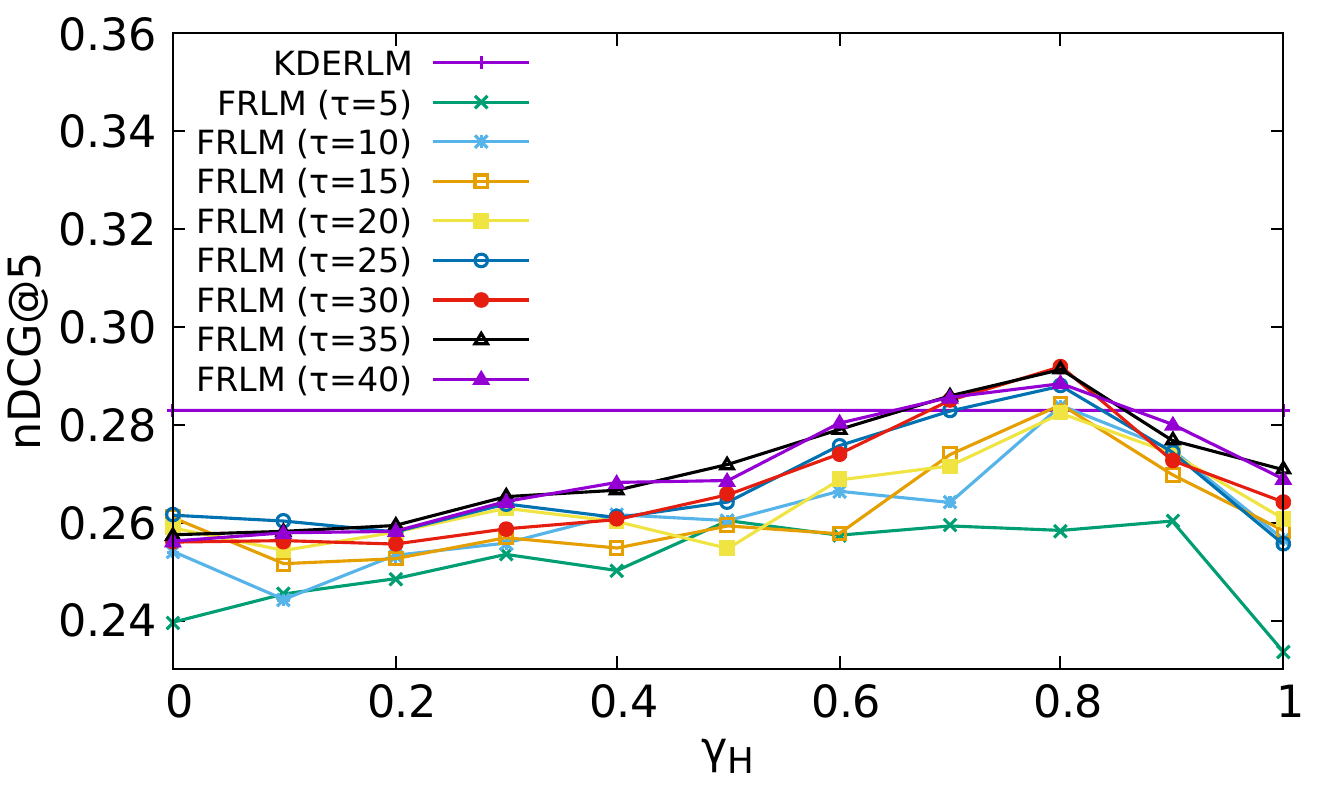} 
     \caption{FRLM nDCG@5 variations}
     \label{fig:ndcg5vary}
\end{subfigure}
\begin{subfigure}[b]{0.49\textwidth}
\centering
    \includegraphics[width=\textwidth]{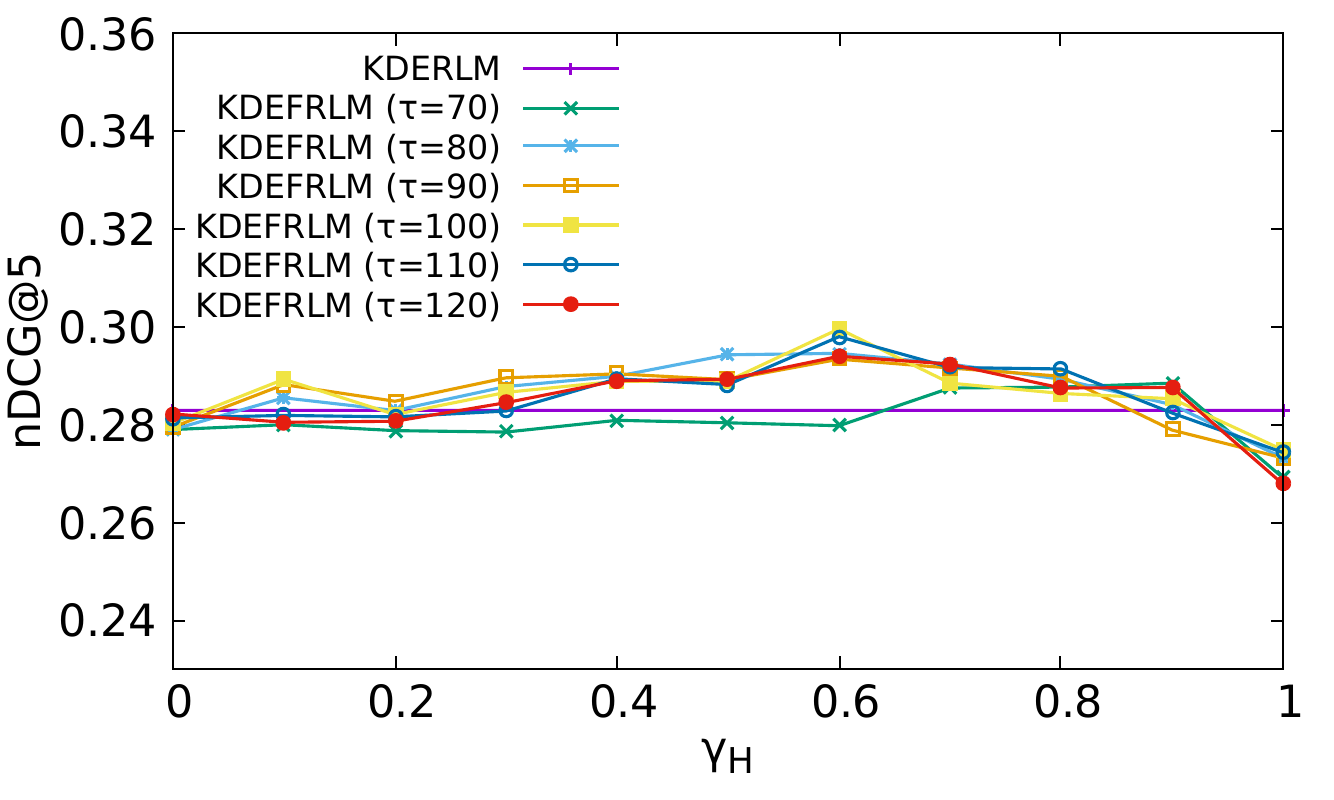} 
     \caption{KDEFRLM nDCG@5 variations}
     \label{fig:KDEFRLM_ndcg5vary}
\end{subfigure}
\quad
\begin{subfigure}[b]{0.49\textwidth}
\centering
     \includegraphics[width=\textwidth]{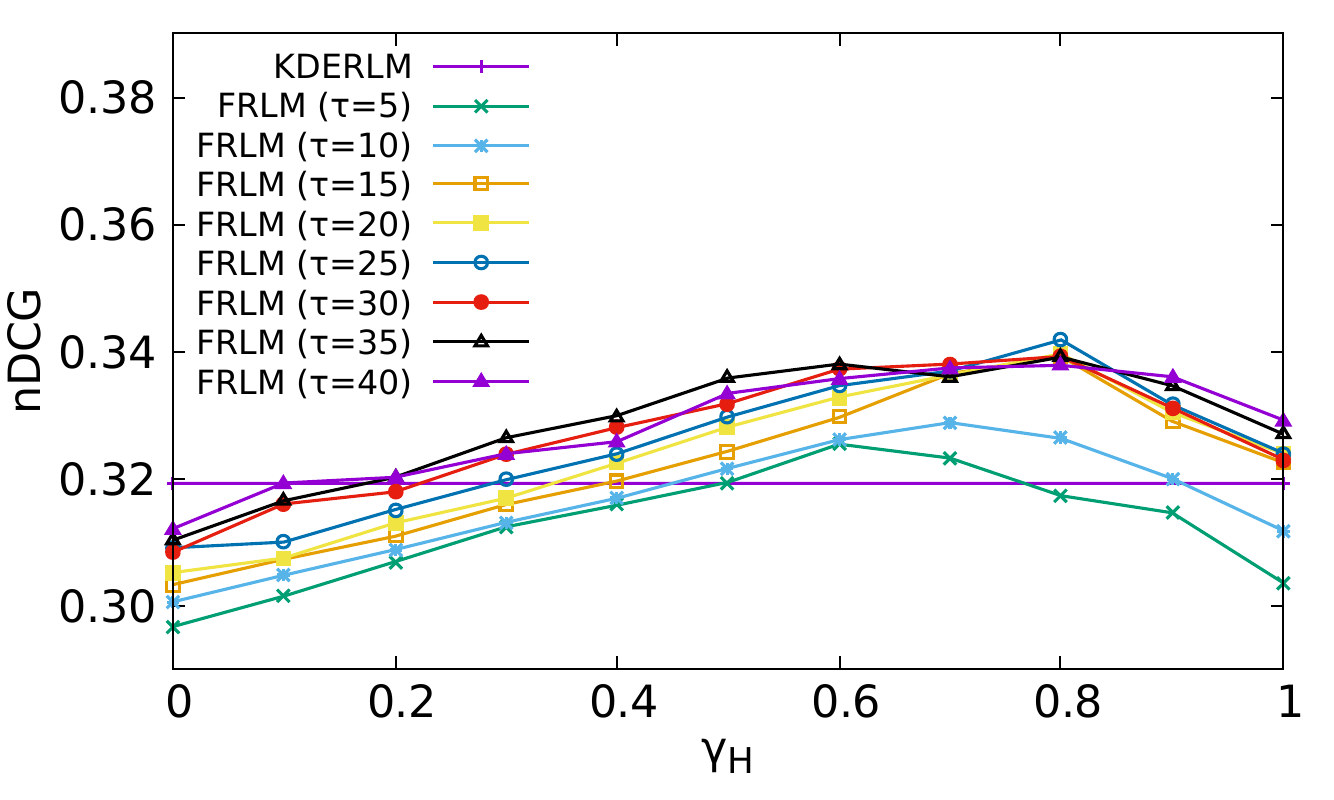} 
     \caption{FRLM nDCG variations}
     \label{fig:ndcgvary}
\end{subfigure}
\begin{subfigure}[b]{0.49\textwidth}
\centering
     \includegraphics[width=\textwidth]{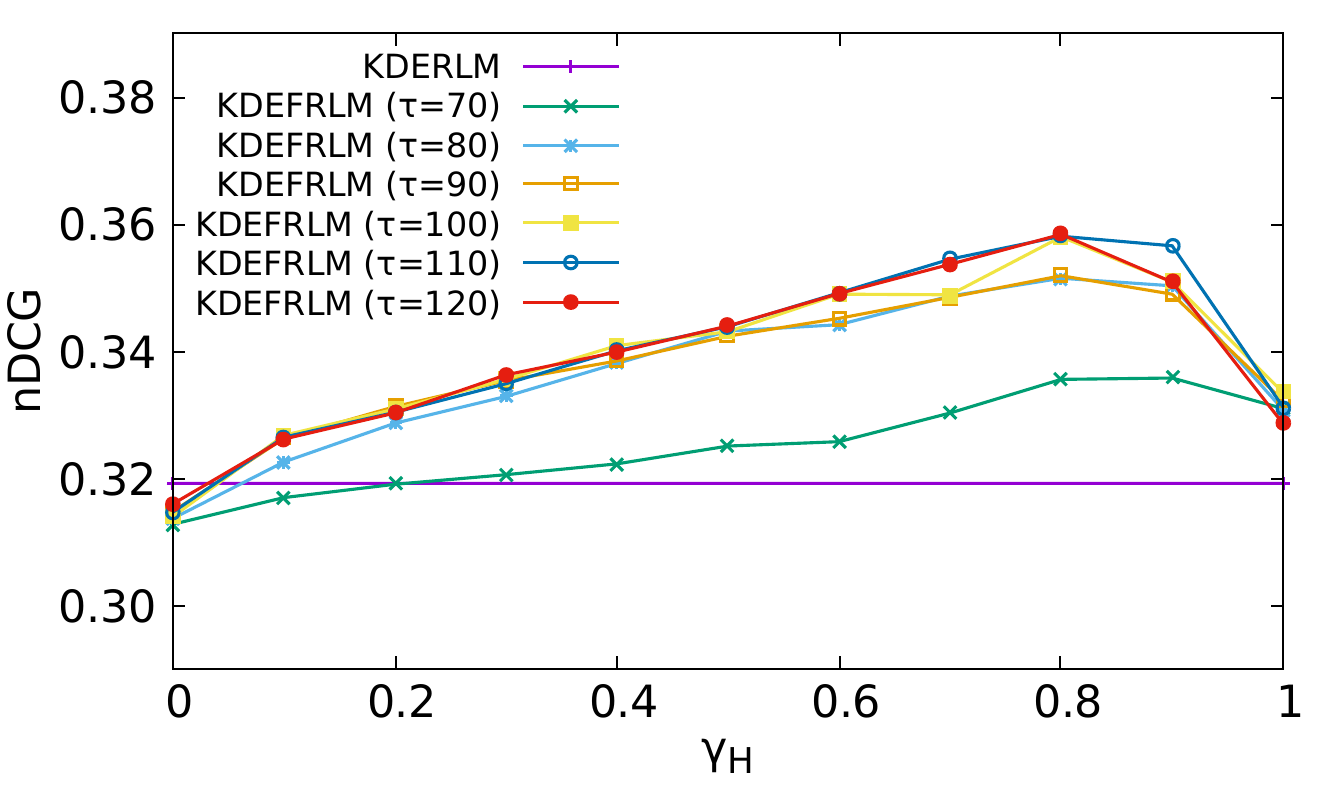} 
     \caption{KDEFRLM nDCG variations}
     \label{fig:KDEFRLM_ndcgvary}
\end{subfigure}
\quad
\begin{subfigure}[b]{0.49\textwidth}
\centering
    \includegraphics[width=\textwidth]{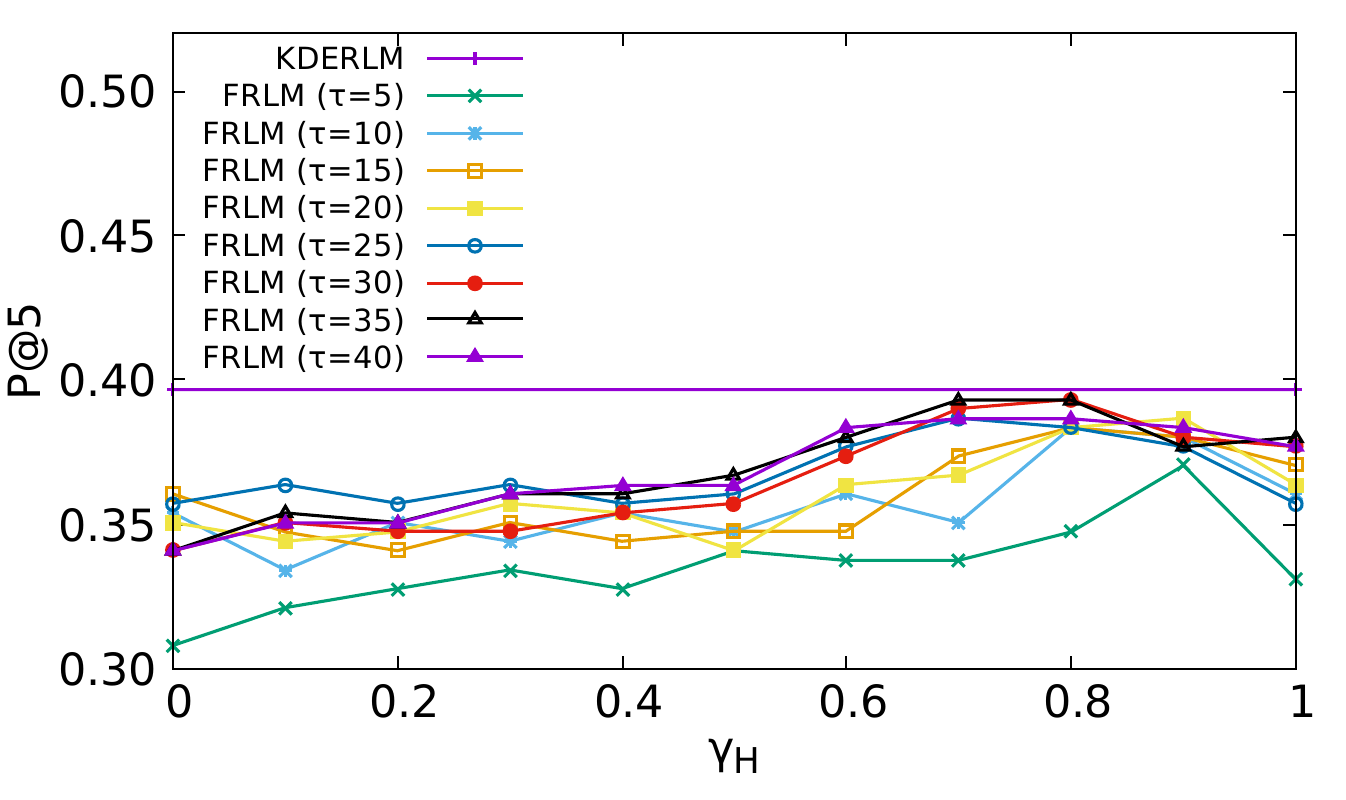} 
     \caption{FRLM P@5 variations}
     \label{fig:P5vary}
\end{subfigure}
\begin{subfigure}[b]{0.49\textwidth}
\centering
    \includegraphics[width=\textwidth]{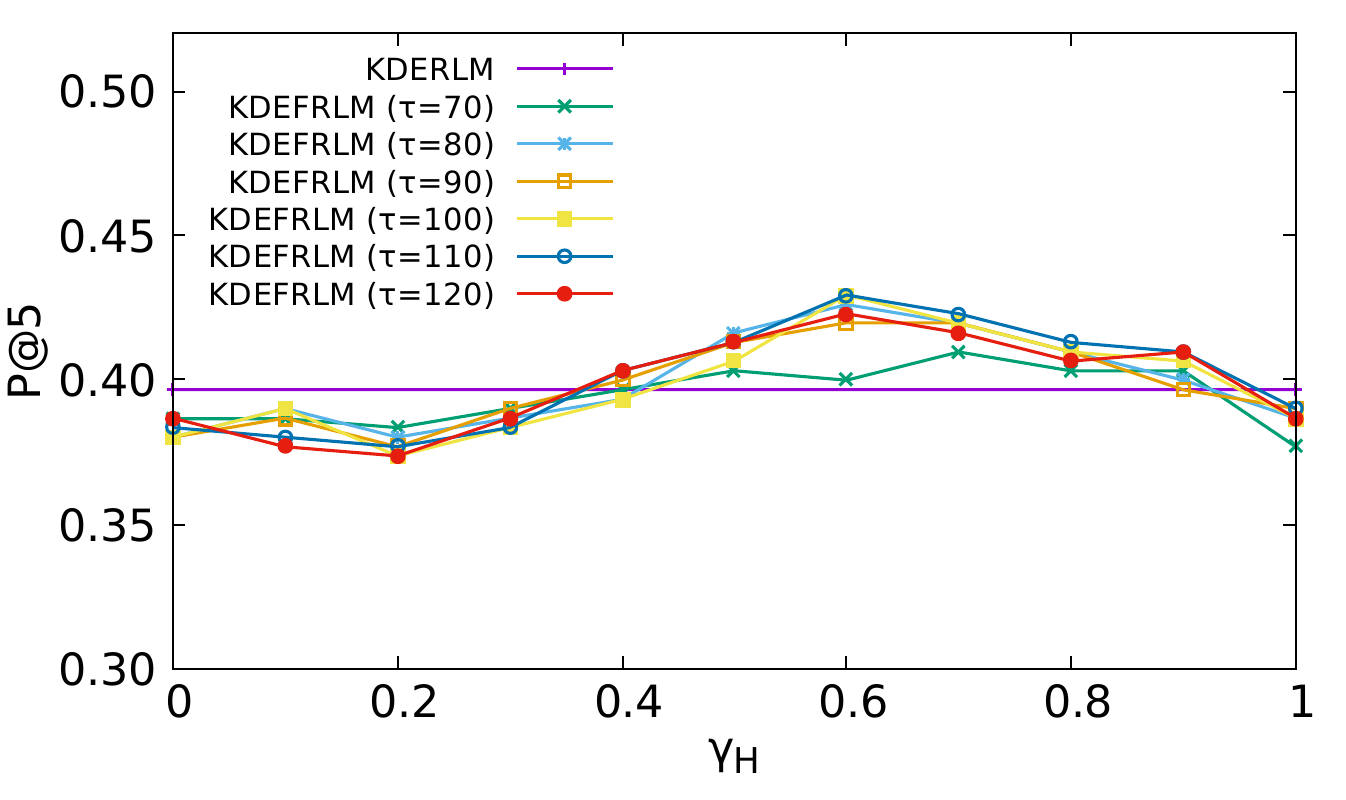} 
     \caption{KDEFRLM P@5 variations}
     \label{fig:KDEFRLM_P5vary}
\end{subfigure}
\begin{subfigure}[b]{0.49\textwidth}
\centering
    \includegraphics[width=\textwidth]{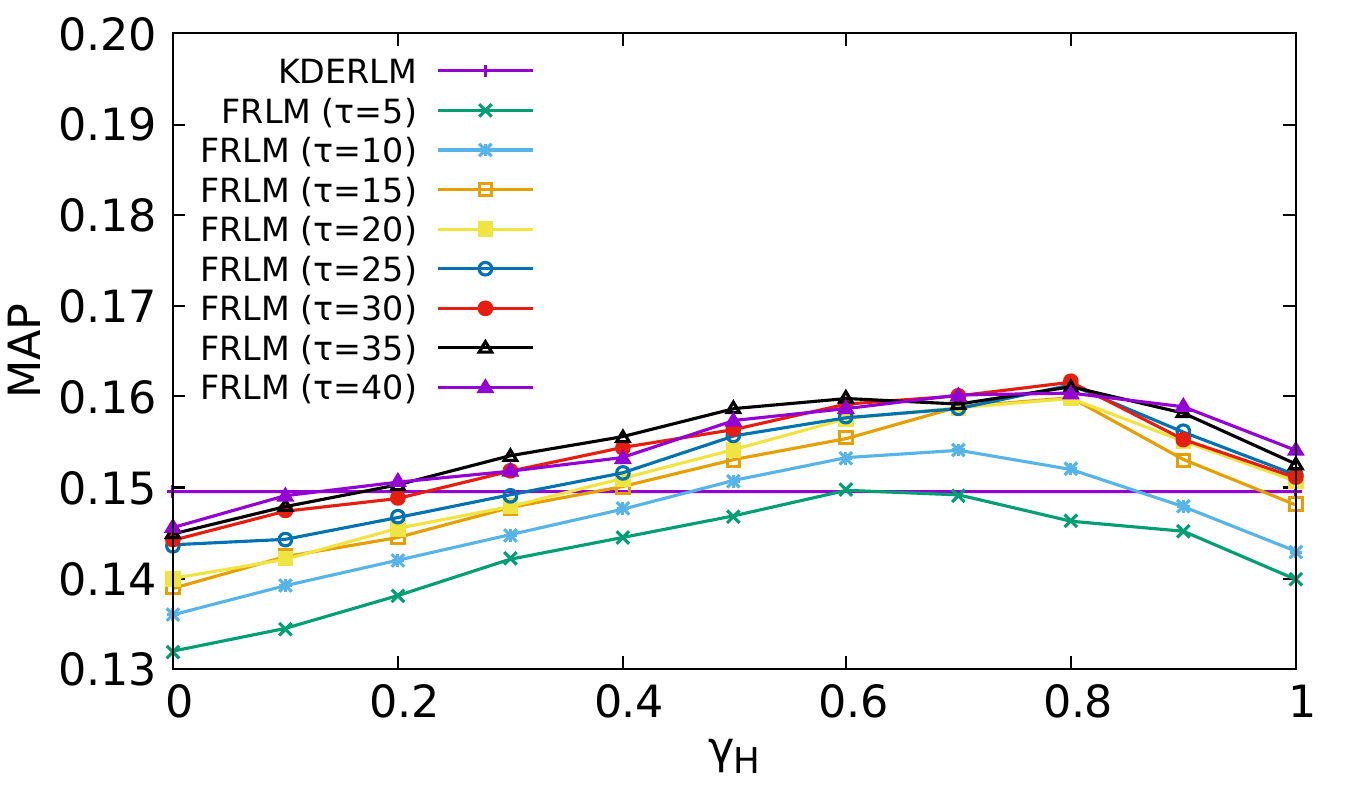} 
     \caption{FRLM MAP variations}
     \label{fig:MAPvary}
\end{subfigure}
\begin{subfigure}[b]{0.49\textwidth}
\centering
    \includegraphics[width=\textwidth]{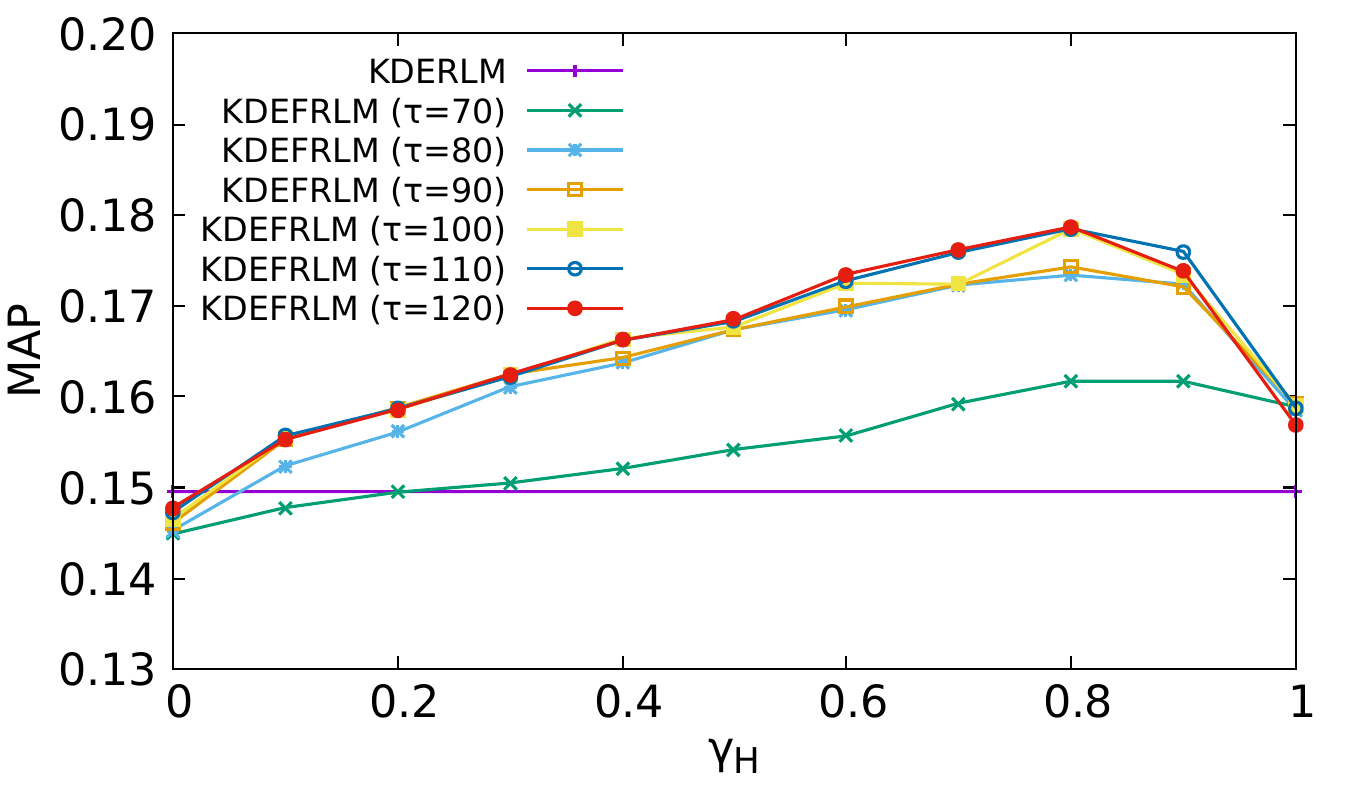} 
     \caption{KDEFRLM MAP variations}
     \label{fig:KDEFRLM_MAPvary}
\end{subfigure}
\caption{Effect of precision at top ranks (nDCG@5 and P@5) and recall (nDCG and MAP) with respect to changes in number of terms used in FRLM ($\psi_l$) Vs KDEFRLM ($\psi_l$) estimation ($\tau$) and the relative importance assigned to user profile information ($\gamma_H$). \hltIRJ{The performance of the strongest baseline (KDERLM) is shown with a straight line. Note that these are the results with location constraint only.}}
\label{fig:FRLM_variation}
\end{figure*}

\begin{figure*}[htp]
\centering
\begin{subfigure}[b]{0.49\textwidth}
\centering
    \includegraphics[width=\textwidth]{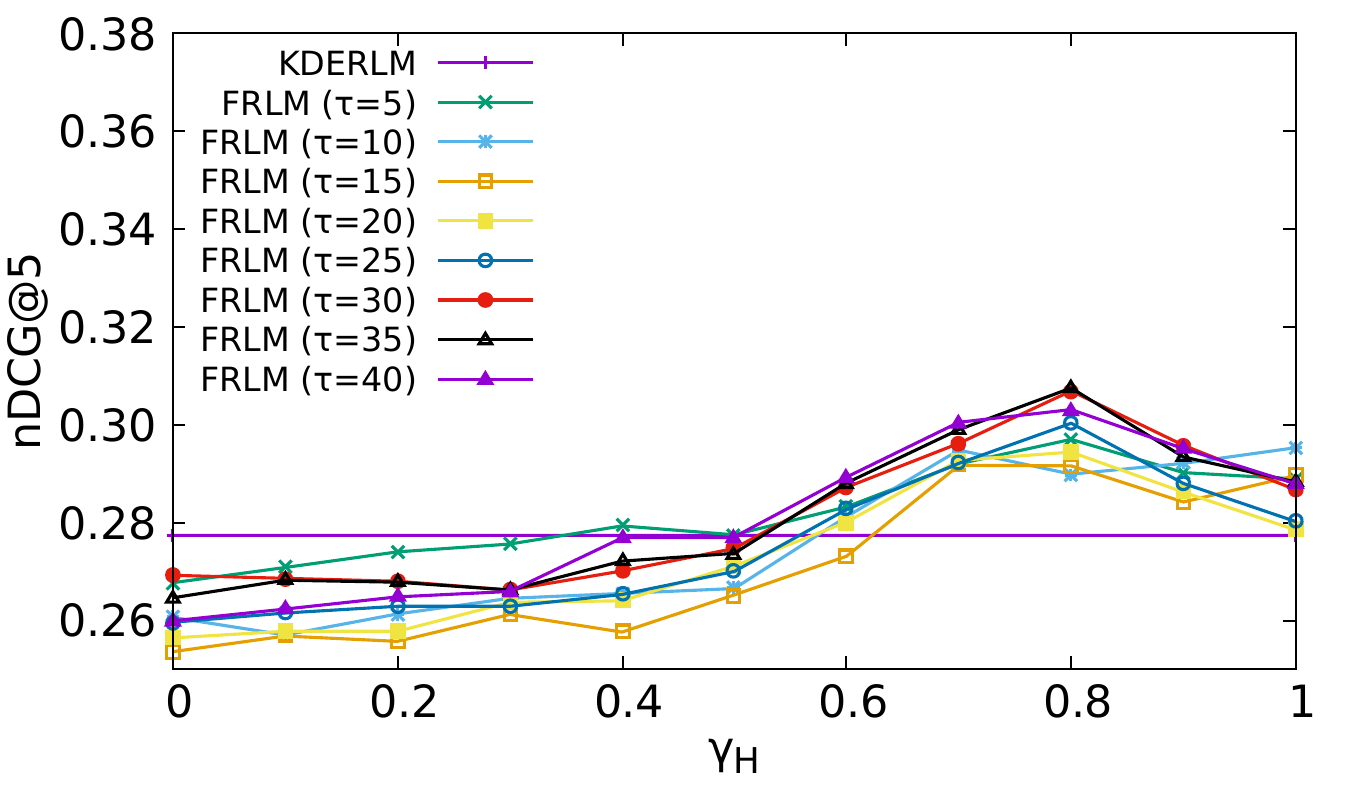} 
     \caption{FRLM nDCG@5 variations}
     \label{fig:ndcg5vary_withPsiJ}
\end{subfigure}
\begin{subfigure}[b]{0.49\textwidth}
\centering
    \includegraphics[width=\textwidth]{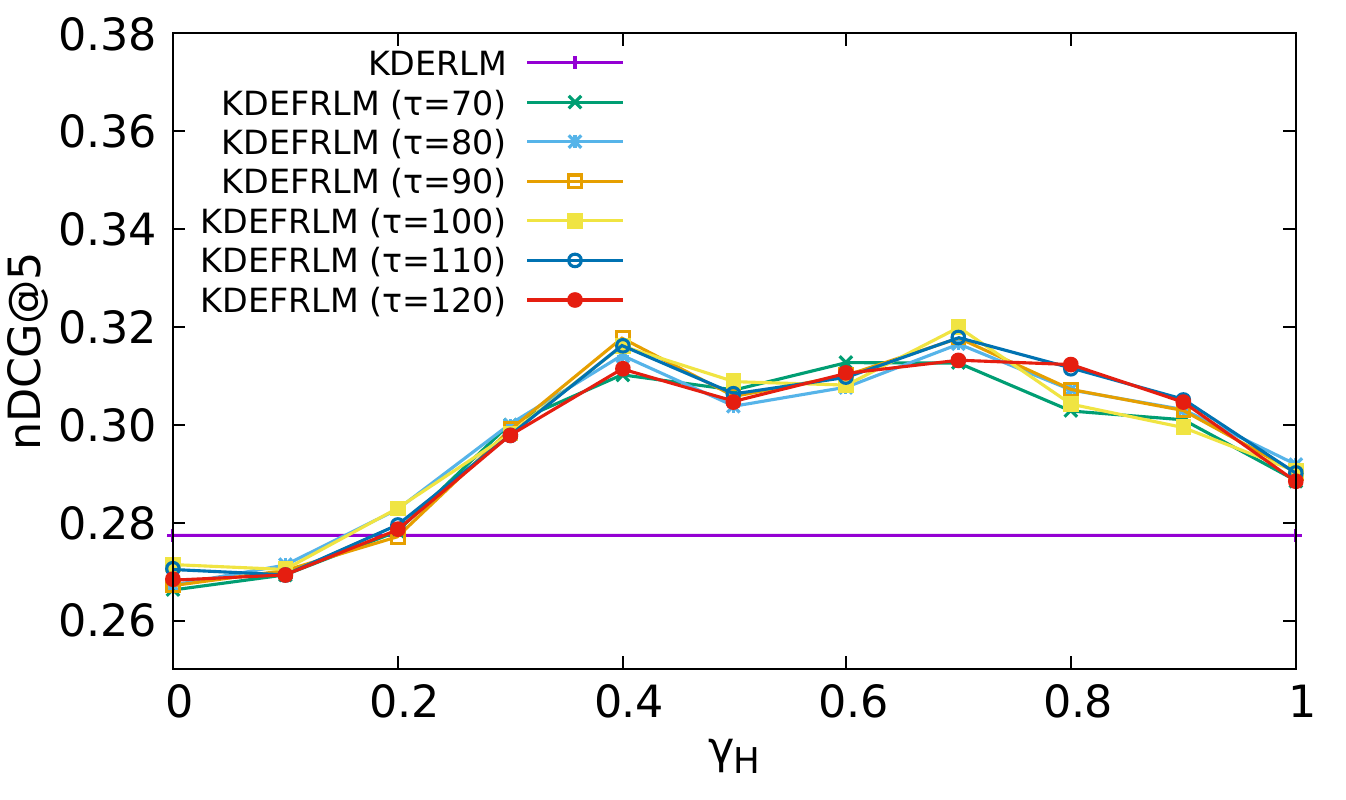} 
     \caption{KDEFRLM nDCG@5 variations}
     \label{fig:KDEFRLM_ndcg5vary_withPsiJ}
\end{subfigure}
\begin{subfigure}[b]{0.49\textwidth}
\centering
     \includegraphics[width=\textwidth]{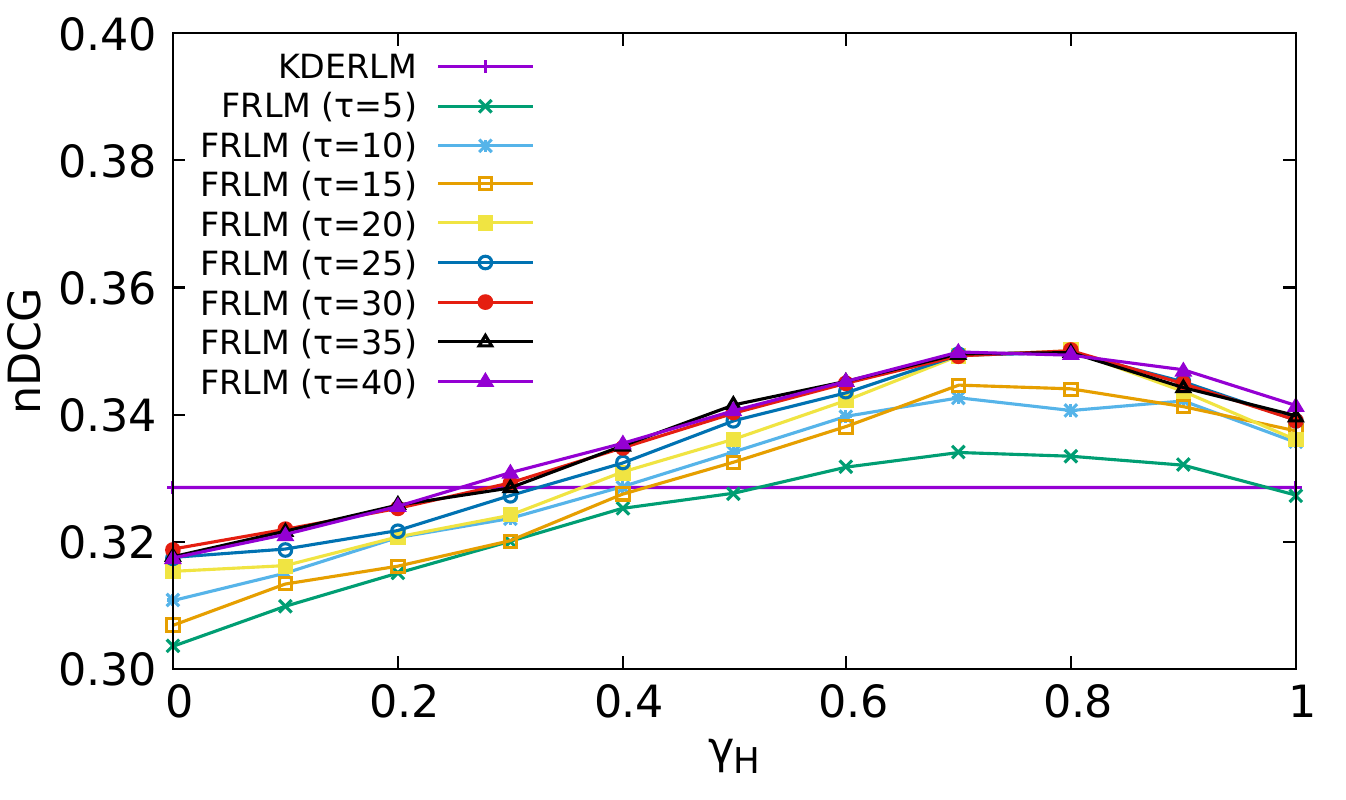} 
     \caption{FRLM nDCG variations}
     \label{fig:ndcgvary_withPsiJ}
\end{subfigure}
\begin{subfigure}[b]{0.49\textwidth}
\centering
     \includegraphics[width=\textwidth]{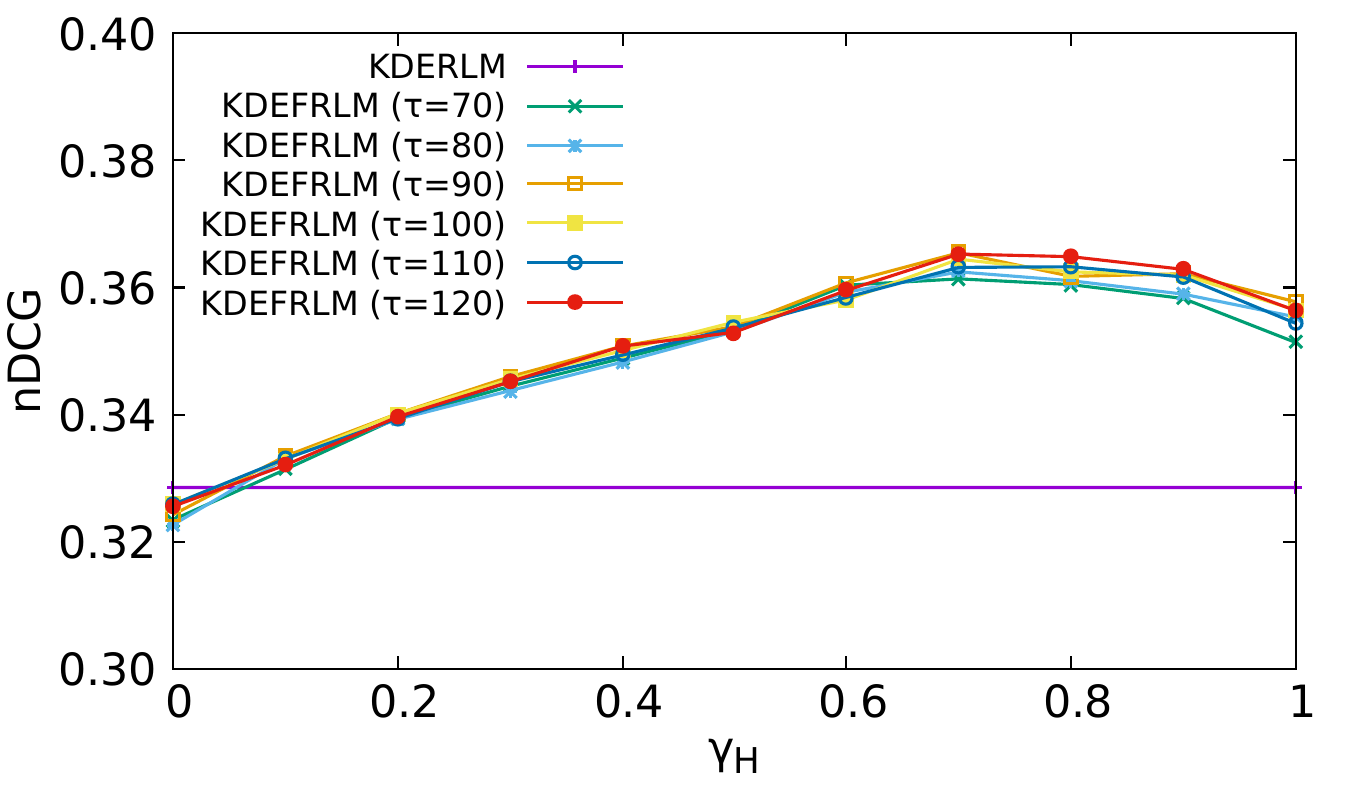} 
     \caption{KDEFRLM nDCG variations}
     \label{fig:KDEFRLM_ndcgvary_withPsiJ}
\end{subfigure}
\begin{subfigure}[b]{0.49\textwidth}
\centering
    \includegraphics[width=\textwidth]{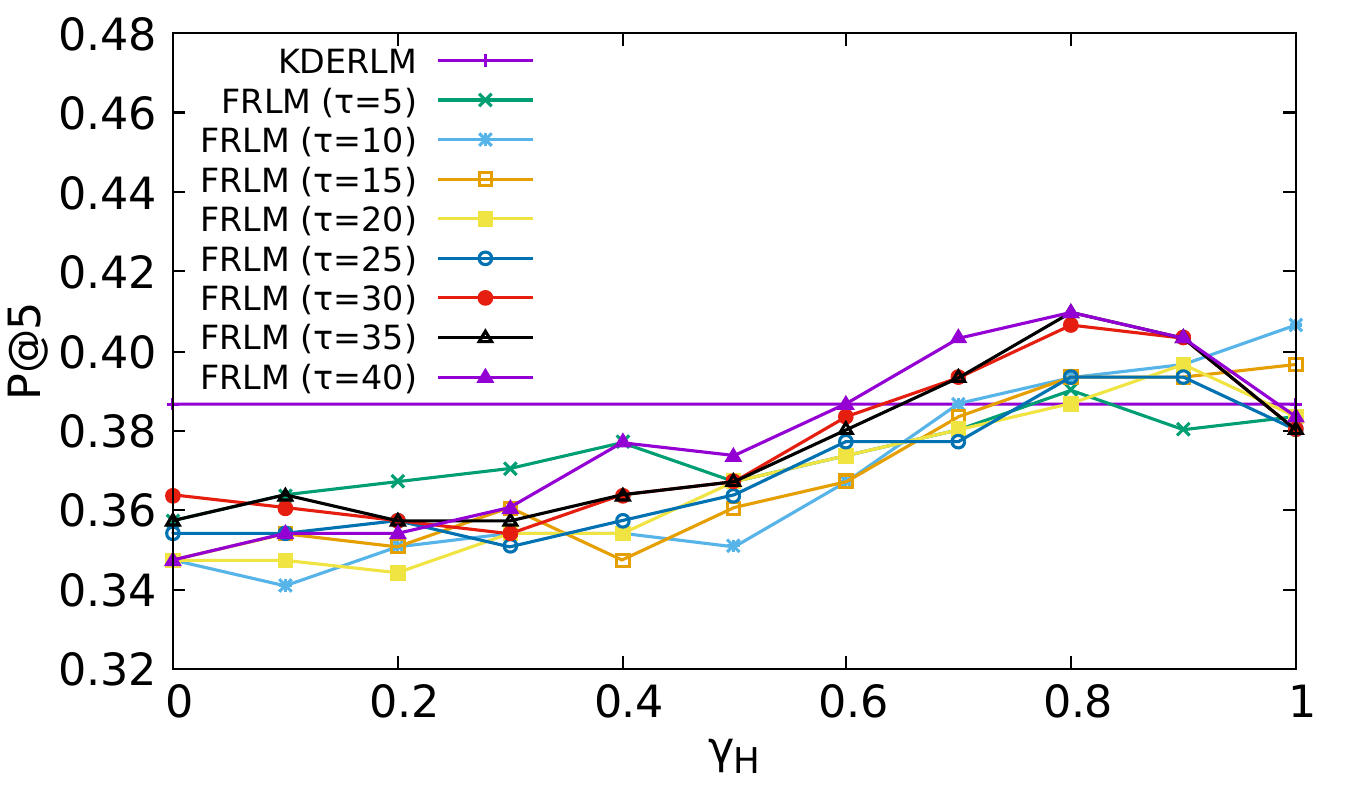} 
     \caption{FRLM P@5 variations}
     \label{fig:P5vary_withPsiJ}
\end{subfigure}
\begin{subfigure}[b]{0.49\textwidth}
\centering
    \includegraphics[width=\textwidth]{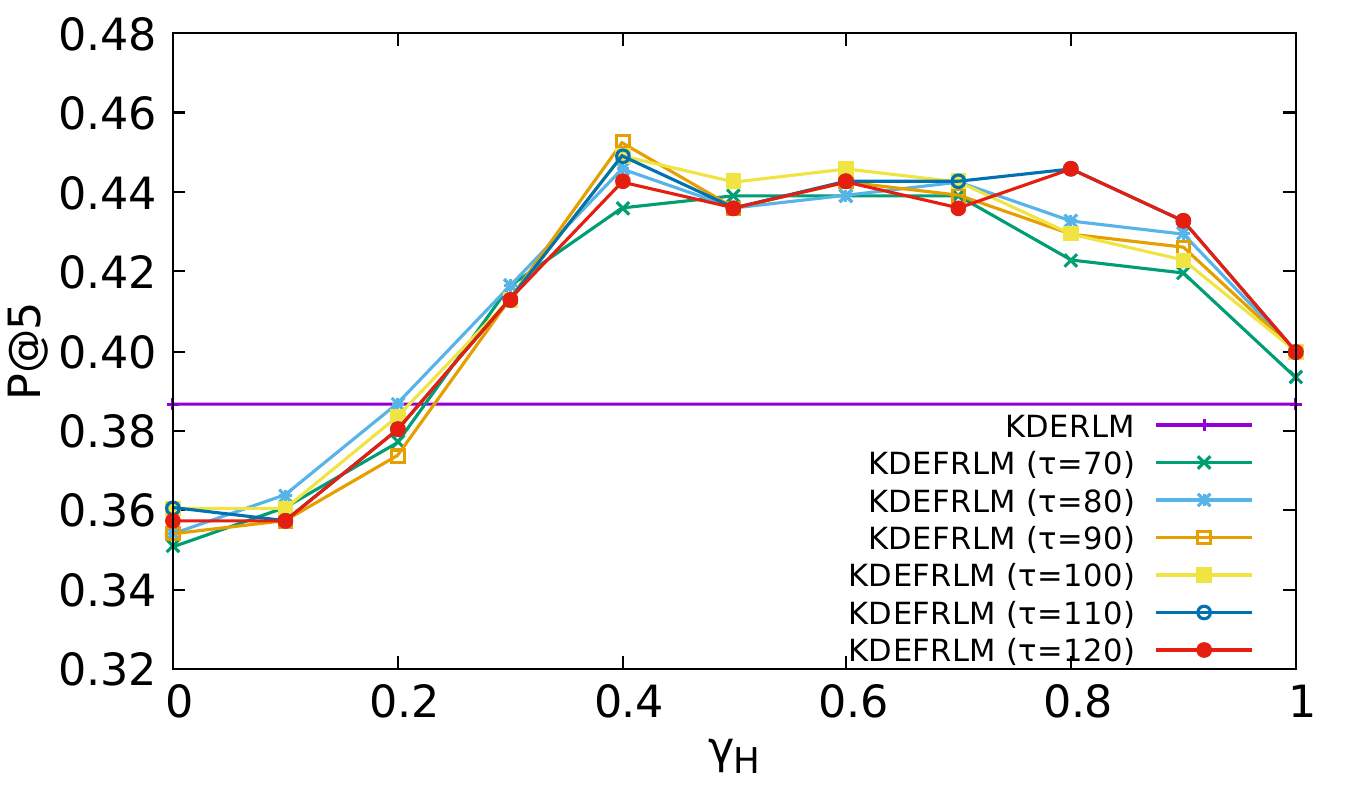} 
     \caption{KDEFRLM P@5 variations}
     \label{fig:KDEFRLM_P5vary_withPsiJ}
\end{subfigure}
\begin{subfigure}[b]{0.49\textwidth}
\centering
    \includegraphics[width=\textwidth]{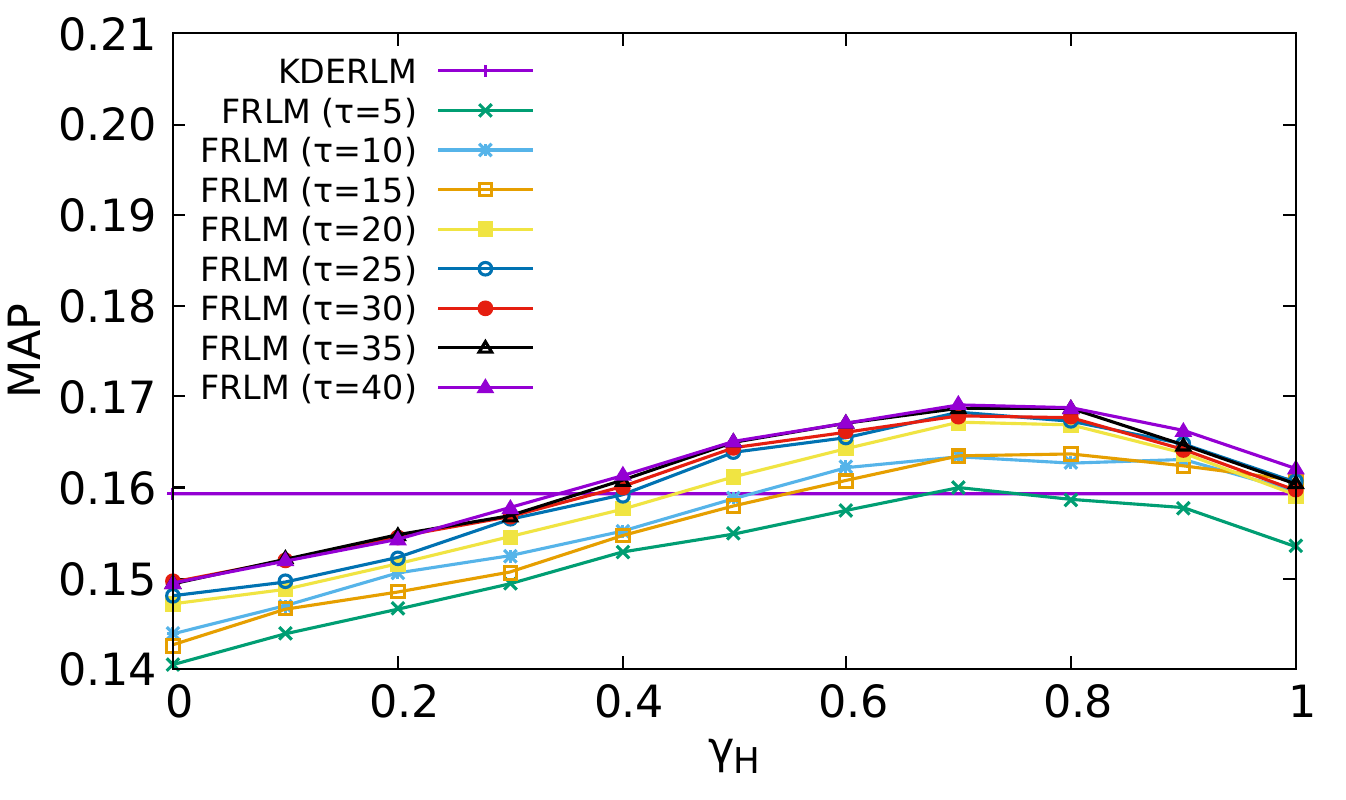} 
     \caption{FRLM MAP variations}
     \label{fig:MAPvary_withPsiJ}
\end{subfigure}
\begin{subfigure}[b]{0.49\textwidth}
\centering
    \includegraphics[width=\textwidth]{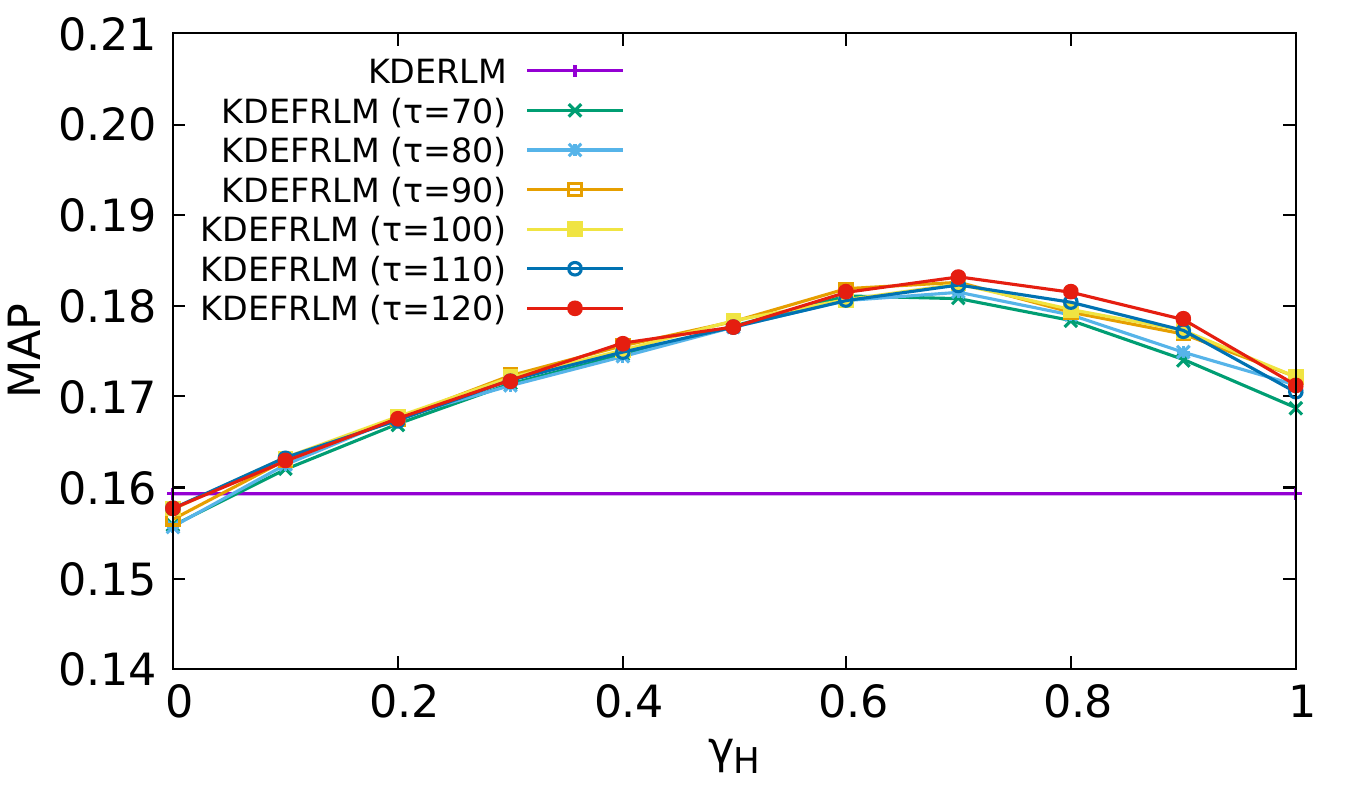} 
     \caption{KDEFRLM MAP variations}
     \label{fig:KDEFRLM_MAPvary_withPsiJ}
\end{subfigure}

\caption{Effect of precision at top ranks (nDCG@5 and P@5) and recall (nDCG and MAP) with respect to changes in number of terms used in FRLM ($\psi_j$) Vs KDEFRLM ($\psi_j$) estimation ($\tau$) and the relative importance assigned to user profile information ($\gamma_H$). \hltIRJ{The performance of the strongest baseline (KDERLM) is shown with a straight line. Note that these are the results with location and soft constraints.}}
\label{fig:FRLM_variation_withPsiJ}
\end{figure*}

\begin{figure*}[htp]
\centering
\begin{subfigure}[t]{0.49\textwidth}
\centering
    \includegraphics[width=\textwidth]{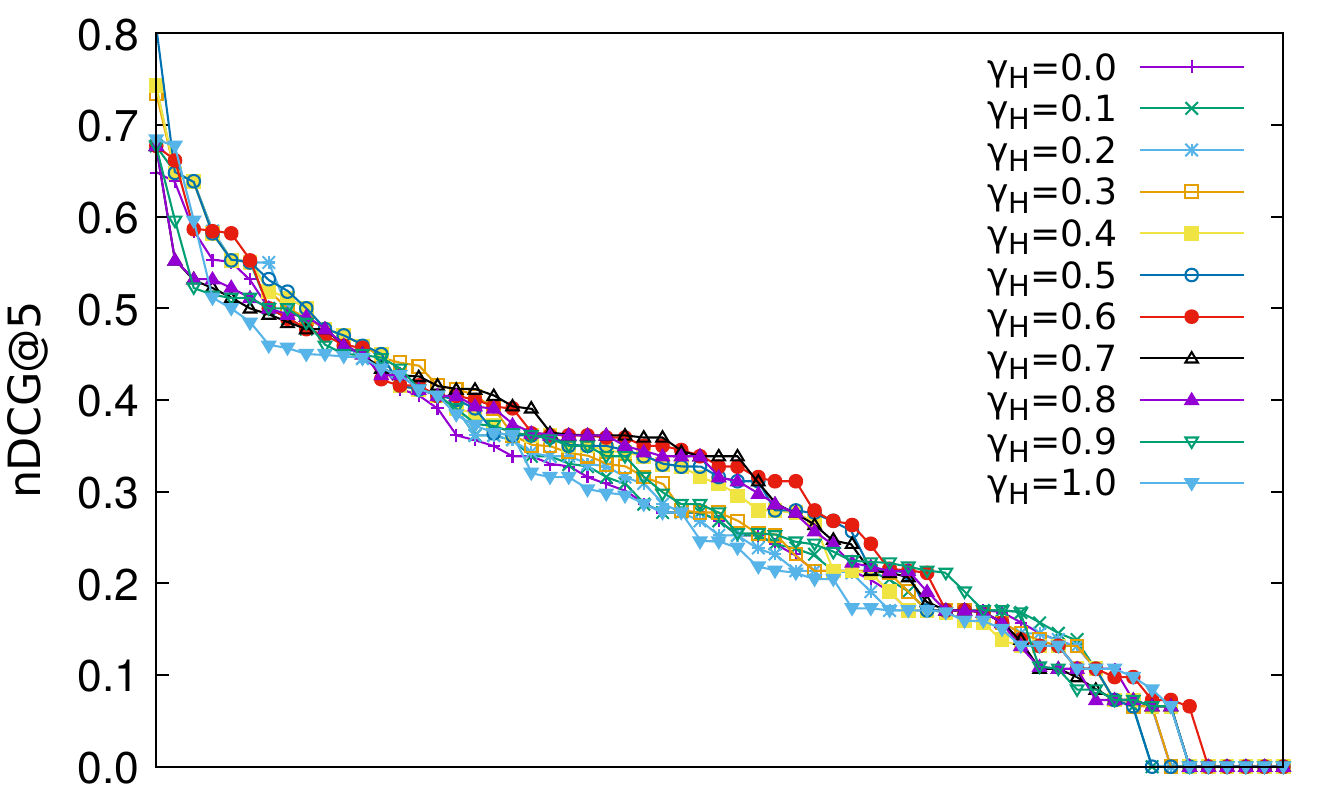} 
     \caption{KDEFRLM ($\psi_l$)}
     \label{fig:KDEFRLM_psi_l_gammaH_sensitivity_nDCG5}
\end{subfigure}
\begin{subfigure}[t]{0.49\textwidth}
\centering
     \includegraphics[width=\textwidth]{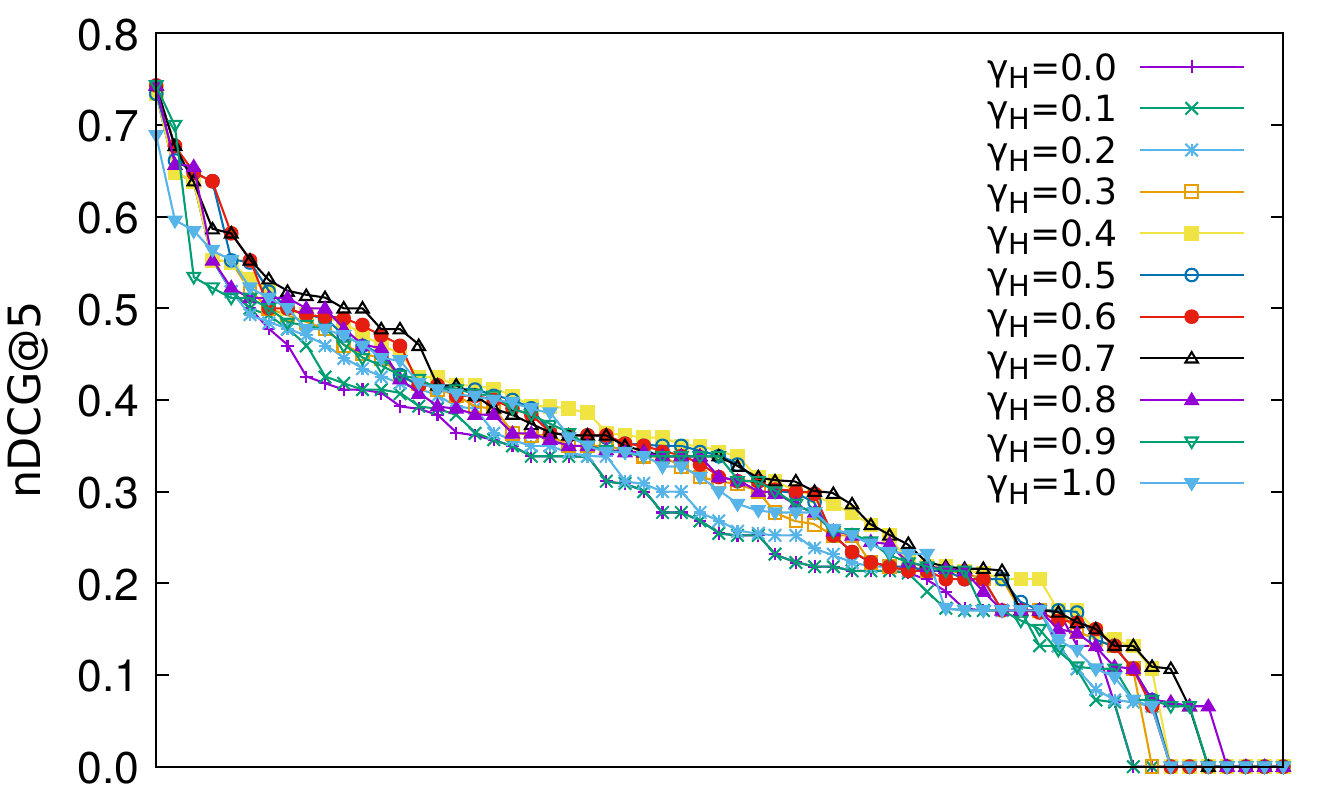} 
     \caption{KDEFRLM ($\psi_j$)}
     \label{fig:KDEFRLM_psi_j_gammaH_sensitivity_nDCG5}
\end{subfigure}
\begin{subfigure}[t]{0.49\textwidth}
\centering
    \includegraphics[width=\textwidth]{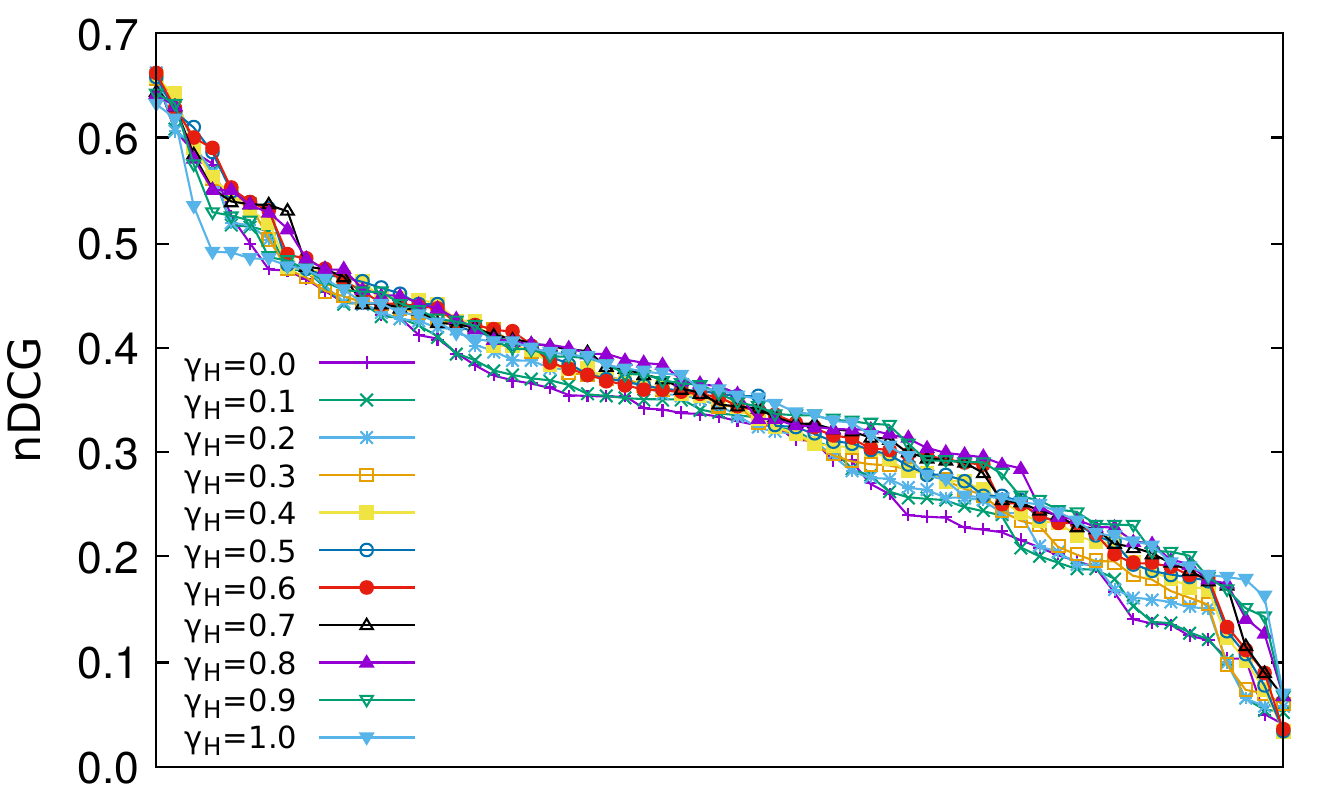} 
     \caption{KDEFRLM ($\psi_l$)}
     \label{fig:KDEFRLM_psi_l_gammaH_sensitivity_nDCG}
\end{subfigure}
\begin{subfigure}[t]{0.49\textwidth}
\centering
     \includegraphics[width=\textwidth]{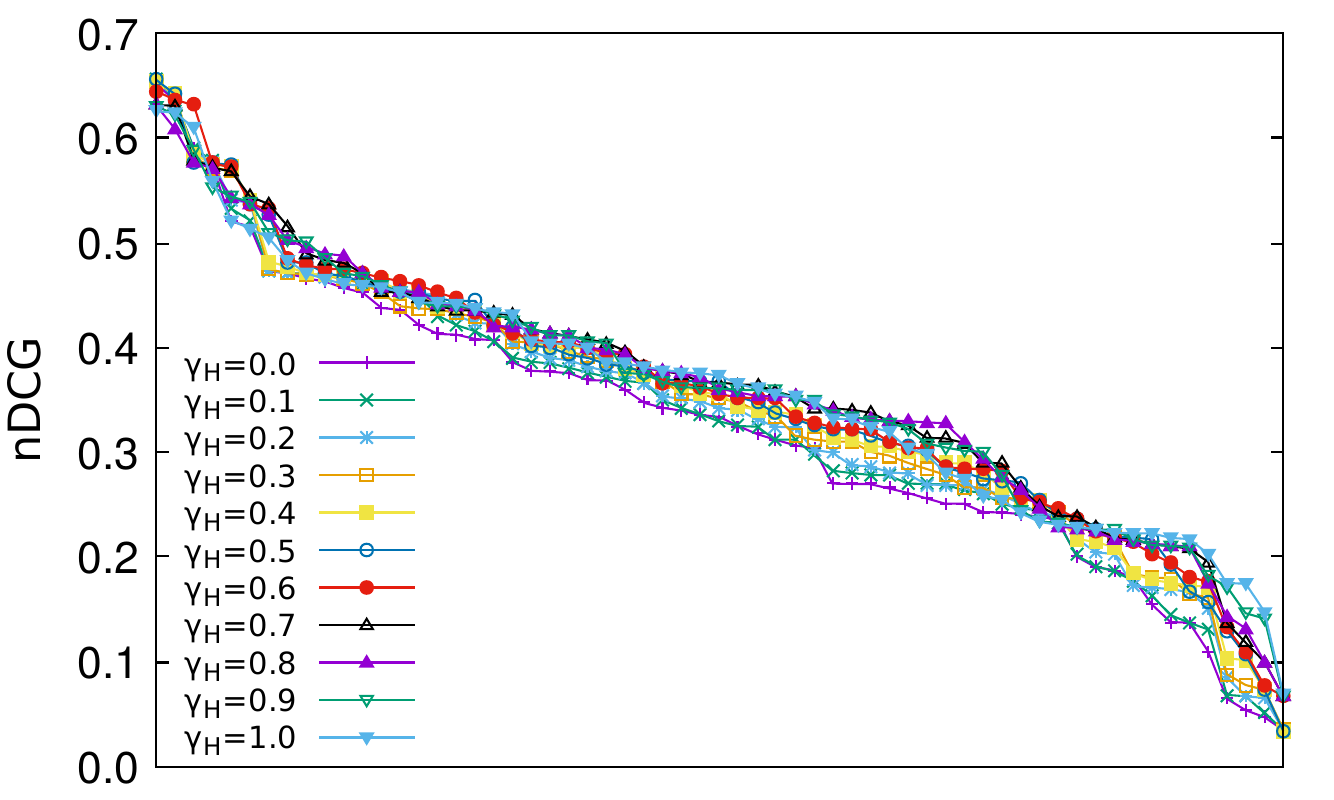} 
     \caption{KDEFRLM ($\psi_j$)}
     \label{fig:KDEFRLM_psi_j_gammaH_sensitivity_nDCG}
\end{subfigure}
\caption{\hlt{User profile based performance of KDEFRLM ($\psi_l$), and KDEFRLM ($\psi_j$) with respect to nDCG@5, and nDCG while varying $\gamma_H$. User profiles (queries) are sorted in decreasing order of per profile nDCG@5 values. Larger the area under a curve, better the overall performance for that specific $\gamma_H$ value.}}
\label{fig:FRLM_Vs_KDE_gammaH_sensitivity}
\end{figure*}

\subsection{Sensitivity Analysis}

\subsubsection{Parameter Sensitivity}
Table \ref{Table:ResultsCombined2_psi_l} and \ref{Table:ResultsCombined2_psi_j} present a summary of the best results obtained with each method (parameters optimized with grid-search for the nDCG@5 metric).
In order to investigate a more wide spectrum of results,
we now investigate the effects of varying the parameter $\gamma_H$ (i.e. the trade-off between exploration and exploitation) on the performance of FRLM and KDEFRLM. To obtain the sensitivity results, we set the value of $M$ (number of top-retrieved documents to consider for the RLM feedback) to \hltd{$5$, and $2$ for FRLM, and KDEFRLM, respectively.}

Figure \ref{fig:FRLM_variation} shows the sensitivity of FRLM versus KDEFRLM (measured with nDCG@5, nDCG, P@5 and MAP) with respect to the number of feedback terms, $\tau$, used to define the term-weight distribution, and the relative importance of the user's historical context with respect to the POIs in the current context, i.e. $\gamma_H$.
%
An interesting observation is that factored relevance models perform best with a balanced trade-off between exploitation and exploration. In particular, the optimal results for FRLM (both in terms of precision oriented measure nDCG@5 and recall oriented measure nDCG) are achieved when $\gamma_H=0.8$. Moreover, the effectiveness of FRLM degrades with the user profile history only ($\gamma_H=1$), which indicates that the history information itself is likely to contain noise in the form of topical diversity. This also demonstrates the benefit of selectively extracting chunks of information from the preference history that are contextually appropriate in the present state. We observe a similar trend in the kernel density based extension of FRLM.

\hltIRJ{While Figure} \ref{fig:FRLM_variation} \hltIRJ{shows the sensitivity of FRLM versus KDEFRLM in location-only setting ($\psi_l$), Figure }\ref{fig:FRLM_variation_withPsiJ} \hltIRJ{shows the sensitivity of FRLM versus KDEFRLM in location + soft constraints setting ($\psi_j$).}
It is also observed that a very small or a very large number of feedback terms tends to decrease retrieval performance. While the former case is unable to sufficiently capture the relevant semantics required to match the user profile with the present context, the latter introduces noise from pieces of profile that are not contextually relevant to the present state in the estimated FRLM or KDEFRLM distributions.
While FRLM achieves the optimal results with a smaller number of expansion terms, $\tau$, KDEFRLM being a more complex model requires a larger number of expansion terms to perform well. However, KDEFRLM is less sensitive to the number of terms and hence a more robust model as compared to FRLM \hltIRJ{(observed in both Figures }\ref{fig:FRLM_variation} and \ref{fig:FRLM_variation_withPsiJ}).

\subsubsection{Per User-Profile Sensitivity Analysis}

\hlt{Instead of a relatively simple approach of employing a constant value for the linear combination parameter $\gamma_H$}, in this section we investigate if individually choosing the values of this parameter based on the user profiles (queries) can lead to better results. In particular, we conduct a grid-based exploration of the parameter $\gamma_H$ for each query separately.
%
Fig. \ref{fig:FRLM_Vs_KDE_gammaH_sensitivity} \hltd{plots the distribution of the nDCG@5, and nDCG values (arranged in a decreasing order) as obtained for a total of 11 possible choices of $\gamma_H$ for each user profile (query). A larger area under the curve corresponding to a particular value of $\gamma_H$ indicates that for a higher number of queries this value of $\gamma_H$ yields optimal retrieval effectiveness. A large number of cross-over points (as seen from Fig.} \ref{fig:FRLM_Vs_KDE_gammaH_sensitivity}\hltd{) of the distribution lines indicates that, generally speaking, different queries achieve optimal results with different values of the exploration-exploitation parameter. This in turn indicates that for some user profiles it is better to rely to a greater degree on the historical preferences (exploitation) whereas for some other ones it is better to allow provision for more exploration into the POI descriptors. Our results also suggests that automatically estimating the value of the exploration-exploitation trade-off can potentially improve results further. This we leave as a future exercise.}

\begin{table}[tp]
\centering
\caption{\hlt{Variations in IR effectiveness} with respect to different choices of pre-trained (out-domain) embedding vectors in comparison to skipgram trained on the target collection (i.e. the result of Table \ref{Table:ResultsCombined2_psi_l} and \ref{Table:ResultsCombined2_psi_j} which is reproduced in this table for convenience) \hltIRJ{ in location only setting ($\psi_l$).}
}
\label{Table:ResultsOutdomain_psi_l}
\tabfitpagew{
\begin{tabular}
    {@{}l@{\ \ }l@{\ }l@{\ }l@{\ }l@{\ }l@{\ }lllllr@{}}
    \toprule
    & & & Context & \multicolumn{3}{c}{Graded Evaluation Metrics} & \multicolumn{4}{c}{Binary Evaluation Metrics}\\
    \cmidrule(r){5-7}
    \cmidrule(l){8-11}
    {Method} & Embedding & Domain & $(\psi)$ & nDCG@5 & nDCG@10 & nDCG & P@5 & P@10 & MAP & MRR \\
    \midrule
    \multirow{4}{*}{KDEFRLM} & word2vec & In & $\psi_l$ & \textbf{0.2996} & \textbf{0.2868} & 0.3490 & 0.4295 & 0.3656 & 0.1725 & \textbf{0.6553}\\
    & word2vec & Out & $\psi_l$ & 0.2993 & 0.2840 & 0.3485 & \textbf{0.4328} & 0.3656 & 0.1735 & 0.6367\\
    & GloVe & Out & $\psi_l$ & 0.2963 & 0.2863 & \textbf{0.3533} & 0.4262 & \textbf{0.3705} & \textbf{0.1750} & 0.6451\\
    & RoBERTa & Out & $\psi_l$ & 0.2971 & 0.2842 & 0.3521 & \textbf{0.4328} & \textbf{0.3705} & 0.1736 & 0.6393\\
    \bottomrule
\end{tabular}
}
\end{table}

\begin{table}[tp]
\centering
\caption{\hlt{Variations in IR effectiveness} with respect to different choices of pre-trained (out-domain) embedding vectors in comparison to skipgram trained on the target collection (i.e. the result of Table \ref{Table:ResultsCombined2_psi_l} and \ref{Table:ResultsCombined2_psi_j} which is reproduced in this table for convenience)\hltIRJ{ in location + \emph{soft} constraints setting ($\psi_j$).}
}
\label{Table:ResultsOutdomain_psi_j}
\tabfitpagew{
\begin{tabular}
    {@{}l@{\ \ }l@{\ }l@{\ }l@{\ }l@{\ }l@{\ }lllllr@{}}
    \toprule
    & & & Context & \multicolumn{3}{c}{Graded Evaluation Metrics} & \multicolumn{4}{c}{Binary Evaluation Metrics}\\
    \cmidrule(r){5-7}
    \cmidrule(l){8-11}
    {Method} & Embedding & Domain & $(\psi)$ & nDCG@5 & nDCG@10 & nDCG & P@5 & P@10 & MAP & MRR \\
    \midrule
    \multirow{8}{*}{KDEFRLM} & \multirow{2}{*}{word2vec} & \multirow{2}{*}{In} & $\psi_s$ & 0.3079 & 0.2852 & 0.3502 & 0.4361 & 0.3557 & 0.1729 & 0.6648\\
    & & & $\psi_j$ & \textbf{0.3199} & 0.2980 & 0.3645 & \textbf{0.4426} & 0.3623 & 0.1824 & \textbf{0.7143}\\
    & \multirow{2}{*}{word2vec} & \multirow{2}{*}{Out} & $\psi_s$ & 0.2990 & 0.2879 & 0.3539 & 0.4393 & 0.3672 & 0.1758 & 0.6424\\
    & & & $\psi_j$ & 0.3044 & 0.2959 & \textbf{0.3653} & \textbf{0.4426} & 0.3787 & \textbf{0.1844} & 0.6582\\
    & \multirow{2}{*}{GloVe} & \multirow{2}{*}{Out} & $\psi_s$ & 0.3107 & \textbf{0.3027} & 0.3623 & 0.4361 & \textbf{0.3852} & 0.1802 & 0.6880\\
    & & & $\psi_j$ & 0.3064 & 0.2959 & 0.3638 & 0.4328 & 0.3754 & 0.1815 & 0.6803\\
    & \multirow{2}{*}{RoBERTa} & \multirow{2}{*}{Out} & $\psi_s$ & 0.2904 & 0.2894 & 0.3558 & 0.4230 & 0.3754 & 0.1767 & 0.6506\\
    & & & $\psi_j$ & 0.2957 & 0.2855 & 0.3572 & 0.4295 & 0.3721 & 0.1785 & 0.6423\\
    \bottomrule
\end{tabular}
}
\end{table}

\subsection{Investigating variations in Embedding methodology}

The KDEFRLM results reported in Table \ref{Table:ResultsCombined2_psi_l} and \ref{Table:ResultsCombined2_psi_j} used word embeddings trained on the domain specific target collection. In this section, we investigate whether alternative embedding choices (e.g. using a larger and more general external corpora) leads to improvements in results as reported in previous studies \cite{embeddingsForIR_Doi_CIKM2018}.
\hltd{In particular, we investigate three different choices for the embedding algorithm, namely \texttt{word2vec}} \cite{Word2vec_NIPS2013}\hltd{, \texttt{GloVe}} \cite{GloVe} \hltd{and \texttt{RoBERTa}} \cite{RoBERTa}\hltd{, the latter being a context embedding model employing a transformer-based architecture to learn a masked language model.}
\hltIRJ{Recall from Section }\ref{embedding}\hltIRJ{ that `In-domain' refers to the setup when the word vectors were trained on the target corpus (POI description), whereas `out-domain' refers to the use of pre-trained word embeddings.}
%

\hlt{In the KDEFRLM framework of Equation} \ref{eq:kdeflm_withPsi}\hlt{, we provide as inputs pre-trained word vectors} instead of word vectors trained on the target collection. While the \texttt{word2vec} (skipgram) and the \texttt{GloVe} vectors are both 300 dimensional (trained respectively on GoogleNews and CommonCrawl), the \texttt{RoBERTa} vectors for each word is $768$ dimensional. To obtain vector representations of stemmed words we follow the methodology of \cite{embeddingsForIR_Doi_CIKM2018} \hltd{which involves first partitioning words into equivalence classes of identical stemmed representations and then consider the average vector of each class as the vector representation of the stem. To illustrate with an example, the vector representation of a word, such as `comput' in the target collection is given by the average over vectors for words in the pre-trained vocabulary, such as `computing', `computer' etc.}

\hlt{For location only retrieval ($\psi_l$), the best retrieval results are obtained} (as measured with the precision oriented metrics - nDCG@5 and nDCG@10) with \texttt{word2vec} embedding trained on the target (TREC-CS) corpus itself (Table \ref{Table:ResultsOutdomain_psi_l}). A likely reason for this is that training on the target collection is possibly able to capture domain specific term semantics in a better way than a generic (domain-agnostic) representation.
An interesting observation is that the pre-trained \texttt{GloVe} model (trained on an external data of generic web-pages, namely CommonCrawl) leads to
KDEFRLM's best performance with respect to the other metrics such as nDCG, P@10, and MAP.
One advantage of using pre-trained vectors on external large corpora is that it offers a generalized way of learning word semantics, and may turn out to be effective when the target corpus is not large enough to learn adequate semantic relationships between words.

\hlt{For multi-contextual retrieval ($\psi_s$ or $\psi_j$ from Table \ref{Table:ResultsOutdomain_psi_j}), again the best performance is achieved by KDEFRLM} (with respect to the metrics - nDCG@5, and P@5) for the \texttt{word2vec} (In-domain) setting. Here, \texttt{word2vec} (Out-domain) embedding also turns out to be effective in contributing to the best performance of KDEFRLM with respect to the metrics - nDCG, P@5, and MAP. It turns out that context embedding (specifically \texttt{RoBERTa}) is not as effective as the shallow word-level embedding methodologies, specially for the \emph{soft} contextual constraints case.
\hltIRJ{A likely reason for this is that context embeddings have been shown to be particularly suited for downstream NLP tasks} \cite{BERT_devlin}\hltIRJ{, and these may not be well suited to model the lexical semantics across words} \cite{DG_PR2020_VariableLenEmbeddings}.
\hltIRJ{In addition, an interesting observation is that our proposed model is fairly stable and not overly dependent on the choice of embeddings, which yields similar performance with \texttt{word2vec}, \texttt{GloVe}, and \texttt{RoBERTa} embeddings.}

\section{Conclusions and Future Work} \label{sec:conclusions}
This paper proposes a generic relevance feedback based framework for contextual POI recommendation.
We gradually build up the overall framework of our proposed model, in increasing order of complexity, by incorporating the following three aspects:
i) \emph{factored relevance modeling} to achieve an optimal combination of
the user's \emph{preference history in past contexts (exploitation)}, and the \emph{relevance of top-retrieved POIs in the user's current context (exploration)},
ii) \emph{word semantics} in the form of \emph{kernel density} estimates computed by distances between embedded word vectors of the user tags and the POI descriptors,
and
iii) \emph{soft (trip-qualifier) constraints} modeled by leveraging information from a knowledge-base of manually assessed contextual appropriateness of words under the pretext of a given context category, either in separate or in joint forms.

Our experiments on the TREC-CS 2016 dataset show that even the simplest of our proposed class of models (i.e. the factored relevance model) outperforms a range of different baseline approaches involving standard IR or recommender system methodologies. Moreover, it is shown that the additional generalizations in our proposed framework, i.e. including word semantics and information from a knowledge base, further improves POI effectiveness.


In future, we aim to extend our experiments to include additional information as a part of a user's context, e.g. the
fine-grained location of a user in terms of GPS coordinates (instead of simply a city name), environmental context (e.g., if a user is indoors or outdoors), traveling amenities context (e.g. if the user has private transport) etc. \hltIRJ{One possible way to obtain such additional contextual information} would be to apply simulation techniques seeking to model the travel behaviour of simulated user agents.
\hltIRJ{It may then be possible to employ an end-to-end joint learning to rank framework for estimating a POI's relevance by leveraging information from both content and other fine-grained contextual information, such as the relative distance between a user and the POI.}
%

\begin{acknowledgements}
This work was supported by the ADAPT Centre for Digital Content Technology, funded under the Science Foundation Ireland Research Centres Programme (Grant 13/RC/2106) and is co-funded under the European Regional Development Fund.
\end{acknowledgements}

%
%


\bibliographystyle{spbasic}

\bibliography{refs}


\end{document}